\documentclass{sig-alternate}
\usepackage{algorithm}
\usepackage{algorithmic}
\usepackage{amsmath}
\usepackage{amssymb}
\usepackage{amsthmWithoutProof}
\usepackage{array}
\usepackage{balance}  % for  \balance command ON LAST PAGE  (see vldb_sample.tex) 
\usepackage{color}
\usepackage{listings}
\usepackage{url}
\usepackage{paralist}
\usepackage{tabularx}
\usepackage{times}

\definecolor{darkgreen}{rgb}{0,.5,.2}
\definecolor{gray}{rgb}{.5,.5,.5}

{\vspace{5mm}\begin{sloppypar}
  \noindent\hrulefill\, BEGIN \,\hrulefill
  \end{sloppypar}
}%
{\begin{sloppypar}
  \noindent\hrulefill\, END   \,\hrulefill
  \end{sloppypar}\vspace{5mm}
}

\newcommand{\todo}[1]{}

\newcommand{\Todo}[2]{#1}
\newcounter{question}

\newcounter{comment}

\newcommand{\removable}[1]{#1}

\lstdefinelanguage{PseudoCode}{%
  morekeywords={FUNCTION,CALL,LET,RETURN,IF,THEN,ELSE,AND,OR,NOT,WHILE,DO,FOR,EACH},%
  sensitive=true,%
  morecomment=[l]--,%
  morecomment=[s]{/*}{*/},%
  morestring=[d]',%
  morestring=[d]"%
}[keywords,comments,strings]

\lstdefinelanguage{SPARQL}{%
  morekeywords={BASE,PREFIX,SELECT,DISTINCT,DESCRIBE,CONSTRUCT,%
        ASK,FROM,NAMED,WHERE,ORDER,BY,ASC,DESC,LIMIT,OFFSET,OPTIONAL,%
        GRAPH,UNION,FILTER,STR,LANG,DATATYPE,REGEX,BOUND,ISURI,ISBLANK,
        ISLITERAL,TRUE,FALSE},%
  sensitive=false,%
  morecomment=[l]--,%
  morecomment=[s]{/*}{*/},%
  morestring=[d]',%
  morestring=[d]"%
}[keywords,comments,strings]

\lstdefinelanguage{N3}{%
  morekeywords={@prefix},%
  sensitive=true,%
}[keywords,comments,strings]
\newcommand{\code}[1]{\texttt{#1}}

\newcommand{\rdfTermInFigure}[1]{\begin{scriptsize}#1\end{scriptsize}}
\newcommand{\rdfTermInText}[1]{\begin{footnotesize}\textnormal{#1}\end{footnotesize}}

\newcommand{\definedTerm}[1]{\textbf{#1}}
\newcommand{\propositionMain}[1]{#1} %\textbf{#1}}

\theoremstyle{plain}
\newtheorem{theorem}{Theorem}
\newtheorem{proposition}{Proposition}
\newtheorem{lemma}{Lemma}
\newtheorem{corollary}{Corollary}
\newtheorem{fact}{Fact}
\newtheorem{example}{Example}

\theoremstyle{definition}
\newtheorem{definition}{Definition}

\newenvironment{myproof}[1]{%

	\noindent \textbf{Proof #1.}%
}{%
	\qed \vspace{2ex}

}

\newcommand{\webproblem}[4]{%
	\noindent
	\begin{tabularx}{\columnwidth}{|p{17mm} >{\normalsize}X|} \hline
		\textbf{Problem:} &
			\problemName{#1}
		\\ Web Input: &
			#2
		\\ Ordin.~Input: &
			#3
		\\ Question: &
			#4
		\\ \hline
	\end{tabularx}%
}

\newcommand{\problemName}[1]{\textsc{#1}}

\newcommand{\IFF}{iff } %{\textsl{iff}}
\newcommand{\LDdoc}{LD document}
\newcommand{\ID}{identifier}
\newcommand{\triple}{data triple}
\newcommand{\TP}{triple pattern}
\newcommand{\AEtask}{AE~task}

\newcommand{\true}{\mathrm{true}} % the symbol for the boolean value of TRUE
\newcommand{\false}{\mathrm{false}} % the symbol for the boolean value of FALSE
\newcommand{\fctDomName}{\mathrm{dom}}
\newcommand{\fctDom}[1]{\fctDomName(#1)}

\newcommand{\symAllURIs}{\mathcal{I}} % the symbol for the set of all URIs / identifiers
\newcommand{\symAllLiterals}{\mathcal{L}} % the symbol for the set of all literals
\newcommand{\symAllVariables}{\mathcal{V}} % the symbol for the set of all query variables
\newcommand{\symURI}{id} % the symbol for a URI / identifier
\newcommand{\symBQP}{B} % the symbol for a BQP
\newcommand{\symBQPpart}{P} % the symbol for a BQP that is part of another BQP
\newcommand{\symSeedURIs}{S} % the symbol for a set of seed URIs / identifiers
\newcommand{\ReachPartScBW}[4]{#4^{(#1,#3)}_{#2}} % the symbol for the (#1,#2,#3)-reachable part of #4
\newcommand{\queryFctBSc}[3]{\mathcal{Q}^{#1,#2}_{#3}}
\newcommand{\queryRsltBScW}[4]{\queryFctBSc{#1}{#2}{#3}\!\bigl(#4\bigr)}

\newcommand{\fctIDsName}{\mathrm{ids}} % the function that maps triples/TPs/BGPs to the set of IDs mentioned in them

\newcommand{\fctVarsName}{\mathrm{vars}} % the function that maps TPs/BGPs to the set of query variables mentioned in them

\newcommand{\WoD}{W\!} % the symbol for a Web of Data
\newcommand{\dataFct}{data} % the function that maps LD docs to the RDF triples in them
\newcommand{\adocFct}{adoc} % the function that maps HTTP URIs to authoritative LD docs
\newcommand{\fctAllDataName}{\mathrm{AllData}} % the function that maps a Web of Data to the set of all data in that Web
\newcommand{\fctEncName}{\mathrm{enc}}
\newcommand{\fctEnc}[1]{\fctEncName(#1)}

\newcommand{\Reach}[1]{#1_\mathfrak{R}} % the symbol for a component #1 of a reachable part of a Web
\newcommand{\DiscPartXX}[1]{#1_\mathfrak{D}} % the symbol for a discovered part of #1
\newcommand{\NewDiscPartXX}[1]{#1_\mathfrak{D}'} % the symbol for a newly discovered part of #1
\newcommand{\Disc}[1]{#1_\mathfrak{D}} % the symbol for a component #1 of a discovered part of a Web
\newcommand{\NewDisc}[1]{#1_\mathfrak{D}'} % the symbol for a component #1 of a newly discovered part of a Web
\newcommand{\expMWD}[3]{exp_{#1}^{#2}\bigl( #3 \bigr)}
\newcommand{\ExpMDeltaMW}[2]{\Delta^{\!#2}(#1)}
\newcommand{\Aug}[4]{aug_{#1,#2}^{#4}\bigl( #3 \bigr)} % the symbol for the (#1,#2)-augmentation of #3 in #4
\newcommand{\AllOpen}[2]{Open\bigl( #1,#2 \bigr)} % the symbol for all open tasks for execution state (#1,#2)
\newcommand{\cNone}{c_\mathsf{None}}
\newcommand{\cAll}{c_\mathsf{All}}
\newcommand{\cMatch}{c_\mathsf{Match}}

\renewcommand{\emptyset}{\varnothing}
\definecolor{darkgray}{rgb}{.5,.5,.5}

\begin{document}
\clubpenalty=10000 
\widowpenalty = 10000

\title{Foundations of Traversal Based Query Execution \\ over Linked Data}
\subtitle{Extended Version*} %\titlenote{The extended version contains proofs for all propositions and lemmas in the paper (cf.~Appendix~\ref{Section:Proofs}).}}

\toappear{*This report presents an extended version of a paper published in HT 2012~\cite{Hartig12:MainPaper}. The extended version contains proofs for all propositions, lemmas, and theorems in the paper (cf.~Appendix~\ref{Appendix:Proofs}).}

% --- Author Metadata here ---
% \conferenceinfo{HT'12,} {June 25--28, 2012, Milwaukee, Wisconsin, USA.} 
% \CopyrightYear{2012} 
% \crdata{978-1-4503-1335-3/12/06}
% --- End of Author Metadata ---

\numberofauthors{2}
\author{
 \alignauthor Olaf Hartig\\
       \affaddr{Humboldt-Universit\"at zu Berlin}\\
       \affaddr{Unter den Linden 6, 10099 Berlin, Germany}\\
       \email{hartig@informatik.hu-berlin.de}
 \alignauthor Johann-Christoph Freytag\\
       \affaddr{Humboldt-Universit\"at zu Berlin}\\
       \affaddr{Unter den Linden 6, 10099 Berlin, Germany}\\
       \email{freytag@informatik.hu-berlin.de}
}

\maketitle

\begin{abstract}
		%The emergence of the Web of Linked Data has spawned research ...
	Query execution over the Web of Linked Data has attracted much attention recently.
A particularly interesting approach is link traversal based query execution which
	%--in contrast to traditional query execution paradigms-- does not assume a fixed set of relevant data sources beforehand; instead, this approach discovers data on the fly and, thus, enables applications to tap the full potential of the Web. The idea of this approach is to integrate the traversal of data links into the query execution process. \par While several authors study possibilities to implement this idea and to optimize query execution in this context,
	proposes to integrate the traversal of data links into the
		%query execution process.
		creation of query results.
	Hence --in contrast to traditional query execution paradigms-- this does not assume a fixed set of relevant data sources beforehand; instead, the traversal process discovers data and data sources on the fly and, thus, enables applications to tap the full potential of the Web. \par While several authors have studied possibilities to implement the idea of link traversal based query execution and to optimize query execution in this context,
no work exists that discusses
	%the
theoretical foundations of the approach in general. Our paper fills this gap.

We introduce a well-defined semantics for queries that may be executed using a link traversal based approach. Based on this semantics we formally analyze properties of such queries. In particular, we study the computability of queries as well as the implications of querying a potentially infinite Web of Linked Data. Our results show that query computation in general is not guaranteed to terminate and that for any given query it is undecidable whether the execution terminates.
Furthermore, we define an abstract execution model that captures the integration of link traversal into the query execution process.
	%We use this model to
	Based on this model we
prove the soundness and completeness of link traversal based query execution and analyze an existing implementation approach.
\end{abstract}


\vspace{-1mm} % Layout Adjustment

\vspace{-1mm} % Layout Adjustment

\section{Introduction} \label{Section:Introduction}
\noindent
During recent years an increasing number of data providers adop\-ted the Linked Data principles for publishing and interlinking structured data on the
	%Web~\cite{BernersLee07:LinkedData,Bizer09:TheStorySoFar}.
	%Web~\cite{Heath11:LDBook}.
	World Wide Web (WWW)~\cite{Heath11:LDBook}.
The
	Web of Linked Data
	%dataspace
that emerges from this process 
	enables users to benefit from a virtually unbounded set of data sources and, thus, opens possibilities not conceivable before.
	%can be understood as a single, globally distributed dataspace.
Consequently, the Web of Linked Data has
	spawned 
	%lead to
research
	%that studies possibilities for executing expressive queries over this dataspace and for optimizing such an execution.
	to execute declarative queries over multiple Linked Data sources. % and to optimize these executions.
%In~\cite{Hartig10:DBPerspectiveOnConsumingLD} we provide an overview of different approaches and refer to the relevant literature.
	%Some
	Most
approaches adapt
	%traditional
techniques that are known from the database
	literature
	%community
(e.g.~data warehousing or query federation). %; other approaches are more tailored towards the characteristics of the Web of Linked Data.
However, the Web of Linked Data is different from traditional data\-base systems; distinguishing characteristics are its unbounded nature and the
	%limited data access capabilities.
	lack of a database catalog.
Due to these characteristics it is impossible to know all data sources that might contribute to the answer of a query.
	%To tap the full potential of the Web,
	In this context,
traditional query execution paradigms are insufficient because those assume a fixed set of potentially relevant data sources beforehand. This assumption presents a restriction that inhibits applications to tap the full potential of the Web; it prevents a serendipitous discovery and utilization of relevant data from unknown sources.

An alternative to traditional query execution paradigms are exploration
	%based
approaches that traverse
	%data links.
	links on the Web of Linked Data.
%
	%\removable{Hence,} these
	These
approaches
	%allow
	enable
a query execution system to automatically discover the most recent data from initially unknown data sources.

The prevalent example of an exploration based approach is link traversal based query execution. The idea of this approach is to intertwine the traversal of data links with the construction of the query result and, thus, to integrate the discovery of data into the query execution process~\cite{Hartig09:QueryingTheWebOfLD}. This general idea may be implemented in various ways. For instance, Ladwig and Tran introduce an asynchronous implementation that adapts the concept of symmetric hash joins~\cite{Ladwig10:LinkedDataQueryProcessingStrategies,Ladwig11:SIHJoin}; Schmedding proposes an implementation that incrementally adjusts the answer to a query each time the execution system retrieves additional data~\cite{Schmedding11:IncrementalSPARQLEvaluationForLinkedData}; our earlier work focuses on an implementation that uses a synchronous pipeline of iterators, each of which is responsible for a particular part of the query~\cite{Hartig11:HeuristicForQueryPlanSelection,Hartig09:QueryingTheWebOfLD}. All existing publications focus on approaches for implementing
	the idea of
	%a
link traversal based query execution and on query optimization in the context of such an implementation. To our knowledge, no work exists that
	%studies the theoretical foundations of
	provides a general foundation for
this new query execution paradigm.
%%%	%With this paper we aim to fill
%%%	Our paper fills
%%%this gap.

We argue that a well-defined query semantics
	%for Linked Data queries
is essential to compare different query execution approaches and to verify implementations. Furthermore, a proper theoretical foundation enables a formal analysis of fundamental properties of queries and query executions. For instance,
	%an understanding of
	studying
the computability of queries may answer whether particular query executions are guaranteed to terminate. In addition to these more
	%fundamental
	theoretical
questions,
	%knowledge of the fundamental 
	an understanding of fundamental properties and limitations may help to gain new insight into challenges and possibilities for query planning and optimization.
Therefore, in this paper we provide such a formal foundation of Linked Data queries
	%with a focus on
	and
link traversal based query execution.
	%Consequently, our
	Our
contributions are:
% \begin{itemize}
% 	\vspace{-2mm} \addtolength{\itemsep}{-0.5\baselineskip} % Layout Adjustment
% 	\item
\vspace{1ex} \\ 1.)
		As a basis, we introduce a
			%\emph{formal framework} that comprises a data model which formalizes the idea of a Web of Linked Data. Furthermore, our framework includes a computation model that captures the limited data access capabilities of computations over the Web.
			\emph{theoretical framework} that comprises a data model and a computation model. The data model formalizes the idea of a Web of Linked Data; the computation model captures the limited data access capabilities of computations over the Web.
			%\emph{data model} that captures the idea of a Web of Linked Data. We model such a Web as a potentially infinite structure of documents that contain triple-based data about entities. We represent these entities via identifiers that serve as data links between documents.
% 	\item
\vspace{1ex} \\ 2.)
		We present a \emph{query model} that introduces a well-defined semantics for
			%\removable{triple-based,}
		conjunctive queries
			%over Linked Data.
			(which is the type of queries supported by existing link traversal based systems).
		Basically, the result of such a query is the set of all valuations that map the query to a subset of all Linked Data that is reachable, starting with entity identifiers mentioned in the query. We emphasize that our model does not prescribe a specific notion of reachability; instead, it is possible to make the notion of reachability applied to answer a query can be made explicit \removable{(by specifying which data links should be followed)}.%
%%% selbst wenn man im Web keine vollst\"andigen Ergebnisse erwaten kann, macht das Verst\"andnis von Vollst\"andigkeit greifbar
% 	\item
\vspace{1ex} \\ 3.)
		We formally \emph{analyze properties} of our query model. In particular, we study the implications of querying a potentially infinite Web and show that it is undecidable whether a query result will be finite or infinite. Furthermore, we analyze the computability of queries by adopting earlier work on Web queries which distinguishes finitely computable queries, eventually computable que\-ries, and queries that are not even eventually computable. We prove that queries in our model are eventually computable. Hence, a \removable{link traversal based} query execution system does not have to deal with queries that are not computable at all. However, we also show that it is undecidable whether a particular query execution terminates.
% 	\item
\vspace{1ex} \\ 4.)
		We define an abstract \emph{\removable{query} execution model} that formalizes the general idea of link traversal based query execution. This model captures the approach of intertwining link traversal and result construction. Based on this model we prove the soundness and completeness of
			the new query execution paradigm.
			%the link traversal based query execution approach.
% 	\item
\vspace{1ex} \\ 5.)
		Finally, we use our execution model to formally analyze a particular implementation of link traversal based query execution.
% \end{itemize}
\vspace{1ex}

\noindent
	%The remainder of this
	This
paper is organized as follows: In Section~\ref{Section:Example} we present an example
	%query execution that applies the
	%	%link traversal approach.
	%	idea of link traversal based query execution.
	that demonstrates the idea of link traversal based query execution.
Section~\ref{Section:ModelingTheWeb} defines our data model and our computation model. We present our query model in Section~\ref{Section:QueryModel} and discuss its properties in Section~\ref{Section:Properties}. Section~\ref{Section:ExecModel} introduces the corresponding execution model. Finally, we discuss related work in Section~\ref{Section:RelatedWork} and conclude
	the paper
in Section~\ref{Section:Conclusion}.%
Appendix~\ref{Appendix:Proofs} provides
	%proofs for all theorems, lemmas and propositions in the paper.
	all proofs.

\begin{figure}[t]
\scriptsize
	\begin{tiny}1\end{tiny} \hspace{2mm}
	\textbf{SELECT ?p ?l WHERE $\lbrace$} \\
	\begin{tiny}2\end{tiny} \hspace{5mm}
	   $<$http://\textbf{bob}.name$>$ \, $<$http://.../\textbf{knows}$>$ \, \textbf{?p} . \\
	\begin{tiny}3\end{tiny} \hspace{5mm}
	   \textbf{?p} \,\,\,  $<$http://.../\textbf{currentProject}$>$ \, \textbf{?pr} . \\
	\begin{tiny}4\end{tiny} \hspace{5mm}
	   \textbf{?pr} \, $<$http://.../\textbf{label}$>$  \,         \textbf{?l} . \textbf{$\rbrace$}
	\vspace{-1mm} % Layout Adjustment
	\caption{Sample query presented in the language SPARQL.} \label{Figure:SampleQuery}
	\vspace{-3mm} % Layout Adjustment
\end{figure}

\section{Example Execution} \label{Section:Example}
\noindent
Link traversal based query execution is a novel query execution paradigm tailored to the Web of Linked Data. Since adhering to the Linked Data principles is the minimal requirement for publishing Linked Data  on the WWW, the link traversal approach relies solely on these principles; it does not assume that each data source provides a data-local query interface (as would be required for query federation). The only way to obtain data is via URI look-ups.

Usually,
	%we express queries over the Web of Linked Data
	Linked Data on the WWW is represented using the RDF data model~\cite{Klyne04:RDF} and queries are expressed
using SPARQL~\cite{PrudHommeaux08:SPARQLLanguage}. %, the query language for RDF data.
SPARQL queries consist of RDF
	graph
patterns that contain query variables, denoted with the symbol '?'.
The semantics of SPARQL is based on pattern matching~\cite{Perez09:SemanticsAndComplexityOfSPARQL}.
Figure~\ref{Figure:SampleQuery} provides a SPARQL representation of a query that asks for projects of acquaintances of user Bob, who is identified by URI \rdfTermInText{http://bob.name}. In lines~2 to~4 the query contains a conjunctive query represented as a set of three SPARQL triple patterns. In the following we outline a link traversal based execution of this conjunctive query.

Link traversal based query execution
	%typically
	usually
starts with an emp\-ty, query-local dataset. We obtain some seed data
	%for pattern matching
by looking up the URIs mentioned in the query:
	%Let $G_{\mathrm{bob}}$ (cf.~Figure~\ref{Figure:SampleData}) be the set of RDF triples that we retrieve by looking up the URI \rdfTermInText{http://bob.name}.
	For the URI \rdfTermInText{http://bob.name} in our sample query we may retrieve a set $G_{\mathrm{b}}$ of RDF triples (cf.~Figure~\ref{Figure:SampleData}), which we add to the local dataset.
Now, we alternate between i)~constructing
	%partial solutions
	valuations
from RDF triples that match a pattern of our query in the query-local dataset, and ii)~augmenting the dataset by looking up URIs which are part of these
	%partial solutions.
	valuations.
For the triple pattern in line~2 of our sample query the local dataset contains a matching triple, originating from $G_{\mathrm{b}}$. Hence, we can construct a
	%partial solution
	valuation
$\mu_1 = \lbrace \text{\rdfTermInText{?p}} \rightarrow \text{\rdfTermInText{http://alice.name}} \rbrace$ that maps query variable \rdfTermInText{?p} to the URI \rdfTermInText{http://alice.name}. By looking up this URI we may retrieve a set $G_{\mathrm{a}}$ of RDF triples, which we also add to the query-local dataset. Based on the augmented dataset we can extend $\mu_1$ by adding a binding for \rdfTermInText{?pr}. We obtain $\mu_2 = \lbrace \text{\rdfTermInText{?p}} \rightarrow \text{\rdfTermInText{http://alice.name}}, \text{\rdfTermInText{?pr}} \rightarrow \text{\rdfTermInText{http://.../AlicesPrj}} \rbrace$, which already covers the pattern in line~2 and~3. Notice, constructing $\mu_2$ is only possible because we retrieved $G_{\mathrm{a}}$. However, before we discovered and resolved the URI \rdfTermInText{http://alice.name}, we neither knew about $G_{\mathrm{a}}$ nor about the existence of the data source from which we retrieved $G_{\mathrm{a}}$. Hence, the traversal of data links enables us to answer queries based on data from initially unknown sources.

\begin{figure}[t]
	\vspace{-3mm} % Layout Adjustment
	\begin{align*}
		( \text{ \rdfTermInFigure{http://\textbf{bob}.name} , \rdfTermInFigure{http://.../\textbf{knows}} , \rdfTermInFigure{http://\textbf{alice}.name} } ) & \in G_{\mathrm{b}} \\
		( \text{ \rdfTermInFigure{http://\textbf{alice}.name} , \rdfTermInFigure{http://.../\textbf{name}} ,  \rdfTermInFigure{"Alice"} } ) & \in G_{\mathrm{a}} \\
		( \text{ \rdfTermInFigure{http://\textbf{alice}.name} , \rdfTermInFigure{http://.../\textbf{currentProject}} ,  \rdfTermInFigure{http://.../\textbf{AlicesPrj}} } ) & \in G_{\mathrm{a}} \\
% 		( \text{ \rdfTermInFigure{http://alice.name} , \rdfTermInFigure{http://.../\textbf{topic\_interest}} ,  \rdfTermInFigure{http://.../\textbf{Tennis}} } ) & \in G_{\mathrm{a}} \\
		( \text{ \rdfTermInFigure{http://.../\textbf{AlicesPrj}} , \rdfTermInFigure{http://.../\textbf{label}} , \rdfTermInFigure{"Alice's Project"} } ) & \in G_{\mathrm{p}}
% \\
% 		( \text{ \rdfTermInFigure{http://.../\textbf{Tennis}} , \rdfTermInFigure{http://.../\textbf{label}} , \rdfTermInFigure{"Tennis"} } ) & \in G_{\mathrm{t}}
	\end{align*}
	\vspace{-7mm} % Layout Adjustment
	\caption{Excerpts from Linked Data retrieved from the Web.} \label{Figure:SampleData}
	\vspace{-3mm} % Layout Adjustment
\end{figure}

We proceed with our execution strategy as follows: We discover and retrieve $G_{\mathrm{p}}$ by looking up the URI \rdfTermInText{http://.../AlicesPrj} and extend $\mu_2$ to $\mu_3 \!=\! \lbrace \text{\rdfTermInText{?p}} \rightarrow \text{\rdfTermInText{http://alice.name}}, \text{\rdfTermInText{?pr}} \rightarrow \text{\rdfTermInText{http://.../AlicesPrj}}, $ $\text{\rdfTermInText{?l}} \rightarrow \text{\rdfTermInText{"Alice's Project"}} \rbrace$, which now covers the whole, conjunctive query. Hence, $\mu_3$ can be reported as the result of that query.

% \removable{As the example demonstrates, link traversal based query execution evaluates a query over a dataset that is continuously augmented with data from the Web. The discovery of this data is driven by the URIs in intermediate solutions. Thus, the idea of the approach is not to follow arbitrary links in discovered data, but to follow only those links that correspond to triple patterns in the executed query.}
\section{Modeling a Web of Linked Data} \label{Section:ModelingTheWeb}
\noindent
In this section we introduce theoretical foundations which shall allow us to define and to analyze queries over Linked Data. %\removable{on the Web}.
%In particular, we propose a data model that
% 	%formally captures
% 	formalizes
% the idea of
% 	%a %%% removed for concrete, WWW Linked Data model
% 	the
% Web of Linked Data; furthermore, we introduce a machine based computation model which
% 	%formally
% captures the limited data access capabilities of computations over the Web.
%
In particular, we propose a data model and a computation model.
For these models we assume a static view of the Web; that is, no changes are made to the data on the Web during the execution of a query.

	\vspace{-1mm} % Layout Adjustment
\subsection{Data Model} \label{Subsection:DataModel}
\noindent
The
	%World Wide Web (WWW)
	WWW
is the most prominent implementation of a Web of Linked Data and it shows that the idea
	of Linked Data
scales to a virtually unlimited dataspace. Nonetheless, other implementations are possible (e.g.~within the boundaries of a closed, globally distributed corporate network). Such an implementation may be based on the same technologies used for
	%Linked Data in
the WWW (i.e. HTTP,
	%HTTP scheme based
URIs, RDF, etc.) or it may use other, similar technologies. 
	%For this reason, our data model abstracts from the concrete technologies that implement Linked Data in the WWW.
	Consequently, our data model abstracts from the concrete technologies that implement Linked Data in the WWW and, thus, enables us to study queries over any Web of Linked Data.

As a basis for our model we use a simple, triple based data model for representing the data that is distributed over a Web of Linked Data (similar to the RDF data model that is used for Linked Data on the WWW). We assume a countably infinite set $\symAllURIs$ of possible {\ID}s (e.g.~all URIs)
	%for entities
and a countably infinite set $\symAllLiterals$ of all possible constant literals (e.g.~all possible strings, natural numbers, etc.). $\symAllURIs$ and $\symAllLiterals$ are disjoint. A \emph{{\triple}} is a tuple $t \in \symAllURIs \times \symAllURIs \times (\symAllURIs \cup \symAllLiterals)$. To denote the set of all {\ID}s
	%that occur
in a {\triple} $t$ we write $\fctIDsName(t)$.

We model a Web of Linked Data as a potentially infinite structure of interlinked documents. Such documents, which we call Linked Data documents, or \emph{{\LDdoc}}s for short, are accessed via identifiers in $\symAllURIs$ and contain data that is represented as a set of {\triple}s.
The following definition captures our approach:
\begin{definition} \label{Definition:WebOfData}
	%Let $\mathcal{T} = \symAllURIs \times \symAllURIs \times (\symAllURIs \cup \symAllLiterals)$ be the infinite set of all possible {\triple}s.
	A \definedTerm{Web of Linked Data}
		%is a tuple $\WoD = ( D,\dataFct,\adocFct )$
		$\WoD$ is a tuple $( D,\dataFct,\adocFct )$
	where:
	\begin{itemize}
		\vspace{-2mm} \addtolength{\itemsep}{-0.5\baselineskip} % Layout Adjustment
		\item $D$ is a set of symbols that represent {\LDdoc}s; $D$ may be finite or countably infinite.
		\item $\dataFct : D \rightarrow 2^{\symAllURIs \times \symAllURIs \times (\symAllURIs \cup \symAllLiterals)}$ is a total mapping such that
			%$\forall \, d \in D : \dataFct(d) \text{ is finite}$.
			$\dataFct(d)$ is finite for all $d \in D$.
		\item $\adocFct : \symAllURIs \rightarrow D$ is a partial, surjective mapping.
	\end{itemize}
\end{definition}

\noindent
While the three elements $D$, $\dataFct$, and $\adocFct$ completely define a Web of Linked Data in our model, we point out that these elements
	%are virtual concepts that
are not directly available to a query execution system. However, by retrieving {\LDdoc}s, such a system may gradually obtain information about the Web. Based on this information the system may (partially) materialize these three elements.
In the remainder of this section we discuss the three elements and introduce additional concepts that we need to define our query model.

We say a Web of Linked Data $\WoD = ( D,\dataFct,\adocFct )$ is \emph{infinite}
	%iff
	if and only if
$D$ is infinite; otherwise, we say $\WoD$ is \emph{finite}.
Our model allows for infinite Webs to cover the possibility that Linked Data about an infinite number of identifiable entities is generated on the fly. The following example illustrates such a case:
\begin{example} \label{Example:LinkedOpenNumbers}
	Let $u_i$ denote an HTTP scheme based URI that identifies the natural number $i$. There is a countably infinite number of such URIs. The WWW server which is responsible for these URIs may be set up to provide a document for each natural number. These documents may be generated upon request and may contain RDF data including the RDF triple $( u_i , \text{\rdfTermInFigure{http://.../next}} , u_{i+1} )$. This triple associates
		the natural number
	$i$ with its successor $i$\textnormal{+1} and, thus, links to the data about $i$\textnormal{+1}~\textnormal{\cite{Vrandecic09:LinkedGeoData}}. An example for such a server is provided by the Linked Open Numbers project\footnote{http://km.aifb.kit.edu/projects/numbers/}.
\end{example}

\noindent
Another example were data about an infinite number of entities may be generated is the LinkedGeoData project\footnote{http://linkedgeodata.org} which provides Linked Data about any circular and rectangular area on Earth~\cite{Auer09:LinkedGeoData}. These examples illustrate that an infinite Web of Linked Data is possible in practice. Covering these cases
	%in our model
enables us to
	%model queries
	model~que\-ries
over such data and analyze the effects of executing such queries.

Even if a Web of Linked Data is infinite, we require countability for $D$. We shall see that this requirement has nontrivial consequences: It limits the potential size of Webs of Linked Data in our
	%model. Such a limit
	model and, thus,
allows us to use a Turing machine based model for analyzing computability of queries over Linked Data (cf.~Section~\ref{Subsection:Properties:Computability}).
We emphasize that the requirement of countability does not restrict us in modeling the WWW as a Web of Linked Data: In the WWW we use URIs to locate documents that contain Linked Data. Even if URIs are not limited in length, they are words over a finite alphabet. Thus, the infinite set of all possible URIs is countable, as is the set of all documents that may be retrieved
	%from the WWW.
	using URIs.

The mapping $\dataFct$ associates each {\LDdoc} $d \!\in\! D$ in a Web of Linked Data $\WoD \!=\! \left( D,\dataFct,\adocFct \right)$ with a finite set of {\triple}s. In practice, these
	%{\triple}s
	triples
are obtained by parsing $d$ after $d$ has been retrieved from the Web. The actual retrieval mechanism depends on the technologies that are used to implement the Web of Linked Data.
%The actual access mechanism is not relevant for our model.
%
To denote the potentially infinite (but countable) set of \emph{all {\triple}s} in $\WoD$ we write $\fctAllDataName( W )$;
	%i.e. $\fctAllDataName( W ) = \bigcup_{d \in D} \dataFct(d)$.
	i.e.~it holds: \begin{equation*}\fctAllDataName( W ) = \bigcup_{d \in D} \dataFct(d)\end{equation*}
	%i.e.~it holds: \begin{equation*}\fctAllDataName( W ) = \big\lbrace \dataFct(d) \,|\, d \in D \big\rbrace\end{equation*}

\noindent
Since we use elements in the set $\symAllURIs$ as {\ID}s for entities, we say that an {\LDdoc} $d \in D$ \emph{describes} the entity identified by an {\ID} $\symURI \in \symAllURIs$ if $\exists (s,p,o) \in \dataFct(d) : \left( s=\symURI \lor o=\symURI \right)$.
%
	% Each {\ID} $\symURI \in \IDs$ is an {\ID} for an entity which is described in the Web of Linked Data, that is, there is at least one {\LDdoc} $d \in D$ such that $\exists t \in \dataFct(D) : \symURI \in \fctIDsName(t)$.
%
Notice, while there might be multiple {\LDdoc}s in $D$ that describe an entity identified by $\symURI$, we do not assume that we can enumerate the set of all these documents; i.e., we cannot discover and retrieve all of them. The possibility to query search engines is out of scope of this paper. It is part of our future work to extend the semantics in our query model in order to take data into account, that is reachable by utilizing search engines.
However, according to the Linked Data principles, each
	%identifier
$\symURI \in \symAllURIs$ may also serve as a reference to a specific {\LDdoc} which is considered as an authoritative source of data about the entity identified by $\symURI$. We model the relationship between {\ID}s and authoritative {\LDdoc}s by mapping $\adocFct$. Since some {\LDdoc}s may be authoritative for multiple entities, we do not require injectivity for $\adocFct$. The ``real world'' mechanism for dereferencing {\ID}s (i.e.~learning about the location of the corresponding, authoritative {\LDdoc}) depends on the implementation of the Web of Linked Data and is not relevant for our model. For each identifier $\symURI \in \symAllURIs$ that cannot be dereferenced (i.e.~``broken links'') or that is not used in the Web it holds $\symURI \notin \fctDom{\adocFct}$.

		% A specific subset of the {\ID}s in $\IDs$ identify {\LDdoc}s. The total function $\urlFct : D \rightarrow \IDs$ maps each {\LDdoc} to its {\ID}. The authoritative {\LDdoc} for an {\LDdoc} is the document itself; hence, it must hold: $\forall d \in D : \adocFct( \urlFct(d) ) = d$.

		% Each such document is identified by a globally unique URL. To denote the URL of a document $d \in V$ we write $\urlFct(d)$. In addition to its role as an identifier, such a URL also serves as an address to access the corresponding document on the Web. Hence, by dereferencing $\urlFct(d)$ using the HTTP protocol we retrieve a serialization of $d$.
		% Such serializations of Web documents may encode
		%	%data. In our work we focus on Web documents with a serialization that can be p ...
		%	data in the form of RDF triples.
		% For each Web document $d \in V$ we write $\dataFct(d)$ to denote the finite set of RDF triples we may extract from the serialization of $d$ that we retrieve by dereferencing $\urlFct(d)$.

An {\ID} $\symURI \in \symAllURIs$ with $\symURI \in \fctDom{\adocFct}$ that is used in the data of an {\LDdoc} $d_1 \in D$ constitutes a \emph{data link} to the {\LDdoc} $d_2 = \adocFct(\symURI) \in D$.
To formally represent the graph structure that is formed by such data links, we introduce the notion of a \emph{Web link graph}. The vertices in such a graph represent the {\LDdoc}s of the corresponding Web of Linked Data; the edges represent data links and are labeled with a {\triple} that denotes the corresponding link in the source document. Formally:
\begin{definition} \label{Definition:WebLinkGraph}
	Let $\WoD = (D,\dataFct,\adocFct)$ be a Web of Linked Data.
	The \definedTerm{Web link graph for $\WoD$}, denoted by $G^\WoD$, is a directed, edge-labeled multigraph $(V,E)$ where $V=D$ and
	\begin{align*}
		E = \big\lbrace
			(d_\mathrm{h},d_\mathrm{t},t)
			\, | \,
			& d_\mathrm{h},d_\mathrm{t} \in D \text{ and }
			t \in \dataFct( d_\mathrm{h} ) \text{ and} \\
			& \exists \, \symURI \in \fctIDsName(t) : \adocFct(\symURI) = d_\mathrm{t}
		\big\rbrace
	\end{align*}
\end{definition}

\noindent
% \removable{Based on the assumption that {\LDdoc}s in a Web of Linked Data actually describe the entity (or entities) for which they are authoritative, it holds that any Web link graph has loops (i.e.~edges that connect vertices to themselves). Furthermore, there is no guarantee that a Web link graph is strongly connected; it does not even have to be weakly connected.}
%
In our query model we
	%%% No! % use Web link graphs to
introduce the concept of reachable parts of a Web of Linked Data that are relevant for answering queries; similarly, our execution model introduces a concept for those parts of a Web of Linked Data that have been discovered at a certain point in the query execution process. To provide a formal foundation for these concepts we define the notion of an induced subweb which resembles the concept of induced subgraphs in graph theory.
\begin{definition} \label{Definition:InducedSubWeb}
	Let $\WoD = \left( D,\dataFct,\adocFct \right)$ be a Web of Linked Data.
	A Web of Linked Data $W' = \left( D' \!, \dataFct' \!, \adocFct' \right)$ is an \definedTerm{induced subweb of $\WoD$} if:
	\begin{enumerate}
		\vspace{-2mm} \addtolength{\itemsep}{-0.5\baselineskip} % Layout Adjustment
		\item $D' \subseteq D$, \label{DefinitionRequirement:InducedSubWeb:D} 
			%\item $\IDs' = \lbrace \symURI \in \IDs | \exists t \in \bigcup_{d \in D'} \dataFct(d) : \symURI \in \fctIDsName(t) \rbrace$, \label{DefinitionRequirement:InducedSubWeb:IDs}
		\item $\forall \, d \in D' : \dataFct'(d) = \dataFct(d)$, and \label{DefinitionRequirement:InducedSubWeb:data}
		\item
			$\forall \, \symURI \in \big\lbrace \symURI \in \symAllURIs \,\big|\, \adocFct(\symURI) \in D' \big\rbrace : \adocFct'(\symURI) = \adocFct(\symURI)$.
			%$\forall \, \symURI \in \symAllURIs_{D'} : \adocFct'(\symURI) = \adocFct(\symURI)$ where $\symAllURIs_{D'} = \lbrace \symURI \in \symAllURIs \,|\, \adocFct(\symURI) \in D' \rbrace$.
 		\label{DefinitionRequirement:InducedSubWeb:adoc}
	\end{enumerate}
\end{definition}

\noindent
It can be easily seen from Definition~\ref{Definition:InducedSubWeb} that
	specifying $D'\!$ is sufficient to define an induced subweb $\left( D' \!, \dataFct' \!, \adocFct' \right)$ of a given Web of Linked Data unambiguously.
	%it is sufficient to define $D'$ in order to unambiguously define an induced subweb $\left( D' \!, \dataFct' \!, \adocFct' \right)$ of a given Web of Linked Data.
Furthermore, it is easy to verify that
	%i)~%
for an induced subweb $W'$ of a Web of Linked Data $\WoD$ it holds $\fctAllDataName(W') \subseteq \fctAllDataName(W)$%\removable{~and
%ii)~the induced subweb relationship is transitive}%
	.
	%, and iii)~the Web link graph for an induced subweb of a Web of Linked Data $\WoD$ is an induced subgraph\footnote {We understand a directed, edge-labeled multigraph $(V'\!,E')$ as an \emph{induced subgraph} of a directed, edge-labeled multigraph $(V,E)$ iff $V' \!\subseteq\! V$ and $E' \!=\! \big\lbrace e \, | \, e = (v_1,v_2,l) \in E \, \land \, v_1 \in V' \, \land \, v_2 \in V' \big\rbrace$.} of the Web link graph for $\WoD$.

\subsection{Computation Model} \label{Subsection:ComputationModel}
\noindent
	%Our data model formally captures Webs of Linked Data by three elements: $D$, $\dataFct$ and $\adocFct$. While these three elements completely define a Web of Linked Data in our model, we point out ...
Usually, functions are computed over structures that are assumed to be fully (and directly) accessible.
	A Web of Linked Data, in contrast, is a structure
	%In contrast, we focus on a Web of Linked Data as a structure
	%In contrast, we focus on a Web of Linked Data
	%In contrast, we focus on a structure
in which accessibility is limited:
	To discover {\LDdoc}s and access their data we have to dereference {\ID}s, but the full set of those {\ID}s for which we may retrieve documents is unknown.
	%In a Web of Linked Data the full set of those {\ID}s for which we may retrieve {\LDdoc}s (and, thus, some data) is unknown.
Hence, to properly analyze queries over a Web of Linked Data we require a model for computing functions on such a Web. This section introduces such a model.
	%	for which we adapt earlier work
	%	%which is an adaptation of earlier work
	%of Abiteboul and Vianu~\cite{Abiteboul00:QueriesAndComputationOnTheWebArticle} and of Mendelzon and Milo~\cite{Mendelzon98:FormalModelsOfWebQueries}.

	%In the context of queries over a hypertext-centric view of
	In earlier work about computation on
the WWW, Abiteboul and Vianu
	%model computation using
	introduce
a specific Turing machine
	%(TM)
called Web machine~\cite{Abiteboul00:QueriesAndComputationOnTheWebArticle}. Mendelzon and Milo
	%introduce
	propose
a similar machine model~\cite{Mendelzon98:FormalModelsOfWebQueries}. These machines formally capture the limited data access capabilities
	%of computation on the WWW.
	on the WWW and thus present an adequate abstraction for computations over a structure such as the WWW.
%
	%We adopt the ideas of Abiteboul and Vianu and Mendelzon and Milo for our work. More precisely, we adapt the idea of
	We adopt the idea of such
a Web machine to our
	%scenario.
	scenario of a Web of Linked Data.
We call our machine a \emph{Linked Data machine} (or LD machine, for short).
% Based on this machine we shall define finite
% 	%computability
% and eventual computability for Linked Data queries.

Encoding (fragments of) a Web of Linked Data
	%$\WoD = ( D,\dataFct,\adocFct )$
	$\WoD = ( D,\dataFct,$ $\adocFct )$
on the tapes of such a machine is straightforward because all relevant structures, such as the sets $D$ or $\symAllURIs$, are countably infinite. In the remainder of this paper we write $\fctEnc{x}$ to denote the encoding of some element $x$ (e.g.~a single {\triple}, a set of triples, a full Web of Linked Data, etc.). For a detailed definition of the encodings we use in this paper, we refer to
	Appendix~\ref{Appendix:Encoding}.
% 	the appendix in~\cite{Hartig11:ExtendedVersion}.

We
	now define our adaptation of the idea of Web machines: %, which we call Linked Data machine (or LD machine, for short):
	%now define
	%	%our notion of
	%LD machine:

\begin{definition} \label{Definition:LDMachine}
	An \definedTerm{LD machine} is a multi-tape
		Turing machine
		%TM
	with five tapes and a finite set of states, including a special state called \emph{expand}.
	%
		%The five tapes are:
		%i)~an ordinary, read-only input tape,
		%ii)~a right-infinite, read-only \emph{Web} (input) tape which can only be accessed in the expand state,
		%iii)~a right-infinite \emph{link traversal} work tape,
		%iv)~an ordinary, two-way infinite work tape,
	%
		The five tapes include two, read-only input tapes:
		i)~an ordinary input tape and
		ii)~a right-infinite \emph{Web} tape which can only be accessed in the expand state;
		two work tapes:
		iii)~an ordinary, two-way infinite work tape and
		iv)~a right-infinite \emph{link traversal} tape;
	and v)~a right-infinite, append-only output tape.
	Initially, the work tapes and the output tape are empty, the Web tape contains a (potentially infinite) word that encodes a Web of Linked
		%Data (Appendix~\ref{Appendix:Encoding} defines the encoding in detail),
		Data,
	and the ordinary input tape contains an encoding of further input (if any).
	Any LD machine operates like an ordinary
		multi-tape
		Turing machine
		%TM
	except when it reaches the expand state. In this case
		%the machine inspects 
		LD machines perform the following \emph{expand procedure}: The machine inspects
	the word currently stored on the link traversal tape. If the suffix of this word is the encoding $\fctEncName(\symURI)$ of some {\ID} $\symURI \in \symAllURIs$ and the word on the Web tape contains \,$\sharp \, \fctEncName(\symURI) \, \fctEncName( \adocFct(\symURI) ) \, \sharp$\,, then the machine appends \,$\fctEncName( \adocFct(\symURI) ) \, \sharp$\, to the (right) end of the word on the link traversal tape by copying from the Web tape; otherwise, the machine appends \,$\sharp$\, to the word on the link traversal tape.
\end{definition}

\noindent
% As mentioned before, the notion of an LD machine is a direct adaptation of Web machines as defined in~\cite{Abiteboul00:QueriesAndComputationOnTheWebArticle} and~\cite{Mendelzon98:FormalModelsOfWebQueries}. Consequently, the expand state in our LD machine corresponds to the expand state in Abiteboul and Vianu's browser machine~\cite{Abiteboul00:QueriesAndComputationOnTheWebArticle} and to the oracle in Mendelzon and Milo's Web machine~\cite{Mendelzon98:FormalModelsOfWebQueries}.
%
Notice how an LD machine is limited in the way it may access a Web of Linked Data that is encoded on its Web (input) tape: Any {\LDdoc} and its data is only available for the computation after the machine performed the expand procedure using a corresponding {\ID}. Hence, the expand procedure models a URI based lookup which
	%conforms to
	is
the (typical) data access method on the WWW.

% \removable{
% We note that the word on the Web tape of an LD machine is infinitely large if (and only if) the encoded Web of Linked Data $\WoD = ( D,\dataFct,\adocFct )$ is infinite. However, the words on the other tapes are always finite at any step in any possible computation of an LD machine. Furthermore, we note that the Web tape may contain data of an {\LDdoc} $d \in D$ multiple times due to our encoding; more precisely, the data of $d$ is stored as many times as there exist {\ID}s $\symURI \in \symAllURIs$ with $\adocFct(\symURI)=d$. We emphasize that such type of a duplication does not present a problem for
% 	%analyzing
% 	an analysis of
% set-based query semantics as we do in this paper.
% }

In the following sections we use the notion of an LD machine for analyzing
	%the theoretical properties of queries over a Web of Linked Data.
	properties of our query model.
In this context we aim to discuss decision problems that shall have a Web of Linked Data $\WoD$ as input. For these problems we assume that the computation may only be performed by an LD machine with $\fctEnc{\WoD}$ on its Web tape:

\begin{definition}
	Let $\mathcal{W}$ be a (potentially infinite) set of Webs of Lin\-ked Data;
	let $\mathcal{X}$ be an arbitrary (potentially infinite) set of finite structures;
	and let $DP \subseteq \mathcal{W} \times \mathcal{X}$. % \removable{be a subset of pairs from $\mathcal{W}$ and~$\mathcal{X}$}.
	The decision problem for $DP$, that is, to decide for any $(\WoD,X) \in \mathcal{W} \times \mathcal{X}$ whether $(\WoD,X) \in DP$, is \definedTerm{LD machine decidable} if there exist an LD machine whose computation on any $\WoD \in \mathcal{W}$ encoded on the Web tape and any $X \in \mathcal{X}$ encoded on the ordinary input tape, has the following property: The machine halts in an accepting state if $(\WoD,X)\in DP$; otherwise the machine halts in a rejecting state.
\end{definition}

\noindent
Obviously, any (Turing) decidable problem that does not have a Web of Linked Data as input, is also LD machine decidable because LD machines are Turing machines%
	\removable{; for these problems the corresponding set $\mathcal{W}$ is empty}
.
\section{Query Model} \label{Section:QueryModel}
\noindent
This section introduces our query model by defining semantics for conjunctive queries over Linked Data.
%
% 	%We define the semantics of BLD queries using a
% 	This definition uses a
% two-phase approach: First, we 
% 	%introduce the concept of reachability to
% define the part of a Web of Linked Data that is
% 	%discovered
% 	reached
% by traversing links using the identifiers in a query as a starting point. Then, we formalize the result of such a query as the set of all valuations that map the query to a subset of all data in the reachable part of the Web. Notice, while this two-phase approach provides for a straightforward definition of the query semantics in our model, it does not correspond to the actual query execution strategy of
% 	%intertwining pattern matching and the traversal of data links which is characteristic for the link traversal based query execution illustrated in Example~\ref{Example:QueryExecution}.
% 	integrating the traversal of data links into the query execution process as illustrated in Example~\ref{Example:QueryExecution}.

\subsection{Preliminaries} \label{Subsection:QueryModel:Preliminaries}
\noindent
We assume an infinite set $\symAllVariables$ of possible query variables that is disjoint from the sets $\symAllURIs$ and $\symAllLiterals$ introduced in the previous section. These variables will be used to range over elements in $\symAllURIs \cup \symAllLiterals$. Thus, \emph{valuation}s in our context are total mappings from a finite subset of $\symAllVariables$ to the set $\symAllURIs \cup \symAllLiterals$. We denote the domain of a particular valuation $\mu$ by $\fctDom{\mu}$.
Using valuations we define our general understanding of queries over a Web of Linked Data as follows:
\begin{definition} \label{Definition:LDQuery}
	Let $\mathcal{W}$ be a set of all possible Webs of Linked Data (i.e. all 3-tuples that correspond to Definition~\ref{Definition:WebOfData})
	and let $\Omega$ be a set of all possible valuations.
	A \definedTerm{Linked Data query} $q$ is a total
		% generic % ATTENTION: you may wish to define genericity for LD queries!
	function $q : \mathcal{W} \rightarrow 2^{\Omega}$.
\end{definition}

\noindent
To express \emph{conjunctive} Linked Data queries we adapt the notion of a
	%basic graph pattern, introduced by the query language SPARQL~\cite{PrudHommeaux08:SPARQLLanguage},
	SPARQL basic graph pattern~\cite{PrudHommeaux08:SPARQLLanguage}
to our data model:
\begin{definition} \label{Definition:BasicQueryPattern}
	A ~ \definedTerm{basic query pattern (BQP)} ~ is a finite set $\symBQP = \lbrace tp_1 ,... \, , tp_n \rbrace$ of tuples $tp_i \in (\symAllVariables \cup \symAllURIs) \times (\symAllVariables \cup \symAllURIs) \times (\symAllVariables \cup \symAllURIs \cup \symAllLiterals)$ (for $1 \!\leq\! i \!\leq\! n$).
	We call such a tuple a \definedTerm{triple pattern}.
\end{definition}

\noindent
	In comparison to traditional notions of conjunctive queries, triple patterns are the counterpart of atomic formulas; furthermore, BQPs have no head, hence no bound variables.
	%Notice, BQPs are conjunctive queries with binary predicates and no bound variables.
%The correspondence of BQPs to basic graph patterns in SPARQL makes it easy to consider more expressive queries by adapting operands from the SPARQL algebra (which resembles the relational algebra~\cite{Perez09:SemanticsAndComplexityOfSPARQL}). However, additional expressiveness is out of scope of this paper.
%
To denote the set of variables and {\ID}s that occur in a triple pattern $tp$ we write $\fctVarsName(tp)$ and $\fctIDsName(tp)$, respectively. Accordingly, the set of variables and {\ID}s that occur in all triple patterns of a BQP $\symBQP$ is denoted by $\fctVarsName(\symBQP)$ and $\fctIDsName(\symBQP)$, respectively.
For a triple pattern $tp$ and a valuation $\mu$ we write $\mu[tp]$ to denote the triple pattern that we obtain by replacing the variables in $tp$ according to $\mu$. Similarly, a valuation $\mu$ is applied to a BQP $\symBQP$ by $\mu[\symBQP] = \lbrace \mu[tp] \mid tp \in \symBQP \rbrace$.
	The result of $\mu[tp]$ is a {\triple} if $\fctVarsName(tp) \subseteq \fctDom{\mu}$.
	%If and only if $\fctVarsName(\symBQP) \subseteq \fctDom{\mu}$, then $\mu[\symBQP]$ is a set of {\triple}s.
Accordingly, we introduce the notion of \emph{matching {\triple}s}:
\begin{definition} \label{Definition:MatchingTriple}
		%Let $t$ be a {\triple} and let $tp $ be a {\TP}. \definedTerm{$t$ matches $tp$}
		A {\triple} $t$ \definedTerm{matches} a {\TP} $tp$ 
	if there exists a valuation $\mu$ such that $\mu[tp] = t$.
\end{definition}

\noindent
While BQPs are syntactic objects, we shall use them
	%to express
	as a representation of
Linked Data queries which have a certain semantics. In the remainder of this section we define this semantics.
Due to the openness and distributed nature of Webs such as the WWW we cannot guarantee query results that are complete w.r.t.~all Linked Data on a Web. Nonetheless, we aim to provide a well-defined semantics. Consequently, we have to limit our understanding of completeness. However, instead of restricting ourselves to data from a fixed set of sources selected or discovered beforehand, we introduce an approach that allows a query to make use of previously unknown data and sources. Our definition of query semantics is based on a two-phase approach: First, we 
	%introduce the concept of reachability to
define the part of a Web of Linked Data that is
	%discovered
	reached
by traversing links using the identifiers in a query as a starting point. Then, we formalize the result of such a query as the set of all valuations that map the query to a subset of all data in the reachable part of the Web. Notice, while this two-phase approach provides for a straightforward definition of the query semantics in our model, it does not correspond to the actual query execution strategy of
	%intertwining pattern matching and the traversal of data links which is characteristic for the link traversal based query execution illustrated in Example~\ref{Example:QueryExecution}.
	integrating the traversal of data links into the query execution process as illustrated in
		%Example~\ref{Example:QueryExecution}.
		Section~\ref{Section:Example}.

\subsection{Reachability} \label{Subsection:QueryModel:Reachability}
\noindent
To introduce the concept of a reachable part of a Web of Linked Data we first define reachability of {\LDdoc}s. Informally, an {\LDdoc} is reachable if there exists a (specific) path in the Web link graph of a Web of Linked Data to the document in question; the potential starting points for such a path are {\LDdoc}s that are authoritative for entities mentioned (via their {\ID}) in the queries. However, allowing for arbitrary paths might be questionable in practice because it would require following \emph{all} data links (recursively) for answering a query completely. A more restrictive approach is the notion of \emph{query pattern based reachability} where a data link only qualifies as a part of paths to reachable {\LDdoc}s, if that link corresponds to a triple pattern in the executed query.
The link traversal based query execution illustrated in Section~\ref{Section:Example} applies
	%such a
	this
notion of query pattern based reachability (as we show in Section~\ref{Subsection:ExecModel:Traversing}).
Our experience in developing a link traversal based query execution system\footnote{http://squin.org} suggests that
	query pattern based
	%this notion of
reachability is a good compromise for answering queries without crawling large portions of the Web that are likely to be irrelevant for the queries. However, other
	%reachability
criteria for specifying which data links should be followed might prove to be more suitable in certain use cases. For this reason, we do not prescribe a specific
	%criterion;
	criterion in our query model;
instead, we enable our
	%query model
	model
to support any possible criterion by making this concept part of the model.
\begin{definition} \label{Definition:ReachabilityCriterion}
	Let
		$\mathcal{T}$
		%$\mathcal{T}= \symAllURIs \times \symAllURIs \times (\symAllURIs \cup \symAllLiterals)$
	be the infinite set of all possible {\triple}s;
	let
		$\mathcal{\symBQP}$
		%$\mathcal{\symBQP}= \mathcal{P}\bigl( (\symAllVariables \cup \symAllURIs) \times (\symAllVariables \cup \symAllURIs) \times (\symAllVariables \cup \symAllURIs \cup \symAllLiterals) \bigr)$
	be the infinite set of all possible BQPs.
	A \definedTerm{reachability criterion} $c$ is a
		%partial,
		total
	computable function
$%	\begin{equation*}
		c : \mathcal{T} \times \symAllURIs \times \mathcal{\symBQP} \rightarrow \lbrace \true, \false \rbrace
$.
%	\end{equation*}
% 	that must be defined for each
% 		% tuple
% 	$(t,\symURI,\symBQP)\in \mathcal{T} \times \symAllURIs \times \mathcal{\symBQP}$ for which $\symURI\in\fctIDsName(t)$.
\end{definition}

\noindent
An example for such a reachability criterion is $\cAll$ which corresponds to the approach of allowing for arbitrary paths to reach {\LDdoc}s; hence, for each tuple $(t,\symURI,\symBQP)\in \mathcal{T} \times \symAllURIs \times \mathcal{\symBQP}$ it holds $\cAll(t,\symURI,\symBQP) = \true$.
The complement of $\cAll$ is $\cNone$ which \emph{always} returns $\false$%
	%\removable{, independent of $t$, $\symURI$, and $\symBQP$}%
.
	Another example is
	%As a last example \removable{for reachability criteria} we introduce
$\cMatch$ which corresponds to the aforementioned query pattern based reachability.
	%The definition of $\cMatch$ requires the notion of matching {\triple}s:
	%\begin{definition} \label{Definition:MatchingTriple}
	%	Let $t = (s,p,o)$ be a {\triple} and
	%	let $tp = (\tilde{s},\tilde{p},\tilde{o})$ be a {\TP}.
	%	\definedTerm{$t$ matches $tp$} iff:
	%	\begin{equation*}
	%		\left( \tilde{s} \notin V \Rightarrow \tilde{s} = s \right) \land
	%		\left( \tilde{p} \notin V \Rightarrow \tilde{p} = p \right) \land
	%		\left( \tilde{o} \notin V \Rightarrow \tilde{o} = o \right)
	%	\end{equation*}
	%\end{definition}
	%
	%Based on matching {\triple}s we define $c_{\mathsf{Match}}$ as follows.
	We define $\cMatch$ based on the notion of matching {\triple}s:
\begin{equation} \label{Equation:CMatch}
	\cMatch\Bigl( t,\symURI, \symBQP \Bigr) =
	\begin{cases}
		\true & \text{if $\exists \, tp \in \symBQP : t \text{ matches } tp$}, \\
		\false & \text{else}.
	\end{cases}
\end{equation}

\noindent
We call a reachability criterion $c_1$ \emph{less restrictive than} another criterion $c_2$ if i)~for each
	tuple
$(t,\symURI,\symBQP)\in \mathcal{T} \times \symAllURIs \times \mathcal{\symBQP}$ for which $c_2(t,\symURI,\symBQP)=\true$, also holds $c_1(t,\symURI,\symBQP)=\true$ and ii)~there exist a $(t',\symURI',\symBQP')\in \mathcal{T} \times \symAllURIs \times \mathcal{\symBQP}$ such that $c_1(t',\symURI',\symBQP')=\true$ but $c_2(t',\symURI',\symBQP')=\false$.
% Two reachability criteria $c_1$ and $c_2$ are \emph{equally restrictive} if for each $(t,\symURI,\symBQP)\in \mathcal{T} \times \symAllURIs \times \mathcal{\symBQP}$ holds $c_1(t,\symURI,\symBQP) = c_2(t,\symURI,\symBQP)$.
%%% defining \emph{more restrictive} as done here is wrong because there is no total order
% $c_1$ is \emph{more restrictive than} $c_2$ if $c_1$ is neither equally restrictive as $c_2$ nor is it less restrictive than $c_2$.
It can be
	%easily
seen that $\cAll$ is the least restrictive
	%reachability
criterion, whereas $\cNone$ is the most restrictive criterion.

Using the concept of reachability criteria for data links we formally define
	%what it means that an {\LDdoc} is reachable:
	reachability of {\LDdoc}s:
\begin{definition} \label{Definition:QualifiedReachability}
	Let $\WoD=(D,\dataFct,\adocFct)$ be a Web of Linked Data;
% 	let $G^\WoD = (D,L)$ be the Web link graph for $\WoD$;
	let $\symSeedURIs \subset \symAllURIs$ be a finite set of seed {\ID}s;
	let $c$ be a reachability criterion;
	and let $\symBQP$ be a BQP.
	An {\LDdoc} $d \in D$ \definedTerm{is $(c,\symBQP)$-reachable from $\symSeedURIs$ in $\WoD$} if either
	\begin{enumerate}
		\vspace{-2mm} \addtolength{\itemsep}{-0.5\baselineskip} % Layout Adjustment
		\item
			there exists an $\symURI \in \symSeedURIs$ such that $\adocFct(\symURI) = d$; or%
			\label{DefinitionCase:QualifiedReachability:IndBegin}
		\item
			there exist
				another {\LDdoc}
				%an
			$d' \in D$,
				%a {\triple}
				a
			$t \in \dataFct(d')$, and
				%an {\ID}
				an
			$\symURI \in \fctIDsName(t)$ such that i)~$d'$ is $(c,\symBQP)$-reachable from $\symSeedURIs$ in $\WoD$, ii)~$c(t,\symURI,\symBQP)=\true$, and iii)~$\adocFct(\symURI)=d$. \label{DefinitionCase:QualifiedReachability:IndStep}
	\end{enumerate}
\end{definition}

\noindent
We note that
	%any {\LDdoc} which is $(\cMatch,\symBQP)$-reachable from some $\symSeedURIs$ in a Web of Linked Data $\WoD$ (for some BQP $\symBQP$), is also $(\cAll,\symBQP)$-reachable from some $\symSeedURIs$ in $\WoD$ because $\cAll$ is less restrictive than $\cMatch$. Furthermore, from case~\ref{DefinitionCase:QualifiedReachability:IndBegin} in Definition~\ref{Definition:QualifiedReachability} we see that
each {\LDdoc} which is authoritative for an entity mentioned (via its identifier) in a
	finite set of seed {\ID}s $\symSeedURIs$,
	%$\symSeedURIs \subset \symAllURIs$,
is always reachable from $\symSeedURIs$ in the corresponding Web of Linked Data, independent of the reachability criterion and the BQP used.

	Based on reachability of {\LDdoc}s we now define reachable parts of a Web of Linked Data.
	%Based on the reachability of {\LDdoc}s we now define the part of a Web of Linked Data that can be discovered by link traversal using the identifiers in a given BQP.
Informally, such a part is an induced subweb covering all reachable {\LDdoc}s. Formally:
\begin{definition} \label{Definition:ReachablePart}
	Let $\WoD = (D,\dataFct,\adocFct)$ be a Web of Linked Data;
	let $\symSeedURIs \subset \symAllURIs$ be a finite set of seed {\ID}s;
	let $c$ be a reachability criterion;
	and let $\symBQP$ be a BQP.
	The \definedTerm{$(\symSeedURIs,c,\symBQP)$-reachable part of
		%$\WoD$}, denoted by $\ReachPartScBW{\symSeedURIs}{c}{\symBQP}{W}$, is the induced subweb $( \Reach{D} , \Reach{\dataFct} ,\Reach{\adocFct} )$
		$\WoD$} is the induced subweb $\ReachPartScBW{\symSeedURIs}{c}{\symBQP}{W} = ( \Reach{D} , \Reach{\dataFct} ,\Reach{\adocFct} )$
	of $\WoD$ that is defined by
	\begin{equation*}
		\Reach{D}=\big\lbrace d \in D \, | \, d \text{ is $(c,\symBQP)$-reachable from $\symSeedURIs$ in $W$} \big\rbrace
	\end{equation*}
% 	To denote the $(\symSeedURIs,c,\symBQP)$-reachable part of $\WoD$ we write $\ReachPartScBW{\symSeedURIs}{c}{\symBQP}{W}$.
\end{definition}

\subsection{Query Results}
\noindent
Based on the previous definitions we define the semantics of conjunctive Linked Data queries that are expressed via BQPs. Recall that Linked Data queries map from a Web of Linked Data to a set of
	valuations. 
	%valuations \removable{(cf.~Definition~\ref{Definition:LDQuery})}.
Our interpretation of BQPs as Linked Data queries requires that each valuation $\mu$ in the result for a particular BQP $\symBQP$ satisfies the following requirement: If we replace the variables in $\symBQP$ according to $\mu$ (i.e.~we compute $\mu[\symBQP]$), we obtain a set of {\triple}s and this set must be a subset of all data in the part of the Web that is reachable according to the notion of reachability that we apply. Since our model supports a virtually unlimited number of notions of reachability, each of which is defined by a particular reachability criterion, the actual result of a query
	%and, thus, the semantics of queries in our model depends
	must depend
on such a reachability criterion. The following definition formalizes our understanding of conjunctive Linked Data queries:

% Informally, we require for each of these valuations that if we replace the variables in the BQP according to the valuation (i.e.~we apply the valuation to the query, cf.~Section~\ref{Subsection:QueryModel:Preliminaries}), we obtain a set of {\triple}s; this set must be a subset of all data in the part of the Web that is reachable according to the notion of reachability that we apply. Since our model supports a virtually unlimited number of notions of reachability, each of which is defined by a particular reachability criterion for data links, the actual result of a query
% 	%and, thus, the semantics of queries in our model depends
% 	must depend
% on such a reachability criterion.
% %
% The following definition captures our understanding of query results formally:
\begin{definition} \label{Definition:Solution}
	Let $\symSeedURIs \subset \symAllURIs$ be a finite set of seed {\ID}s;
	let $c$ be a reachability criterion;
	and let $\symBQP$ be a BQP;
	let $\WoD$ be a Web of Linked Data;
		%and
	let $\ReachPartScBW{\symSeedURIs}{c}{\symBQP}{W}$ denote the $(\symSeedURIs,c,\symBQP)$-reach\-able part of $\WoD$.
	The \definedTerm{conjunctive Linked Data query (CLD query)} that uses $\symBQP$, $\symSeedURIs$, and
		%$c$
		$c$, denoted by $\queryFctBSc{\symBQP}{\symSeedURIs}{c}$,
	is a Linked Data query
		%that is defined as follows:
		defined as:
	\begin{align*}
		\queryRsltBScW{\symBQP}{\symSeedURIs}{c}{\WoD} = \big\lbrace \mu \,\big|\, \mu \text{ is a valuation with } & \fctDom{\mu} = \fctVarsName(\symBQP) \\ \text{ and } \mu[\symBQP] & \subseteq \fctAllDataName\bigl( \ReachPartScBW{\symSeedURIs}{c}{\symBQP}{W} \bigr) \big\rbrace
	\end{align*}
	Each $\mu \in \queryRsltBScW{\symBQP}{\symSeedURIs}{c}{\WoD}$ is a \definedTerm{solution for $\queryFctBSc{\symBQP}{\symSeedURIs}{c}$ in $\WoD$}.
\end{definition}

\noindent
Since we define the result of queries w.r.t.~a reachability criterion, the semantics of such queries depends on this criterion. Thus, strictly speaking, our query model introduces a family of query semantics, each of which is characterized by a reachability criterion. Therefore, we refer to a CLD query for which we use a particular reachability criterion $c$ as a CLD query \emph{under $c$-semantics}.

% \Todo{Before we discuss properties of}{adjust this to the new structure}
% 	%our query model
% 	CLD queries
% in Section~\ref{Section:Properties}, we introduce our query execution model in the next section.

\section{Properties of the Query Model} \label{Section:Properties}
\noindent
In this section we discuss properties of our query model. In particular, we focus on the implications of querying Webs that are infinite and on the (LD machine based) computability of queries.

\todo{begin with a --(in)finiteness-independent-- section about completeness of query results (w.r.t.~$\fctAllDataName(\WoD)$) and ``degrees of completeness''}

\subsection{Querying an Infinite Web of Linked Data} \label{Subsection:Properties:Infiniteness}
\noindent
From Definitions~\ref{Definition:QualifiedReachability} and~\ref{Definition:ReachablePart} in Section~\ref{Section:QueryModel} it can be easily seen that any reachable part of a \emph{finite} Web of Linked Data must also be finite, independent of the query that we want to answer and the reachability criterion that we use. Consequently, the result of CLD queries over such a finite Web is also guaranteed to be finite.
	%The following two examples show 
	We shall see
that a similarly general statement does not exist when the queried Web
	%of Linked Data
is infinite such as the WWW.

To study the implications of querying an infinite Web we first take a look at some example queries. For these examples we assume an \emph{infinite} Web of Linked Data $\WoD_\mathsf{inf} \!=\! \left( D_\mathsf{inf},\dataFct_\mathsf{inf},\adocFct_\mathsf{inf} \right)$ that contains {\LDdoc}s for all natural numbers (similar to the documents in Example~\ref{Example:LinkedOpenNumbers}). The data in these documents refers to the successor of the corresponding number and to all its divisors. Hence, for each natural number\footnote{In this paper we write
	%$\mathbb{N}^0$ to denote the set of all natural numbers, including zero. $\mathbb{N}^+$ denotes all natural numbers without zero.}
	$\mathbb{N}^+$ to denote the set of all natural numbers without zero. $\mathbb{N}^0$ denotes all natural numbers, including zero.}
$k \in \mathbb{N}^+$, identified by $\mathsf{no}_k \! \in \symAllURIs$, exists an {\LDdoc} $\adocFct_\mathsf{inf}(\mathsf{no}_k) = d_k \in D_\mathsf{inf}$ such that
% .  Formally, let , then
\begin{equation*}
	\dataFct_\mathsf{inf}( d_k )
	=
	\Big\lbrace ( \mathsf{no}_k, \mathsf{succ} , \mathsf{no}_{k\!+\!1} ) \Big\rbrace
	\, \cup \hspace{-2mm}
	\bigcup_{y \in \mathrm{Div}(k)} \hspace{-2mm} \Big\lbrace ( \mathsf{no}_k, \mathsf{div} , \mathsf{no}_y ) \Big\rbrace 
\end{equation*}
where $\mathrm{Div}(k)$ denotes the set of all divisors of $k \in \mathbb{N}^+$\!, $\mathsf{succ} \in \symAllURIs$ identifies the successor relation
	%for natural numbers,
	for $\mathbb{N}^+$,
and \,$\mathsf{div} \!\in\! \symAllURIs$ identifies the relation that associates a number $k \!\in\! \mathbb{N}^+$ with a divisor
	%of $k$.
	$y \!\in\! \mathrm{Div}(k)$.
%
% \removable{In the following we investigate several queries and discuss, for each case, what the reachable part of $\WoD_\mathsf{inf}$ and the query result is.}

\begin{example} \label{Example:Infiniteness2}
	Let $\symBQP_1 = \big\lbrace ( \mathsf{no}_2, \mathsf{succ} , ?x ) \big\rbrace$ be a BQP ($?x \in \symAllVariables$) that asks for the successor of 2. Recall,
		%\removable{$\adocFct(\symURI_2)=d_2$ and}
	$\dataFct_\mathsf{inf}( d_2 )$ contains three
		{\triple}s:
		%triples:
	$( \mathsf{no}_2, \mathsf{succ} , \mathsf{no}_3 )$, $( \mathsf{no}_2, \mathsf{div} , \mathsf{no}_1 )$, and $( \mathsf{no}_2, \mathsf{div} , \mathsf{no}_2 )$. We consider reachability criteria $\cAll$, $\cMatch$, and $\cNone$ (cf.~Section~\ref{Subsection:QueryModel:Reachability}) and $\symSeedURIs_1 = \lbrace \mathsf{no}_2 \rbrace$: The $(\symSeedURIs_1,\cAll,\symBQP_1)$-reachable part of $\WoD_\mathsf{inf}$ is infinite and consists of\footnote{We assume $\mathsf{succ} \notin \fctDom{\adocFct_\mathsf{inf}}$ and $\mathsf{div} \notin \fctDom{\adocFct_\mathsf{inf}}$.} the {\LDdoc}s $d_1, ... \, , d_k , ...$~. In contrast, the $(\symSeedURIs_1,\cMatch,\symBQP_1)$-reach\-able part
		%$\ReachPartScBW{\symSeedURIs_1}{\cMatch}{\symBQP_1}{\WoD_\mathsf{inf}}$
		$\ReachPartScBW{\symSeedURIs_1}{\cMatch}{\symBQP_1}{\WoD}$
	and the $(\symSeedURIs_1,\cNone,\symBQP_1)$-reach\-able part
		%$\ReachPartScBW{\symSeedURIs_1}{\cNone}{\symBQP_1}{\WoD_\mathsf{inf}}$
		$\ReachPartScBW{\symSeedURIs_1}{\cNone}{\symBQP_1}{\WoD}$
	are finite:
		%$\ReachPartScBW{\symSeedURIs_1}{\cMatch}{\symBQP_1}{\WoD_\mathsf{inf}}$
		$\ReachPartScBW{\symSeedURIs_1}{\cMatch}{\symBQP_1}{\WoD}$
	consists of $d_2$ and $d_3$, whereas
		%$\ReachPartScBW{\symSeedURIs_1}{\cNone}{\symBQP_1}{\WoD_\mathsf{inf}}$
		$\ReachPartScBW{\symSeedURIs_1}{\cNone}{\symBQP_1}{\WoD}$
	only consists of $d_2$. The query result in all three cases
% 		is $\lbrace \mu \rbrace$
% 		%is $\symBQP_1^{\cAll}(W)=\symBQP_1^{\cMatch}(W)=\symBQP_1^{\cNone}(W) = \lbrace \mu \rbrace$
% 	where $\mu = \lbrace ?x \rightarrow \mathsf{no}_3 \rbrace$; i.e.~$\fctDom{\mu} = \lbrace ?x \rbrace$ and $\mu(?x) = \mathsf{no}_3$.
contains a single solution $\mu$ for which $\fctDom{\mu} = \lbrace ?x \rbrace$ and $\mu(?x) = \mathsf{no}_3$;
	%for the sake of readability we write
	i.e.~%
$\mu = \lbrace ?x \rightarrow \mathsf{no}_3 \rbrace$.
\end{example}

\begin{example} \label{Example:Infiniteness4NewText}
	We now consider the BQP $\symBQP_2 = \big\lbrace ( \mathsf{no}_2, \mathsf{succ} , ?x ) , $ $( ?x, \mathsf{succ} , ?y ) , ( ?z, \mathsf{div} , ?x ) \big\rbrace$ with $?x,?y,?z \!\in\! \symAllVariables$ and $\symSeedURIs_2 \!=\! \lbrace \mathsf{no}_2 \rbrace$.
	Under $\cNone$-semantics the query result is empty because the
		%$(\symSeedURIs_2,\cNone,\symBQP_2)$%
		$(\symSeedURIs_2,$ $\cNone,\symBQP_2)$%
	-reach\-able part of $\WoD_\mathsf{inf}$ only consists of
		{\LDdoc}
		%document
	$d_2$ (as in the previous example).
	For $\cAll$ and $\cMatch$ the reachable parts are infinite (and equal):
		Both
		%Both,
		%	%$\ReachPartScBW{\symSeedURIs_2}{\cMatch}{\symBQP_2}{\WoD_\mathsf{inf}}$
		%	$\ReachPartScBW{\symSeedURIs_2}{\cMatch}{\symBQP_2}{\WoD}$
		%and
		%	%$\ReachPartScBW{\symSeedURIs_2}{\cAll}{\symBQP_2}{\WoD_\mathsf{inf}}$,
		%	$\ReachPartScBW{\symSeedURIs_2}{\cAll}{\symBQP_2}{\WoD}$,
	consist of the
		%{\LDdoc}s
		documents
	$d_1, ... \, , d_k , ...$ (as was the case for
		%$\ReachPartScBW{\symSeedURIs_1}{\cAll}{\symBQP_1}{\WoD_\mathsf{inf}}$,
		%$\ReachPartScBW{\symSeedURIs_1}{\cAll}{\symBQP_1}{\WoD}$
		$\cAll$ but not for $\cMatch$
	in the previous example).
	While the query result is also equal for both criteria, it differs significantly from the previous example because it is infinite: $\queryRsltBScW{\symBQP_2}{\symSeedURIs_2}{\cMatch}{\WoD_\mathsf{inf}} = \queryRsltBScW{\symBQP_2}{\symSeedURIs_2}{\cAll}{\WoD_\mathsf{inf}} = \lbrace \mu_1, \mu_2, ... \, \mu_i, ... \rbrace$ where
	\begin{align*}
		\mu_1 &= \lbrace ?x \rightarrow \mathsf{no}_3 , ?y \rightarrow \mathsf{no}_4 , ?z \rightarrow \mathsf{no}_3 \rbrace ,\\
		\mu_2 &= \lbrace ?x \rightarrow \mathsf{no}_3 , ?y \rightarrow \mathsf{no}_4 , ?z \rightarrow \mathsf{no}_6 \rbrace ,\\
		 \text{and, in general: ~ } \mu_i &= \lbrace ?x \rightarrow \mathsf{no}_3 , ?y \rightarrow \mathsf{no}_4 , ?z \rightarrow \mathsf{no}_{(3 i)} \rbrace .
	\end{align*}
\end{example}

\noindent
A special type of CLD queries not covered by the examples are queries that use an empty set of seed {\ID}s. However, it is easily verified that answering such queries is trivial:
\begin{fact} \label{Fact:FinitenessQueryResult:TrivialCase}
	Let $\WoD$ be
		%an arbitrary
		a
	Web of Linked Data.
	For each CLD query $\queryFctBSc{\symBQP}{\symSeedURIs}{c}$ for which $\symSeedURIs = \emptyset$, it holds:
	The set of {\LDdoc}s in the $(\symSeedURIs,c,\symBQP)$-reachable part of $\WoD$ is empty and, thus,
	$\queryRsltBScW{\symBQP}{\symSeedURIs}{c}{\WoD} = \emptyset$.
\end{fact}

\noindent
Due to
	%their triviality, CLD queries with an empty set of seed {\ID}s are
	its triviality, an empty set of seed {\ID}s presents
a special case that we exclude from most of our results.
We now summarize the conclusions that we draw from
	%Examples~\ref{Example:Infiniteness2} to~\ref{Example:Infiniteness4}:
	Examples~\ref{Example:Infiniteness2} and~\ref{Example:Infiniteness4NewText}:

\begin{proposition} \label{Proposition:Finding}
	Let $\symSeedURIs \subset \symAllURIs$ be a finite but nonempty set of seed {\ID}s;
	let $c$ and $c'$ be reachability criteria;
	let $\symBQP$ be a
		%BQP \removable{such that $\queryFctBSc{\symBQP}{\symSeedURIs}{c}$ and $\queryFctBSc{\symBQP}{\symSeedURIs}{c'}$ are CLD queries};
		BQP;
	and let $\WoD$ be an \emph{infinite} Web of Linked Data.
	It holds:
	\begin{enumerate}
		\vspace{-2mm} %\addtolength{\itemsep}{-0.4\baselineskip} % Layout Adjustment
		\item $\ReachPartScBW{\symSeedURIs}{\cNone}{\symBQP}{\WoD}$ is always finite; so is $\queryRsltBScW{\symBQP}{\symSeedURIs}{\cNone}{\WoD}$. \label{Proposition:Finding:Case1}
% 		\item If $c$ is less restrictive than $\cNone$, then $\ReachPartScBW{\symSeedURIs}{c}{\symBQP}{\WoD}$ is either finite or infinite; the same holds for $\queryRsltBScW{\symBQP}{\symSeedURIs}{c}{\WoD}$. \label{Proposition:Finding:Case2}
		\item If $\ReachPartScBW{\symSeedURIs}{c}{\symBQP}{W}$ is finite, then $\queryRsltBScW{\symBQP}{\symSeedURIs}{c}{\WoD}$ is finite. \label{Proposition:Finding:Case3}
		\item If $\queryRsltBScW{\symBQP}{\symSeedURIs}{c}{\WoD}$ is infinite, then $\ReachPartScBW{\symSeedURIs}{c}{\symBQP}{W}$ is infinite. \label{Proposition:Finding:Case4}
		\item If $c$ is less restrictive than $c'$ and $\ReachPartScBW{\symSeedURIs}{c}{\symBQP}{W}$ is finite,
			%then
			~then
		$\ReachPartScBW{\symSeedURIs}{c'}{\symBQP}{W}$ is finite. \label{Proposition:Finding:Case5}
		\item If $c'$ is less restrictive than $c$ and $\ReachPartScBW{\symSeedURIs}{c}{\symBQP}{W}$ is infinite, then $\ReachPartScBW{\symSeedURIs}{c'}{\symBQP}{W}$ is infinite. \label{Proposition:Finding:Case6}
		\item If $c'$ is less restrictive than $c$, then $\queryRsltBScW{\symBQP}{\symSeedURIs}{c}{\WoD} \subseteq \queryRsltBScW{\symBQP}{\symSeedURIs}{c'}{\WoD}$. \label{Proposition:Finding:Case7}
	\end{enumerate}
\end{proposition}

\noindent
Proposition~\ref{Proposition:Finding} provides valuable insight into the dependencies between reachability criteria, the (in)fi\-nite\-ness of reachable parts of an infinite Web, and the (in)fi\-nite\-ness of query results.
%Particularly interesting is the first and the fourth conclusion in Proposition ... because they provide a criterion whether query results are finite or infinite.
In practice, however, we are primarily interested in
	the following questions: Does the execution of a given CLD query reach an infinite number of {\LDdoc}s? Do we have to expect an infinite query result?
	%criteria that allow us to decide if a particular query execution may reach an infinite number of {\LDdoc}s and if we have to expect an infinite query result.
We formalize these questions as (LD machine) decision problems:

\vspace{1ex}
\webproblem{FinitenessReachablePart}
{a (potentially infinite) Web of Linked Data $\WoD$}
{a CLD query $\queryFctBSc{\symBQP}{\symSeedURIs}{c}$ where $\symSeedURIs$ is nonempty and $c$ is less restrictive than $\cNone$}
{Is the $(\symSeedURIs,c,\symBQP)$-reachable part of $\WoD$ finite?}

\vspace{1ex}

\webproblem{FinitenessQueryResult}
{a (potentially infinite) Web of Linked Data $\WoD$}
{a CLD query $\queryFctBSc{\symBQP}{\symSeedURIs}{c}$ where $\symSeedURIs$ is nonempty and $c$ is less restrictive than $\cNone$}
{Is the query result $\queryRsltBScW{\symBQP}{\symSeedURIs}{c}{\WoD}$ finite?}
\vspace{1ex}

\noindent
Unfortunately, it is impossible to define a general algorithm for answering these problems as
	%we show in the following:
	our following
		%results show.
		result shows.
% \begin{theorem} \label{Theorem:Problem:FinitenessReachablePart}
% 	\problemName{FinitenessReachablePart} is not LD machine decidable.
% \end{theorem}
% \begin{theorem} \label{Theorem:Problem:FinitenessQueryResult}
% 	\problemName{FinitenessQueryResult} is not LD machine decidable.
% \end{theorem}
\begin{theorem} \label{Theorem:Problem:Finiteness}
	The problems \problemName{FinitenessReachablePart} and \problemName{Fi\-nitenessQueryResult} are not LD machine decidable.
\end{theorem}

\subsection{Computability of Linked Data Queries} \label{Subsection:Properties:Computability}
\noindent
	%Example~\ref{Example:Infiniteness4}
	Example~\ref{Example:Infiniteness4NewText}
illustrates that some CLD queries may have
	a result that is infinitely large.
	%an infinitely large result.
Even if a query has a finite result it may still be necessary to retrieve infinitely many {\LDdoc}s to ensure that the computed result is complete. Hence, any attempt to answer such queries completely induces a non-terminating computation.

In what follows, we formally analyze feasibility and limitations for computing CLD queries. For this analysis we adopt notions of computability that Abiteboul and Vianu introduce in
	%earlier work on Web queries~\cite{Abiteboul00:QueriesAndComputationOnTheWebArticle}.
	the context of queries over a hypertext-centric view of the WWW~\cite{Abiteboul00:QueriesAndComputationOnTheWebArticle}.
These notions are: \emph{finitely computable queries}, which correspond to the traditional notion of computability; and \emph{eventually computable que\-ries} whose computation may not terminate but each element of the query result will eventually be reported during the computation. While Abiteboul and Vianu define these notions of computability using their concept of a Web machine (cf.~Section~\ref{Subsection:ComputationModel}), our adaptation for Linked Data queries uses an LD machine:

\begin{definition} \label{Definition:FinitelyComputable}
	A Linked Data query $q$  is \definedTerm{finitely computable} if there exists an LD machine which, for any Web of Linked Data $\WoD$ encoded on the Web tape, halts after a finite number of
		%computation
	steps and produces a possible encoding of $q(W)$ on its output tape.
\end{definition}
\begin{definition} \label{Definition:EventuallyComputable}
	A Linked Data $q$ query is \definedTerm{eventually computable} if there exists an LD machine whose computation on any Web of Linked Data $\WoD$ encoded on the Web tape has the following two properties:
	\begin{inparaenum}[1.)]
		\item the word on the output tape at each step of the computation is a prefix of a possible encoding of $q(W)$ and \label{DefinitionRequirement:EventuallyComputable:Prefix}
		\item the encoding $\fctEncName(\mu')$ of any $\mu' \in q(\WoD)$ becomes part of the word on the output tape after a finite number of computation steps. \label{DefinitionRequirement:EventuallyComputable:All}
	\end{inparaenum}
\end{definition}

\todo{Notice, Defs.... about queries in general. Hence, the notion of fin. and evtl. computability also applies to other types of queries that have an other expressivity than CLD queries.}

\noindent
We now analyze the computability of CLD queries. As a preliminary
	%consideration
we identify a dependency between the computation of a CLD query over a particular Web of Linked Data and the (in)fi\-nite\-ness of the corresponding reachable part of that Web:

\begin{lemma} \label{Lemma:ComputabilityCriteria}
	The result of a CLD query $\queryFctBSc{\symBQP}{\symSeedURIs}{c}$ over a (potentially infinite) Web of Linked Data $\WoD$ can be computed by an LD machine that halts after a finite number of computation steps if and only if the $(\symSeedURIs,c,\symBQP)$-reachable part of $\WoD$ is finite.
\end{lemma}

\noindent
	%The following result is an immediate consequence of Lemma~\ref{Lemma:ComputabilityCriteria}, Fact~\ref{Fact:FinitenessQueryResult:TrivialCase}, and~Proposition~\ref{Proposition:Finding}.
	The following, immediate consequence of Lemma~\ref{Lemma:ComputabilityCriteria} is trivial.
	%The following, immediate consequence of Lemma~\ref{Lemma:ComputabilityCriteria} is trivial to show (given Fact~\ref{Fact:FinitenessQueryResult:TrivialCase} and~Proposition~\ref{Proposition:Finding}):
	%As an immediate consequence of Lemma~\ref{Lemma:ComputabilityCriteria} we show the following:

\begin{corollary} \label{Corollary:ComputabilityTrivial}
	CLD queries that use an empty set of seed {\ID}s and CLD queries under $\cNone$-semantics are finitely computable.
\end{corollary}

\noindent
While Corollary~\ref{Corollary:ComputabilityTrivial} covers
	%the special case of CLD queries that use i)~a non-seeding BQP or ii)~reachability criterion $\cNone$,
	some special cases,
the following result identifies the computability of CLD queries in the general case.

\begin{theorem} \label{Theorem:Computability}
	Each CLD query is \Todo{either finitely computable or eventually computable}{This can be made more explicit by introducing the property of \emph{ensuring finiteness} for reachability criteria. Then, all CLD queries that use a reachability criterion which ensures finiteness are finitely computable. All other CLD queries are eventually computable (see ESWC paper).}.
\end{theorem}

\noindent
Theorem~\ref{Theorem:Computability} emphasizes that execution systems for CLD queries do not have to deal with queries that are not
	even eventually
	%(eventually)
computable.
	%The theorem
	Theorem~\ref{Theorem:Computability}
also shows that query computations in the general case are not guaranteed to terminate. The reason for this result is the potential infiniteness of Webs of Linked Data. However, even if a CLD query is only eventually computable, its computation over a particular Web of Linked Data may still terminate (even if this Web is infinite).
	%For instance, the computation of the query in Example~\ref{Example:Infiniteness2} under $\cMatch$-semantics ...
%
	%While Theorem~\ref{Theorem:Computability} provides us with a general statement about the computability of CLD queries,
	Thus,
in practice, we are interested in criteria that allow us to decide whether a particular query execution
	is guaranteed to terminate.
	%terminates.
We formalize this decision problem: % as follows:

\vspace{1ex}
\webproblem{ComputabilityCLD}
{a (potentially infinite) Web of Linked Data $\WoD$}
{a CLD query $\queryFctBSc{\symBQP}{\symSeedURIs}{c}$ where $\symSeedURIs$ is nonempty and $c$ is less restrictive than $\cNone$}
{Does an LD machine exist that
	%computes $\symBQP^c(\WoD)$ and halts?}
	i)~computes $\queryRsltBScW{\symBQP}{\symSeedURIs}{c}{\WoD}$ and ii)~halts?}
\vspace{1ex}

\noindent
Unfortunately:
\begin{theorem} \label{Theorem:Problem:ComputabilityDecision}
	\problemName{ComputabilityCLD} is not LD machine decidable.
\end{theorem}

\noindent
As a consequence of the results in this section we note that any system which executes CLD queries over an infinite Web of Linked Data (such as the WWW) must be prepared for
	%non-terminating queries.
	query executions that do not terminate and that discover an infinite amount of data.
\section{Query Execution Model} \label{Section:ExecModel}
\noindent
In Section~\ref{Section:QueryModel} we use a two-phase approach to define (a family of) semantics for conjunctive queries over Linked Data. A query execution system that would directly implement this two-phase approach would have to retrieve all {\LDdoc}s before it could generate the result for a query. Hence, the first solutions could only be generated after all data links (that qualify according to the used reachability criterion) have been followed recursively. Retrieving the complete set of reachable documents may exceed the resources of the execution system or it may take a prohibitively long time; it is even possible that this process does not terminate at all
	%(cf.~Theorem~\ref{Theorem:Computability}).
	(cf.~Section~\ref{Subsection:Properties:Computability}).
The
	%approach
	link traversal based query execution
that we demonstrate in
	%Example~\ref{Example:QueryExecution}
	Section~\ref{Section:Example}
applies an alternative strategy: % to answer queries over Linked Data:
It intertwines the link traversal based retrieval of data with a pattern matching process that generates solutions incrementally. Due to such an integration of link traversal and result construction it is possible to report first solutions early, even if not all links have been followed and not all data has been retrieved. %Furthermore, this integration enables the execution engine to dynamically (re-)prioritize (i.e.~adapt) URI look-ups based on the discovered data and on the open tasks during the query execution.
To describe
	%this approach
	link traversal based query execution
formally, we introduce an abstract query execution model. In this section we present this model and use it for proving soundness and completeness of the modeled approach.

\subsection{Preliminaries} \label{Subsection:ExecModel:Preliminaries}
\noindent
Usually, queries are executed over a finite structure of data (e.g. an instance of a relational schema or an RDF dataset) that is assumed to be fully available to the execution system. However,
	%the query execution approach that we describe in this paper has to answer
	in this paper we are concerned with
queries over a Web of Linked Data that may be infinite and that is fully unknown at the beginning of a query execution process.
	To learn about such a Web we have to dereference {\ID}s and parse documents that we retrieve. Conceptually, dereferencing an {\ID} corresponds to achieving partial knowledge of the set $D$ and mapping $\adocFct$ with which we model the queried Web of Linked Data $\WoD = \left( D,\dataFct,\adocFct \right)$. Similarly, parsing documents retrieved from the Web corresponds to learning mapping $\dataFct$. To formally represent what we know about a Web of Linked Data at any particular point in a query execution process we introduce the concept of discovered parts.
	%A query execution system obtains data by, first, dereferencing {\ID}s and, then, retrieving and parsing the {\LDdoc}s which it learned about in the first, dereferencing step. Conceptually, these two steps correspond to applying mappings $\adocFct$ and $\dataFct$ with which we model the corresponding Web of Linked Data $\WoD \!=\! \left( D,\dataFct,\adocFct \right)$. Thus, at any point in the query execution process the system only knows about a specific subset of the {\LDdoc}s in $D$; this subset consists of those $d \!\in\! D$ that have already been retrieved by the system. Similarly, the system may not fully know mappings $\adocFct$ and $\dataFct$. To model what the system knows about a Web of Linked Data at an \emph{arbitrary} point in the execution process we introduce the concept of discovered parts.
\begin{definition} \label{Definition:DiscoveredPart}
	A \definedTerm{discovered part} of a Web of Linked Data $\WoD$ is an induced subweb of $\WoD$ that is finite.
\end{definition}

\noindent
We require finiteness for discovered parts of a Web of Linked Data $\WoD$. This requirement models the fact that we obtain information about $\WoD$ only gradually; thus, at any point in a query execution process we only know a finite part of $\WoD$, even if $\WoD$ is infinite.

The (link traversal based) execution of a CLD query $\queryFctBSc{\symBQP}{\symSeedURIs}{c}$ over a Web of Linked Data $\WoD = \left( D,\dataFct,\adocFct \right)$ starts with a discovered part $\mathfrak{D}_\mathsf{init}^{\symSeedURIs,W}$ (of $\WoD$) which contains only those {\LDdoc}s from $\WoD$ that can be retrieved by dereferencing {\ID}s from $\symSeedURIs$; hence, $\mathfrak{D}_\mathsf{init}^{\symSeedURIs,W} = ( D_0,\dataFct_0,\adocFct_0)$ is defined by: \begin{equation} \label{Equation:D_0}
D_0 = \big\lbrace \adocFct(\symURI) \,\big|\, \symURI \in \symSeedURIs \text{ and } \symURI \in \fctDom{\adocFct} \big\rbrace
	%$.%
	\end{equation}

\noindent
In the
	remainder of this section
	%following
we first define how we may use data from a discovered part
	%of a Web
to construct (partial) solutions for a CLD query in an incremental fashion. Furthermore, we formalize how the link traversal approach expands such a discovered part
	in order to construct
	%to enable the construction of
further solutions. Finally, we discuss an abstract procedure that formally captures how the approach intertwines the expansion of discovered parts with the construction of solutions.

\subsection{Constructing Solutions} \label{Subsection:ExecModel:Constructing}
\noindent
The query execution approach that we aim to capture with our query execution model constructs solutions for a query incrementally (cf.~%
	%Example~\ref{Example:QueryExecution}%
	Section~\ref{Section:Example}%
). To formalize the intermediate products of such a construction we introduce the concept of partial solutions.
\begin{definition} \label{Definition:PartialSolution}
	A \definedTerm{partial solution} for
		%a
	CLD query $\queryFctBSc{\symBQP}{\symSeedURIs}{c}$ in a Web of Linked Data $\WoD$ is a pair $(\symBQPpart,\mu)$ where $\symBQPpart \subseteq \symBQP$ and $\mu \in \queryRsltBScW{\symBQPpart}{\symSeedURIs}{c}{\WoD}$.
\end{definition}

\noindent
According to Definition~\ref{Definition:PartialSolution} each partial solution $(\symBQPpart,\mu)$ for a CLD query $\queryFctBSc{\symBQP}{\symSeedURIs}{c}$
	%(in a Web of Linked Data $\WoD$)
is a solution for the CLD query $\queryFctBSc{\symBQPpart}{\symSeedURIs}{c}$ that uses BQP $\symBQPpart$ (instead of $\symBQP$).
Since $\symBQPpart$ is a part of $\symBQP$ we say that partial solutions \emph{cover}
	only
a part of the queries that we want to answer.

The (link traversal based) execution of a CLD query $\queryFctBSc{\symBQP}{\symSeedURIs}{c}$
	over a Web of Linked Data $\WoD$
starts with an \emph{empty partial solution} $\sigma_0 = (\symBQPpart_0,\mu_0)$ which covers the empty part $\symBQPpart_0 = \emptyset$ of $\symBQP$ (i.e. $\fctDom{\mu_0} = \emptyset$). During query execution we (incrementally) extend partial solutions to cover larger parts of $\symBQP$. Those partial solutions that cover the whole query can be reported as solutions for $\queryFctBSc{\symBQP}{\symSeedURIs}{c}$
	in $\WoD$.
However, to extend a partial solution we may use data only from {\LDdoc}s that we have already
	%discovered (and parsed).
	discovered.
Consequently, the following definition formalizes the extension of a partial solution based on a discovered part of a Web of Linked Data.
\begin{definition} \label{Definition:PartSolAugmentation}
	Let $\DiscPartXX{W}$ be a discovered part of a Web of Linked Data $\WoD$;
	let $\queryFctBSc{\symBQP}{\symSeedURIs}{c}$ be a CLD query;
	and let $\sigma = (\symBQPpart,\mu)$ be a partial solution for $\queryFctBSc{\symBQP}{\symSeedURIs}{c}$ in $\WoD$.
	If there exist a {\TP} $tp \in \symBQP \setminus \symBQPpart$ and a {\triple} $t \in \fctAllDataName\bigl( \DiscPartXX{W} \bigr)$ such that $t$ matches $tp$ then
	the \definedTerm{$(t,tp)$-augmentation of $\sigma$ in $\DiscPartXX{W}$}, denoted by $\Aug{t}{tp}{\sigma}{\DiscPartXX{W}}$, is a pair $( \symBQPpart', \mu' )$ such that $\symBQPpart' = \symBQPpart \cup \lbrace tp \rbrace$ and $\mu'$ extends $\mu$ as follows:
	\begin{inparaenum}[1.)]
% 		\vspace{-2mm} \addtolength{\itemsep}{-0.5\baselineskip} % Layout Adjustment
		\item $\fctDom{\mu'} = \fctVarsName(\symBQPpart')$ and
		\item $\mu'[\symBQPpart'] = \mu[\symBQPpart] \cup \lbrace t \rbrace$.
	\end{inparaenum}
% 	To denote the $(t,\! tp)$-augmentation of $\sigma$ in $\DiscPartXX{W}\!$ we write $\Aug{t}{tp}{\sigma}{\DiscPartXX{W}}$.
\end{definition}

\noindent
The following proposition shows that the result of augmenting a partial solution is again a partial solution, as long as the discovered part of the Web that we use for such an augmentation is fully contained in the reachable part of the Web.
\begin{proposition} \label{Proposition:PartSolAugmentationSoundness}
	Let $\DiscPartXX{W}$ be a discovered part of a Web of Linked Data $\WoD$ and let $\queryFctBSc{\symBQP}{\symSeedURIs}{c}$ be a CLD query.
	If $\DiscPartXX{W}$ is an induced subweb of the $(\symSeedURIs,c,\symBQP)$-reachable part of $\WoD$
	and $\sigma$ is a partial solution for $\queryFctBSc{\symBQP}{\symSeedURIs}{c}$ in $\WoD$,
	then
		%any $(t,tp)$-augmentation of $\sigma$ in $\DiscPartXX{W}$
		$\Aug{t}{tp}{\sigma}{\DiscPartXX{W}}$
	is also a partial solution for $\queryFctBSc{\symBQP}{\symSeedURIs}{c}$ in $\WoD$%
		%.
		, for all possible $t$ and $tp$.
\end{proposition}

\subsection{Traversing Data Links} \label{Subsection:ExecModel:Traversing}
\noindent
During query execution we may traverse data links to expand the discovered part. Such an expansion may allow us to compute further augmentations for partial solutions.
	%In Example~\ref{Example:QueryExecution} we achieve 
	The link traversal based approach implements
such an expansion by dereferencing {\ID}s that
	%we find
	occur
in valuations $\mu$ of partial
	%solutions.
	solutions (cf.~%
		%Example~\ref{Example:QueryExecution}).
		Section~\ref{Section:Example}).
Formally, we define
	such a
	%this
valuation based expansion as follows:

% \begin{definition} \label{Definition:MatchBasedExpansion}
% 	Let $\WoD = ( D,\dataFct,\adocFct )$ be a Web of Linked Data;
% 	let $\DiscPartXX{W} = ( \Disc{D},\Disc{\dataFct},\Disc{\adocFct} )$ be a discovered part of $\WoD$;
% 	let $c$ be a reachability criterion;
% 	let $\sigma = (\symBQPpart,\mu)$ be a partial solution for a CLD query $\symBQP^c$ in $\WoD$.
% %
% 	The \definedTerm{$\sigma$-expansion} of $\DiscPartXX{W}$ (in the context of $\symBQP$ and $c$) is an induced subweb $( \NewDisc{D},\NewDisc{\dataFct},\NewDisc{\adocFct} )$ of $\WoD$ such that $\NewDisc{D} = \Disc{D} \cup \ExpMDelta{\Disc{D}}{\symBQP,c}{\sigma}$ with
% 	\begin{align*}
% 		\ExpMDelta{\Disc{D}}{\symBQP,c}{\sigma} \!=\!
% 		\Big\lbrace
% 			\adocFct\bigl( \mu(v) \bigr)
% 			\, \big| \, & 
% 			v \in \fctDom{\mu} \, \land &&\\
% 			& \mu(v) \in \symAllURIs \, \land &&\\
% 			& \mu(v) \in \fctDom{\adocFct} \, \land &&\\
% 			& \exists \, tp \in \symBQP : \exists \, d \in \Disc{D} : \exists \, t \in \dataFct(d) : \mu[tp]=t \land c(d,t,\mu(v),\symBQP) = \true
% 		\Big\rbrace
% 	\end{align*}
% 	To denote the $\sigma$-expansion of $\DiscPartXX{W}$ (in the context of $\symBQP$ and $c$) we write $\expSQcD{\sigma}{\symBQP}{c}{\DiscPartXX{W}}$.
% \end{definition}

\begin{definition} \label{Definition:MatchBasedExpansion}
	Let $\DiscPartXX{W} = ( \Disc{D},$ $\Disc{\dataFct},\Disc{\adocFct} )$ be a discovered part of a Web of Linked Data $\WoD = ( D,\dataFct,\adocFct )$
	and let $\mu$ be a valuation.
	The \definedTerm{$\mu$-expansion} of $\DiscPartXX{W}$ in $\WoD$, denoted by $\expMWD{\mu}{\WoD}{\DiscPartXX{W}}$, is an induced subweb $( \NewDisc{D},\NewDisc{\dataFct},\NewDisc{\adocFct} )$ of $\WoD$, defined by $\NewDisc{D} = \Disc{D} \cup \ExpMDeltaMW{\mu}{\WoD}$ where
	\begin{align*}
		\ExpMDeltaMW{\mu}{\WoD} =
		\big\lbrace
			\adocFct\bigl( \mu(?v) \bigr)
			\, \big| \, 
			& ?v \in \fctDom{\mu} \\
			& \text{and } \mu(?v) \in \fctDom{\adocFct}
		\big\rbrace
	\end{align*}
% 	To denote the $\mu$-expansion of $\DiscPartXX{W}$ in $\WoD$ we write $\expMWD{\mu}{\WoD}{\DiscPartXX{W}}$.
\end{definition}

\noindent
The following propositions show that expanding discovered parts is a
	%monotone
	monotonic
operation (Proposition~\ref{Proposition:MatchBasedExpansionMonotonicity})
and that the set of all possible discovered parts is closed under this operation (Proposition~\ref{Proposition:MatchBasedExpansionClosedness}).
\begin{proposition} \label{Proposition:MatchBasedExpansionMonotonicity}
	Let $\DiscPartXX{W}$ be a discovered part of a Web of Linked Data $\WoD$,
	then $\DiscPartXX{W}$ is an induced subweb of $\expMWD{\mu}{\WoD}{\DiscPartXX{W}}$, for all possible $\mu$.
\end{proposition}
\begin{proposition} \label{Proposition:MatchBasedExpansionClosedness}
	Let $\DiscPartXX{W}$ be a discovered part of a Web of Linked Data $\WoD$, then $\expMWD{\mu}{\WoD}{\DiscPartXX{W}}$ is also a discovered part of $\WoD$, for all possible $\mu$.
\end{proposition}

\noindent
We motivate the expansion of discovered parts of a queried Web of Linked Data by the possibility that data obtained from additionally discovered documents may allow us to construct more (partial) solutions. However, Proposition~\ref{Proposition:PartSolAugmentationSoundness} indicates that the augmentation of partial solutions is only sound if the discovered part that we use for the augmentation is fully contained in the corresponding reachable part of the Web. Thus, in order to use a discovered part that has been expanded based on (previously constructed) partial solutions, it should be guaranteed that the expansion never exceeds the reachable part. Under $c_{\mathsf{Match}}$-semantics we have such a
	%guarantee as the following proposition shows.
	guarantee:
\begin{proposition} \label{Proposition:MatchBasedExpansionBoundedness}
	Let $\sigma = (\symBQPpart,\mu)$ be a partial solution for a CLD query $\queryFctBSc{\symBQP}{\symSeedURIs}{\cMatch}$
		(under $\cMatch$-semantics)
	in a Web of Linked Data $\WoD$;
	and let $\ReachPartScBW{\symSeedURIs}{\cMatch}{\symBQP}{\WoD}$
		%be
		denote
	the $(\symSeedURIs,\cMatch,\symBQP)$-reachable part of $\WoD$.
	If a discovered part $\DiscPartXX{W}$ of $\WoD$ is an induced subweb of $\ReachPartScBW{\symSeedURIs}{\cMatch}{\symBQP}{\WoD}$,
	then
		%the $\mu$-expansion of that discovered part
		$\expMWD{\mu}{\WoD}{\DiscPartXX{W}}$
	is also an induced subweb of $\ReachPartScBW{\symSeedURIs}{\cMatch}{\symBQP}{\WoD}$.
\end{proposition}

\noindent
We explain the restriction to $c_{\mathsf{Match}}$-semantics in Proposition~\ref{Proposition:MatchBasedExpansionBoundedness} as follows:
During link traversal based query execution we expand the discovered part of the queried Web only by using valuations that occur in partial solutions (cf.~%
	%Example~\ref{Example:QueryExecution}).
	Section~\ref{Section:Example}).
Due to this approach, we only dereference {\ID}s for which there exists a {\triple} that matches a triple pattern in our query. 
Hence, this approach indirectly enforces query pattern based reachability (cf.~Section~\ref{Subsection:QueryModel:Reachability}). As a result, link traversal based query execution only supports CLD queries under $c_{\mathsf{Match}}$-semantics; so does our query execution model.

% In the remainder of this paper we focus on $c_{\mathsf{Match}}$-semantics; we abbreviate the term ``partial $c_{\mathsf{Match}}$-solution'' by \emph{partial solution}.

\subsection{Combining Construction and Traversal} \label{Subsection:ExecModel:CombiningConstructionAndTraversal}
\noindent
Although
	%an incremental expansion of the discovered part of a reachable subweb and a recursive augmentation of partial solutions
	incrementally expanding the discovered part of the reachable subweb and recursively augmenting partial solutions
may be understood as separate processes, the
	%main
idea of link traversal based query execution
	%(as demonstrated Example~\ref{Example:QueryExecution})
	%(as demonstrated Section~\ref{Section:Example})
is to combine these two processes.
	%Algorithm~\ref{Procedure:MatchBasedExec} illustrates an abstract procedure
	%	%, called $ltbExec$,
	%which captures this idea formally. In the following we introduce this procedure.
	We now introduce an abstract procedure which captures this idea formally.

As a basis for
	%the procedure we formalize the state of a query execution as
	our formalization we represent the state of a query execution by
a pair $\bigl( \mathfrak{P},\mathfrak{D} \bigr)$; $\mathfrak{P}$ denotes the (finite) set of partial solutions that have already been constructed
	%;
	at the current point in the execution process;
$\mathfrak{D}$ denotes the currently discovered part of the queried Web of Linked Data. As discussed before, we initialize $\mathfrak{P}$ with the empty partial solution $\sigma_0$ (cf.~Section~\ref{Subsection:ExecModel:Constructing}) and $\mathfrak{D}$ with $\mathfrak{D}_\mathsf{init}^{\symSeedURIs,W}$ (cf.~Section~\ref{Subsection:ExecModel:Preliminaries}). During the query execution process $\mathfrak{P}$ and $\mathfrak{D}$ grow monotonically: We augment partial solutions from $\mathfrak{P}$ and add the results back to $\mathfrak{P}$; additionally, we use partial solutions from $\mathfrak{P}$ to expand $\mathfrak{D}$.
However, conceptually we combine these two types of tasks, augmentation and expansion, into a single type: % which we call \emph{{\AEtask}s}.
\begin{definition} \label{Definition:AETask}
	Let $\queryFctBSc{\symBQP}{\symSeedURIs}{\cMatch}$ be a CLD query (under $\cMatch$-se\-man\-tics);
	let $\bigl( \mathfrak{P},\mathfrak{D} \bigr)$ represent a state of a (link traversal based) execution of $\queryFctBSc{\symBQP}{\symSeedURIs}{\cMatch}$.
	An \definedTerm{{\AEtask} for $\bigl( \mathfrak{P},\mathfrak{D} \bigr)$} is a tuple $(\sigma,t,tp)$ for which it holds
	i)~$\sigma = (\symBQPpart,\mu) \in \mathfrak{P}$,
	ii)~$t \in \fctAllDataName\bigl( \mathfrak{D} \bigr)$,
	iii)~$tp \in \symBQP \setminus \symBQPpart$, and
	iv)~$t$ matches $tp$	.
\end{definition}

\noindent
	%We now describe the effect of performing {\AEtask}s formally: Let $(\sigma,t,tp)$ be an {\AEtask} for $\bigl( \mathfrak{P},\mathfrak{D} \bigr)$. Performing this task includes
	Performing an {\AEtask} $(\sigma,t,tp)$ for $\bigl( \mathfrak{P},\mathfrak{D} \bigr)$ comprises two steps: 
1.)~changing $\mathfrak{P}$ to $\mathfrak{P} \cup \lbrace (\symBQPpart'\!,\mu') \rbrace$, where $(\symBQPpart'\!,\mu') = \Aug{t}{tp}{\sigma}{\mathfrak{D}}$ is the $(t,tp)$-augmentation of $\sigma$ in $\mathfrak{D}$, and 2.)~expanding $\mathfrak{D}$ to
	%$\expMWD{\mu'}{\WoD}{\mathfrak{D}}$.
	the $\mu'$-expansion of $\mathfrak{D}$ in $\WoD$.
%
% 	Furthermore, the valuation $\mu'$ from the newly constructed partial solution $(\symBQPpart'\!,\mu')$ must be reported as a solution in $\symBQP^{c_{\mathsf{Match}}}(W)$ if $(\symBQPpart'\!,\mu')$ covers $\symBQP$ as a whole, that is, if $\symBQPpart'=\symBQP$.
%
Notice, constructing the augmentation in the first step is always possible because the prerequisites for {\AEtask}s, as given in Definition~\ref{Definition:AETask}, correspond to the prerequisites for augmentations (cf.~Definition~\ref{Definition:PartSolAugmentation}).
However, not all possible {\AEtask}s may actually change $\mathfrak{P}$ and $\mathfrak{D}$; instead, some tasks $(\sigma,t,tp)$ may produce an augmentation $\Aug{t}{tp}{\sigma}{\mathfrak{D}}$ that turns out to be a partial solution which has already been produced for another task.
Thus, to guarantee progress during a query execution process we must
	%thus avoid performing {\AEtask}s that do not
	only perform those {\AEtask}s that 
produce new augmentations.
	%Formally, we define tasks that guarantee progress as follows:
	To identify such tasks we introduce the concept of \emph{open {\AEtask}s}.
\begin{definition} \label{Definition:OpenTasks}
	An {\AEtask} $(\sigma,t,tp)$ for the state $( \mathfrak{P},\mathfrak{D} )$ of a link traversal based query execution is \definedTerm{open} if $\Aug{t}{tp}{\sigma}{\mathfrak{D}} \notin \mathfrak{P}$.
	To denote the set of all open {\AEtask}s for $( \mathfrak{P},\mathfrak{D} )$ we write $\AllOpen{\mathfrak{P}}{\mathfrak{D}}$.
\end{definition}

\noindent
% \removable{
% 	Due to our definition of open {\AEtask}s it is guaranteed that performing such a task always results in adding a new partial solution to $\mathfrak{P}$. $\mathfrak{D}$, in contrast, may not always be affected by performing an open {\AEtask}. Instead, for
% 		the valuation $\mu'$ of
% 	a newly constructed partial solution $(\symBQPpart'\!,\mu')$ it is possible that all {\LDdoc}s in $\ExpMDeltaMW{\mu'}{\WoD}$ (cf.~Definition~\ref{Definition:MatchBasedExpansion}) have already been discovered based on other partial solutions (which have been constructed before). Hence, in such a case $\expMWD{\mu'}{\WoD}{\mathfrak{D}}$ is the same as $\mathfrak{D}$. However, we shall see that performing (all) open {\AEtask}s is sufficient for discovering all reachable {\LDdoc}s and, thus, for answering a query completely.\par}
%
%
We now use the introduced concepts to present our abstract procedure $ltbExec$ (cf.~Algorithm~\ref{Procedure:MatchBasedExec}) with which we formalize the general idea of link traversal based query execution. After initializing $\mathfrak{P}$ and $\mathfrak{D}$
	%(cf.~lines~\ref{Line:InitP} and~\ref{Line:InitD} in Algorithm~\ref{Procedure:MatchBasedExec}),
	(lines~\ref{Line:InitP} and~\ref{Line:InitD} in Algorithm~\ref{Procedure:MatchBasedExec}),
the procedure amounts to a continuous execution of open {\AEtask}s.
We represent this continuous process by a loop (lines~\ref{Line:WhileBegin} to~\ref{Line:WhileEnd}); each iteration of this loop performs an open {\AEtask} (lines~\ref{Line:Augmentation} to~\ref{Line:ChangeD}) and checks whether the newly constructed partial solution $(\symBQPpart'\!,\mu')$ covers the executed CLD query as a whole, in which case the valuation $\mu'$ in $(\symBQPpart'\!,\mu')$ must be reported as a solution for the query (line~\ref{Line:ReportSolution}).
We emphasize that the set $\AllOpen{\mathfrak{P}}{\mathfrak{D}}$ of all open {\AEtask}s always changes when $ltbExec$ performs such a task.
The loop terminates when no more open {\AEtask}s
	for (the current) $\bigl( \mathfrak{P},\mathfrak{D} \bigr)$
	exist (which may never be the case as we know from Lemma~\ref{Lemma:ComputabilityCriteria}).
	%exist. However, such a termination is only possible if the $(\symSeedURIs,\cMatch,\symBQP)$-reachable part of the queried Web of Linked Data is finite (cf.~Lemma~\ref{Lemma:ComputabilityCriteria}).

\begin{algorithm}[t]
	\caption{\, $ltbExec(\symSeedURIs,\symBQP,W)$ -- Report all $\mu \in \queryRsltBScW{\symBQP}{\symSeedURIs}{\cMatch}{\WoD}$.} \label{Procedure:MatchBasedExec}
	\begin{algorithmic}[1]
		\STATE {$\mathfrak{P} := \lbrace \sigma_0 \rbrace$} \label{Line:InitP}
% 		\hspace{5mm} \COMMENT{$\sigma_0 = (\emptyset,\mu_0)$ with $\fctDom{\mu_0} = \emptyset$ }
		\STATE {$\mathfrak{D} := \mathfrak{D}_\mathsf{init}^{\symSeedURIs,W}$ }
% 		\STATE {$\mathfrak{D} := ( D_0,\dataFct_0,\adocFct_0)$ } \\
% 			\hspace{16mm} \COMMENT{$( D_0,\dataFct_0,\adocFct_0)$ is an induced subweb of $\WoD$ defined \\
% 			\hspace{16mm} // \, by $D_0 \!=\! \lbrace \adocFct(\symURI) \,|\, \symURI \!\in\! \fctIDsName(\symBQP) \land \symURI \in \fctDom{\adocFct} \rbrace$
% 			}
		\label{Line:InitD}

		\medskip
		\WHILE {$\AllOpen{\mathfrak{P}}{\mathfrak{D}} \neq \emptyset$} \label{Line:WhileBegin}
			\STATE {Choose open {\AEtask} $(\sigma,t,tp) \in \AllOpen{\mathfrak{P}}{\mathfrak{D}}$} \label{Line:SelectTask}

			\medskip
			\STATE {$(\symBQPpart',\mu') := \Aug{t}{tp}{\sigma}{\mathfrak{D}}$} \label{Line:Augmentation}
			\STATE {$\mathfrak{P} := \mathfrak{P} \cup \big\lbrace (\symBQPpart',\mu') \big\rbrace$} \hspace{3mm} \COMMENT{indirectly changes $\AllOpen{\mathfrak{P}}{\mathfrak{D}}$} \label{Line:ChangeP}
			\STATE {$\mathfrak{D} := \expMWD{\mu'}{\WoD}{\mathfrak{D}}$} \label{Line:ChangeD}

			\medskip
			\STATE {\textbf{if} $\symBQPpart' = \symBQP$ \textbf{then} report $\mu'$ \textbf{endif}} \label{Line:ReportSolution}

			\medskip
		\ENDWHILE \label{Line:WhileEnd}
	\end{algorithmic}
\end{algorithm}

\todo{add a paragraph (or a whole subsection) that discusses the relationship between this procedure and the theoretical, LD machine based analysis (for instance, what is the relationship between finitely/eventually computable queries and termination of $ltbExec(\symSeedURIs,\symBQP,W)$)}

We emphasize the abstract nature of Algorithm~\ref{Procedure:MatchBasedExec}. The fact that we model
	$ltbExec$
	%link traversal based query execution
as a single loop which performs (open) {\AEtask}s sequentially,
	%should not be understood to
	does not
imply that the
	link traversal based query execution paradigm
	%approach
has to be implemented in such a form. Instead, different implementation approaches are possible, some of which have already been proposed in the
	literature~\cite{Hartig11:HeuristicForQueryPlanSelection, Hartig09:QueryingTheWebOfLD,Ladwig10:LinkedDataQueryProcessingStrategies,Ladwig11:SIHJoin}.
	%literature. For instance, Ladwig and Tran present an implementation that uses asynchronously connected operators, each of which executes a symmetric hash join algorithm~\cite{Ladwig10:LinkedDataQueryProcessingStrategies,Ladwig11:SIHJoin}. In terms of our model each of these operators (implicitly) performs a particular subset of all open {\AEtask}s.
In contrast to
	%these, concrete approaches
	the concrete (implementable) algorithms discussed
		%by Ladwig and Tran~\cite{Ladwig10:LinkedDataQueryProcessingStrategies,Ladwig11:SIHJoin} as well as in our earlier work~\cite{Hartig11:HeuristicForQueryPlanSelection, Hartig09:QueryingTheWebOfLD},
		in this earlier work,
we understand Algorithm~\ref{Procedure:MatchBasedExec} as an instrument for presenting and for studying the general idea that is common to all link traversal based query execution approaches.
% and for analyzing (((and comparing))) implementations of this approach.

\todo{compare Algorithm~\ref{Procedure:MatchBasedExec} / $ltbExec$ to Algorithm~\ref{Algorithm:CLDQueryMachine}}
\todo{discuss why we use ``(arbitrary)'' in the following lemmas}

\subsection{Application of the Model}
\noindent
Based on our query execution model we now show that the idea of link traversal based query execution is sound and complete, that is, the set of all valuations reported by
	%an execution of
$ltbExec(\symSeedURIs,\symBQP,W)$ is equivalent to the query result $\queryRsltBScW{\symBQP}{\symSeedURIs}{\cMatch}{\WoD}$. Formally:
\begin{theorem} \label{Theorem:MatchBasedExec}
	Let $\WoD$ be a Web of Linked Data and let $\queryFctBSc{\symBQP}{\symSeedURIs}{\cMatch}$ be a CLD query (under $\cMatch$-semantics).
	\begin{itemize}
		\vspace{-2mm} \addtolength{\itemsep}{-0.5\baselineskip} % Layout Adjustment
		\item Soundness: \propositionMain{For any valuation $\mu$ reported by an execution of $ltbExec(\symSeedURIs,\symBQP,W)$ holds $\mu \in \queryRsltBScW{\symBQP}{\symSeedURIs}{\cMatch}{\WoD}$.}
		\item Completeness: \propositionMain{Any $\mu \in \queryRsltBScW{\symBQP}{\symSeedURIs}{\cMatch}{\WoD}$ will eventually be reported by any execution of $ltbExec(\symSeedURIs,\symBQP,W)$.}
	%\end{enumerate}
	\end{itemize}
\end{theorem}

\noindent
Theorem~\ref{Theorem:MatchBasedExec} formally verifies the applicability of link traversal based query execution for answering conjunctive queries over a Web of Linked Data. For experimental evaluations that demonstrate the feasibility of link traversal based execution of queries over Linked Data on the WWW
	%(using a system that implements the iterator based approach)
we refer to~\cite{Hartig11:HeuristicForQueryPlanSelection,Hartig09:QueryingTheWebOfLD,Ladwig10:LinkedDataQueryProcessingStrategies,Ladwig11:SIHJoin}.
We note, however, that the implementation approaches used for these evaluations do not
	%consider a set $\symSeedURIs$ of seed {\ID}s that is specified explicitly.
	allow for an explicit specification of seed {\ID}s $\symSeedURIs$.
	%explicitly allows for specifying seed {\ID}s.
Instead, these approaches use the {\ID}s in the BQP of a query as seed and, thus, only support CLD queries $\queryFctBSc{\symBQP}{\symSeedURIs}{c}$ for which $\symSeedURIs = \fctIDsName(\symBQP)$.
Theorem~\ref{Theorem:MatchBasedExec} highlights that this is a limitation of these particular implementation approaches and not a general property of link traversal based query execution.
%\section{Implementations} \label{Section:Implementations}
%\noindent

In the remainder of this section we use our (abstract) execution model to analyze the iterator based implementation of link traversal based query execution that we introduce in~\cite{Hartig11:HeuristicForQueryPlanSelection,Hartig09:QueryingTheWebOfLD}. The analysis of this implementation approach is particularly interesting because this approach trades completeness of query results for the guarantee that all query executions terminate as we shall see.

The implementation approach applies a synchronized pipeline of operators that evaluate the BQP
	$\symBQP = \lbrace tp_1 , ... \, , tp_n \rbrace$
of a CLD query in a fixed order. This pipeline is implemented as a chain of iterators $I_1 , ... \, , I_n$; iterator $I_k$ is responsible for triple pattern $tp_k$ (for all $1 \leq k \leq n$) from the
	%\emph{ordered BQP}\footnote{We represent an ordered BQP as a list, denoted by comma-separated elements enclosed in brackets.} $\bar{\symBQP} = \lbrack tp_1 , ... \, , tp_n \rbrack$.
	\emph{ordered} BQP.
While the selection of an order for the BQP is
	%a query
	an
optimization problem~\cite{Hartig11:HeuristicForQueryPlanSelection}, we assume a given order for the following analysis (in fact, the order is irrelevant for the analysis).
Each iterator $I_k$ provides valuations that are solutions for CLD query $\queryFctBSc{\symBQPpart_k}{\symSeedURIs}{\cMatch}$ where $\symBQPpart_k=\lbrace tp_1 , ... \, , tp_k \rbrace$. To determine these solutions each iterator
	$I_k$
executes the following four steps repetitively:
First, $I_k$ consumes a valuation $\mu'$ from its direct predecessor and applies this valuation to its triple pattern $tp_k$, resulting in a triple pattern $tp_k' = \mu'[tp_k]$; second, $I_k$ (tries to) generate solutions by finding matching triples for $tp_k'$ in the query-local dataset; third, $I_k$
	%ensures that the query-local dataset contains the data from all these {\LDdoc}s that can be retrieved by looking up all $\symURI \!\in\! \fctIDsName( tp_k' )$.
	uses the generated solutions to expand the query-local dataset;
and, fourth, $I_k$ (iteratively) reports each of the generated solutions.
For a more detailed description
	of this implementation approach
we refer to~\cite{Hartig11:HeuristicForQueryPlanSelection}.

In terms of our abstract
	%query
execution model, each iterator performs a particular subset of all possible open {\AEtask}s: For each open {\AEtask} $(\sigma,t,tp)$ performed by iterator $I_k$ it holds i)~$tp=tp_k$ and ii)~$\sigma = (\symBQPpart_{k-1},\mu)$ where $\symBQPpart_{k-1} = \lbrace tp_1 , ... \, , tp_{k-1} \rbrace$. However, $I_k$ may not perform all (open) {\AEtask}s which have these properties.
\begin{lemma} \label{Lemma:IteratorFiniteness}
	During an iterator
		%based
	execution of an arbitrary CLD query $\queryFctBSc{\symBQP}{\symSeedURIs}{\cMatch}$
		%(under $\cMatch$-semantics)
		(that uses $\cMatch$)
	over an arbitrary Web of Linked Data $\WoD$ it holds: The set of {\AEtask}s performed by each iterator is finite.
\end{lemma}

\noindent
Based on Lemma~\ref{Lemma:IteratorFiniteness} we easily see that an iterator
	%based
execution of a CLD query
	$\queryFctBSc{\symBQP}{\symSeedURIs}{\cMatch}$
	%(under $\cMatch$-semantics)
	%(that uses $\cMatch$)
may not perform all possible (open) {\AEtask}s.
	%Instead, the set of {\AEtask}s performed during such an execution is always finite and, thus,
	Thus,
we may show the following result as a corollary of Lemma~\ref{Lemma:IteratorFiniteness}.

\begin{theorem} \label{Theorem:Iterator}
	Any iterator
		based
	execution of a CLD query $\queryFctBSc{\symBQP}{\symSeedURIs}{\cMatch}$ (that uses $\cMatch$) over an arbitrary Web of Linked Data $\WoD$ reports a finite subset of $\queryRsltBScW{\symBQP}{\symSeedURIs}{\cMatch}{\WoD}$ and terminates.
\end{theorem}

\noindent
Theorem~\ref{Theorem:Iterator} shows that the analyzed implementation of link traversal based query execution trades completeness of query results for the guarantee that all query executions terminate. The degree
	%of completeness
	to which the reported subset of a query result is complete
depends on the order selected for the
	%executed BQP.
	BQP of the executed query as our experiments in~\cite{Hartig11:HeuristicForQueryPlanSelection} show.
% Due to the lack of a-priori information about 
% Since information about which portion of the reachable part 
A formal analysis of this dependency is part of our future work.

\todo{emphasize that a formal analysis of this dependency is future work (and that the presented exec.model provides a ``good'' basis for such an analysis)}

\todo{Add a theorem that shows that the completeness (degree) is undecidable because we would have to execute the query using an implementation approach (different from the iterator implementation) that guarantees completeness ?}

\section{Related Work} \label{Section:RelatedWork}
\noindent
Since its emergence the
	%WWW
	World Wide Web
has spawned research to adapt de\-clar\-a\-tive query languages for retrieval of information from the
	%WWW (e.g.~\cite{Konopnicki95:WWWQuerySystemW3QS,Lakshmanan96:QueryingAndRestructuringTheWeb,Mendelzon97:QueryingTheWWW}).
	%Web~\cite{Florescu98:DBTechniquesForWWW}.
	WWW~\cite{Florescu98:DBTechniquesForWWW}.
Most of these works
	%are based on an understanding of
	understand
	%model
the WWW as a graph of objects
	%(e.g.~pages)
interconnected by hypertext links; in some models objects have certain attributes (e.g.~title,
	modification date)%
	%size)%
~\cite{%Mendelzon97:QueryingTheWWW,
Mendelzon98:FormalModelsOfWebQueries}
or
	%some
	an
	%kind of an
internal structure~\cite{Guan98:QueriesQverTheWWW,Konopnicki95:WWWQuerySystemW3QS}. Query languages
	%proposed and investigated
	studied
in this context allow a user to either ask for specific objects~\cite{Konopnicki95:WWWQuerySystemW3QS}, for their
	%attributes~\cite{Mendelzon97:QueryingTheWWW},
	attributes~\cite{Mendelzon98:FormalModelsOfWebQueries},
or for specific object content~\cite{Guan98:QueriesQverTheWWW}.
However, there is no explicit connection between data that may be obtained from different objects (in contrast to the more recent idea of Linked Data).
	%Therefore, approaches to answer queries that combine data from multiple objects have been inherently limited or they required significant effort to
	%	%detect and resolve co-references~\cite{Elmagarmid07:DuplicateRecordDetectionSurvey,Florescu98:DBTechniquesForWWW}.
	%	address the record linkage problem~\cite{Elmagarmid07:DuplicateRecordDetectionSurvey,Florescu98:DBTechniquesForWWW}.
Nonetheless, some of the foundational work such as~\cite{Abiteboul00:QueriesAndComputationOnTheWebArticle} and~\cite{Mendelzon98:FormalModelsOfWebQueries} can be adapted to query execution over a Web of Linked Data. In this paper we analyze the computability of CLD queries by adopting Abiteboul and Vianu's notions of computability~\cite{Abiteboul00:QueriesAndComputationOnTheWebArticle}, for which we have to adapt their machine model of computation on the Web.

In addition to the
	early
	%aforementioned
work on Web queries, query execution over Linked Data on the WWW has attracted much attention recently. In~\cite{Hartig10:DBPerspectiveOnConsumingLD} we provide an overview of different approaches and refer to the relevant literature. However, the only work we are aware of that formally captures the concept of Linked Data and provides a well-defined semantics for queries in this context is Bouquet et al.~\cite{Bouquet09:QueryingWebOfData}.
	In contrast to our more abstract, technology-independent data model, their
	%Their
focus is Linked Data on the WWW, implemented using concrete technologies such as URIs and RDF. They adopt the common understanding of a set of RDF triples as graphs%
	%, that is, the elements in the first and third position of the triples represent vertices that are connected by a directed edge; the same element in different triples denotes the same vertice in the corresponding graph%
~\cite{Klyne04:RDF}. Consequently, Bouquet et al.~model a Web of Linked Data as a ``graph space'', that is, a set of RDF graphs, each of which is associated with a URI that, when dereferenced on the WWW, allows a system to obtain that graph. Hence, RDF graphs in Bouquet et al.'s graph space correspond to the {\LDdoc}s in our data model; the URIs associated with RDF graphs in a graph space have a role similar to that of those {\ID}s in our data model for which the corresponding mapping $\adocFct$ returns an actual {\LDdoc} (i.e.~all {\ID}s in $\fctDom{\adocFct}$). Therefore, RDF graphs in a graph space form another type of (higher level) graph, similar to the Web link graph in our model (although, Bouquet et al.~do not define that graph explicitly). Based on their data model, Bouquet et al.~define three types of query methods for conjunctive queries: a bounded method which only uses those RDF graphs that are referred to in queries, a navigational method which corresponds to
	%CLD queries under 
	our query model,
and a direct access method which assumes an oracle that provides all RDF graphs which are ``relevant'' for a given query. For the navigational method the authors define a notion of reachability that allows a query execution system to follow all data links. Hence, the semantics of queries using this navigational method is equivalent to CLD queries under $\cAll$-semantics in our query model. Bouquet et al.'s navigational query model does not support other, more restrictive notions of reachability, as is possible with our model. Furthermore, Bouquet et al.~do not
	discuss the computability of queries and the infiniteness of the WWW.
	%analyze the computability of queries and they do not discuss the infiniteness of the WWW.
\section{Conclusions and Further Work} \label{Section:Conclusion}
\noindent
Link traversal based query execution is a novel query execution approach tailored to the Web of Linked Data. The ability to discover data from unknown sources is its most distinguishing advantage over traditional query execution paradigms which assume a fixed set of potentially relevant data sources beforehand. In this paper we provide a formal foundation for this new approach.

We introduce
	%a query model that proposes
	a family of
well-defined semantics for conjunctive Linked Data queries, taking into account the limited data access capabilities that are typical for the WWW. We show that the execution of such queries may not terminate
	(cf.~Theorem~\ref{Theorem:Computability})
because --due to the existence of data generating servers-- the WWW is infinite (at any point in time). Moreover, queries may have a result that is infinitely large. We show that it is impossible to provide an algorithm for deciding whether any given query (in our model) has a finite
	result (cf.~Theorem~\ref{Theorem:Problem:Finiteness}).
	%result.
Furthermore, it is also impossible to decide (in general) whether a query execution
	terminates (cf.~Theorem~\ref{Theorem:Problem:ComputabilityDecision}),
	%terminates,
even if the expected result would be known to be finite.

In addition to our query model we introduce an execution model that formally captures the link traversal based query execution par\-a\-digm. This model abstracts from any particular approach to implement this paradigm. Based on this model we prove that the
	%approach
	general idea of link traversal based query execution
is sound and complete for conjunctive Linked Data
	queries (cf.~Theorem~\ref{Theorem:MatchBasedExec}).
	%queries.

Our future work focuses on more expressive types of Linked Data queries. In particular, we aim to study which other features of query languages such as SPARQL are feasible in the context of querying a Web of Linked Data and what the
	%(theoretical)
implications of supporting such features are.
Moreover, we will extend our models to capture the dynamic nature of the Web and, thus, to study the implications of changes in data sources during the execution of a~query.

\section{Acknowledgements} \label{Section:Acknowledgements}
We thank our colleague Matthias Sax for a fruitful discussion that led to one of the proofs in this paper.

% \balance
\bibliographystyle{abbrv}
\bibliography{main}

\newpage
\onecolumn
\appendix
\noindent
The Appendix is organized as follows:
\begin{itemize}
  \item Appendix~\ref{Appendix:Encoding} describes how we encode relevant structures (such as a Web of Linked Data and a valuation) on the tapes of Turing machines.
  \item Appendix~\ref{Appendix:Proofs} contains the full technical proofs for all results in this paper.
\end{itemize}

\section{Encoding} \label{Appendix:Encoding}
\noindent
To encode Webs of Linked Data and query results on the tapes of a Turing machine we assume the existence of a total order $\prec_\symAllURIs$, $\prec_\symAllLiterals$, and $\prec_\symAllVariables$ for the {\ID}s in $\symAllURIs$, the constants in $\symAllLiterals$, and the variables in $\symAllVariables$, respectively; in all three cases $\prec_x$ could simply be the lexicographic order of corresponding string representations. Furthermore, we assume a total order $\prec_t$ for {\triple}s that is based on the aforementioned orders.
% 	\removable{%
% 		as follows: For two distinct {\triple}s $t_1=(s_1,p_1,o_1)$ and $t_2=(s_2,p_2,o_2)$ it holds $t_1 \prec_t t_2$ iff either
% 		\begin{align*}
% 	\text{(i) } \, & s_1 \prec_\symAllURIs s_2, \\
% 	\text{(ii) } \,  & s_1 = s_2 \land p_1 \prec_\symAllURIs p_2, \\
% 	\text{(iii) } \, & s_1 = s_2 \land p_1 = p_2 \land o_1 \in \symAllURIs \land o_2 \in \symAllLiterals, \\
% 	\text{(iv) } \, & s_1 = s_2 \land p_1 = p_2 \land o_1,o_2 \in \symAllURIs \land o_1 \prec_\symAllURIs o_2, \text{ or} \\
% 	\text{(v) } \, & s_1 = s_2 \land p_1 = p_2 \land o_1,o_2 \in \symAllLiterals \land o_1 \prec_\symAllLiterals o_2.
% \end{align*}
% }%

For each $\symURI \in \symAllURIs$, $c \in \symAllLiterals$, and $v \in \symAllVariables$ let $\fctEncName(\symURI)$, $\fctEncName(c)$, and $\fctEncName(v)$ be the binary representation of $\symURI$, $c$, and $v$, respectively.
The encoding of a {\triple} $t=(s,p,o)$, denoted by $\fctEncName(t)$, is a word
\,$%\begin{equation*}
	\langle \, \fctEncName(s) \, , \, \fctEncName(p) \, , \, \fctEncName(o) \, \rangle
$. %\end{equation*}

The encoding of a finite set of {\triple}s $T = \lbrace t_1, ... \, , t_n \rbrace$, denoted by $\fctEncName(T)$, is a word
\,$%\begin{equation*}
	\langle\!\langle \, \fctEncName(t_1) \, , \, \fctEncName(t_2) \, , \, ... \, , \, \fctEncName(t_n) \, \rangle\!\rangle
$\, %\end{equation*}
where the $\fctEncName(t_i)$ are ordered as follows: For each two {\triple}s $t_x,t_y \in T$, $\fctEncName(t_x)$ occurs before $\fctEncName(t_y)$ in $\fctEncName(T)$ if $t_x \prec_t t_y$.

For a Web of Linked Data $\WoD = (D,\dataFct,\adocFct)$, the encoding of {\LDdoc} $d \in D$, denoted by $\fctEncName(d)$, is the word $\fctEncName( \dataFct(d) )$. The encoding of $\WoD$ itself, denoted by $\fctEncName(\WoD)$, is a word
\begin{equation*}
	\sharp \, \fctEncName(\symURI_1) \, \fctEncName( \adocFct(\symURI_1) ) \, \sharp \, ... \, \sharp \, \fctEncName(\symURI_i) \, \fctEncName( \adocFct(\symURI_i) ) \, \sharp \, ...
\end{equation*}
where $\symURI_1,...,\symURI_i,...$ is the (potentially infinite but countable) list of {\ID}s in $\fctDom{\adocFct}$, ordered according to $\prec_\symAllURIs$.

The encoding of a valuation $\mu$ with $dom(\mu) = \lbrace v_1 ,...\,, v_n \rbrace$, denoted by $\fctEncName(\mu)$, is a word
\begin{equation*}
	\langle\!\langle \, \fctEncName(v_1) \rightarrow \fctEncName\bigl( \mu(v_1) \bigr) \, , \, ... \, , \, \fctEncName(v_n) \rightarrow \fctEncName\bigl( \mu(v_n) \bigr) \, \rangle\!\rangle
\end{equation*}
where the $\fctEncName(\mu(v_i))$ are ordered as follows: For each two variables $v_x,v_y \in dom(\mu)$, $\fctEncName(\mu(v_x))$ occurs before $\fctEncName(\mu(v_y))$ in $\fctEncName(\mu)$ if $v_x \!\prec_V\! v_y$.

Finally, the encoding of a (potentially infinite) set of valuations $\Omega = \lbrace \mu_1 , \mu_2 , ... \rbrace$, denoted by $\fctEncName(\Omega)$, is a word
\,$%\begin{equation*}
	\fctEncName( \mu_1 )  \, \fctEncName( \mu_2 ) \, ...
$\, %\end{equation*}
where the $\fctEncName( \mu_i )$ \Todo{may occur in any order}{that might be a problem}.

\section{Proofs} \label{Appendix:Proofs}

\subsection{Additional References for the Proofs}
\noindent
[Pap93] C.~H.~Papadimitriou. \textit{Computational Complexity}. Addison Wesley, 1993.

\subsection{Proof of Proposition~\ref{Proposition:Finding}} \noindent
Let: \begin{itemize} \itemsep1mm
	\item $\symSeedURIs \subset \symAllURIs$ be a finite but nonempty set of seed {\ID}s;
	\item $c$ and $c'$ be reachability criteria;
	\item $\symBQP$ be a BQP such that $\queryFctBSc{\symBQP}{\symSeedURIs}{c}$ and $\queryFctBSc{\symBQP}{\symSeedURIs}{c'}$ are CLD queries;
	\item $\WoD = (D,\dataFct,\adocFct)$ be an \emph{infinite} Web of Linked Data.
\end{itemize}

~\\
\textbf{\ref{Proposition:Finding:Case1}. $\ReachPartScBW{\symSeedURIs}{\cNone}{\symBQP}{\WoD}$ is always finite; so is $\queryRsltBScW{\symBQP}{\symSeedURIs}{\cNone}{\WoD}$.}

\noindent
Let $\Reach{D}$ denote the set of all {\LDdoc}s in $\ReachPartScBW{\symSeedURIs}{\cNone}{\symBQP}{\WoD}$. Since $\cNone$ always returns $\false$ it is easily verified that there is no {\LDdoc} $d \in D$ that satisfies case~\ref{DefinitionCase:QualifiedReachability:IndStep} in Definition~\ref{Definition:QualifiedReachability}. Hence, it must hold $\Reach{D} = \big\lbrace \adocFct(\symURI) \,\big|\, \symURI \in \symSeedURIs \text{ and } \symURI \in \fctDom{\adocFct} \big\rbrace$ (cf.~case~\ref{DefinitionCase:QualifiedReachability:IndBegin} in Definition~\ref{Definition:QualifiedReachability}). Since $\symSeedURIs$ is finite we see that $\Reach{D}$ is guaranteed to be finite (and so is $\ReachPartScBW{\symSeedURIs}{\cNone}{\symBQP}{\WoD}$).
The finiteness of $\queryRsltBScW{\symBQP}{\symSeedURIs}{\cNone}{\WoD}$ can then be shown based on Proposition~\ref{Proposition:Finding}, case~\ref{Proposition:Finding:Case3}.

% ~\\
% \textbf{\ref{Proposition:Finding:Case2}. If $c$ is less restrictive than $\cNone$, then $\ReachPartScBW{\symSeedURIs}{c}{\symBQP}{\WoD}$ is either finite or infinite; the same holds for $\queryRsltBScW{\symBQP}{\symSeedURIs}{c}{\WoD}$.}
% 
% \noindent
% This proposition is easily verified based on Examples~\ref{Example:Infiniteness2} and%
% 	%~\ref{Example:Infiniteness3}:
% 	~\ref{Example:Infiniteness4NewText}:
% Reachability criterion $\cMatch$ is less restrictive than $\cNone$. In Example~\ref{Example:Infiniteness2} the $(\symSeedURIs_1,\cMatch,\symBQP_1)$-reachable part of the infinite Web of Linked Data $\WoD_\mathsf{inf}$ is finite and $\queryRsltBScW{\symBQP_1}{\symSeedURIs_1}{\cMatch}{\WoD_\mathsf{inf}}$ is finite.
% In Example~%
% 	%\ref{Example:Infiniteness3}
% 	\ref{Example:Infiniteness4NewText}
% the $(\symSeedURIs_2,\cMatch,\symBQP_2)$-reachable part of $\WoD_\mathsf{inf}$ is infinite and $\queryRsltBScW{\symBQP_2}{\symSeedURIs_2}{\cMatch}{\WoD_\mathsf{inf}}$ is infinite.

~\\
\textbf{\ref{Proposition:Finding:Case3}. If $\ReachPartScBW{\symSeedURIs}{c}{\symBQP}{W}$ is finite, then $\queryRsltBScW{\symBQP}{\symSeedURIs}{c}{\WoD}$ is finite.}

\noindent
If $\ReachPartScBW{\symSeedURIs}{c}{\symBQP}{W}$ is finite, there exist only a finite number of different possible subsets of $\fctAllDataName( \ReachPartScBW{\symSeedURIs}{c}{\symBQP}{W} )$. Hence, there can only be a finite number of different valuations $\mu$
	%such that for each of them
	with
$\mu[\symBQP] \subseteq \fctAllDataName( \ReachPartScBW{\symSeedURIs}{c}{\symBQP}{W} )$.

~\\
\textbf{\ref{Proposition:Finding:Case4}. If $\queryRsltBScW{\symBQP}{\symSeedURIs}{c}{\WoD}$ is infinite, then $\ReachPartScBW{\symSeedURIs}{c}{\symBQP}{W}$ is infinite.}

\noindent
If $\queryRsltBScW{\symBQP}{\symSeedURIs}{c}{\WoD}$ is infinite, we have infinitely many valuations $\mu \in \queryRsltBScW{\symBQP}{\symSeedURIs}{c}{\WoD}$. For each of them exists a unique subset $\mu[\symBQP] \subseteq \fctAllDataName( \ReachPartScBW{\symSeedURIs}{c}{\symBQP}{W} )$ (cf.~Definition~\ref{Definition:Solution}). Hence, there are infinitely many such subsets. Thus, $\ReachPartScBW{\symSeedURIs}{c}{\symBQP}{W}$ must be infinite.

~\\
\textbf{\ref{Proposition:Finding:Case5}. If $c$ is less restrictive than $c'$ and $\ReachPartScBW{\symSeedURIs}{c}{\symBQP}{W}$ is finite, then $\ReachPartScBW{\symSeedURIs}{c'}{\symBQP}{W}$ is finite.}

\noindent
If $\ReachPartScBW{\symSeedURIs}{c}{\symBQP}{W}$ is finite, then exists finitely many {\LDdoc}s $d\in D$ that are $(c,\symBQP)$-reach\-able from $\symSeedURIs$ in $\WoD$. A subset of them is also $(c',\symBQP)$-reach\-able from $\symSeedURIs$ in $\WoD$ because $c$ is less restrictive than $c'$. Hence, $\ReachPartScBW{\symSeedURIs}{c'}{\symBQP}{W}$ must also be finite.

~\\
\textbf{\ref{Proposition:Finding:Case6}. If $c'$ is less restrictive than $c$ and $\ReachPartScBW{\symSeedURIs}{c}{\symBQP}{W}$ is infinite, then $\ReachPartScBW{\symSeedURIs}{c'}{\symBQP}{W}$ is infinite.}

\noindent
If $\ReachPartScBW{\symSeedURIs}{c}{\symBQP}{W}$ is infinite, then exists infinitely many {\LDdoc}s $d\in D$ that are $(c,\symBQP)$-reach\-able from $\symSeedURIs$ in $\WoD$. Each of them is also $(c',\symBQP)$-reach\-able from $\symSeedURIs$ in $\WoD$ because $c'$ is less restrictive than $c$. Hence, $\ReachPartScBW{\symSeedURIs}{c'}{\symBQP}{W}$ must also be infinite.

~\\
\textbf{\ref{Proposition:Finding:Case7}. If $c'$ is less restrictive than $c$, then $\queryRsltBScW{\symBQP}{\symSeedURIs}{c}{\WoD} \subseteq \queryRsltBScW{\symBQP}{\symSeedURIs}{c'}{\WoD}$.}

\noindent
If $c'$ is less restrictive than $c$, then each {\LDdoc} $d\in D$ that is $(c,\symBQP)$-reach\-able from $\symSeedURIs$ in $\WoD$ is also $(c',\symBQP)$-reach\-able from $\symSeedURIs$ in $\WoD$. Hence, $\fctAllDataName( \ReachPartScBW{\symSeedURIs}{c}{\symBQP}{W} ) \subseteq \fctAllDataName( \ReachPartScBW{\symSeedURIs}{c'}{\symBQP}{W} )$ and, thus, $\queryRsltBScW{\symBQP}{\symSeedURIs}{c}{\WoD} \subseteq \queryRsltBScW{\symBQP}{\symSeedURIs}{c'}{\WoD}$.

% \subsection{Proof of Theorem~\ref{Theorem:Problem:FinitenessReachablePart}} \label{Proof:Theorem:Problem:FinitenessReachablePart}
\subsection{Proof of Theorem~\ref{Theorem:Problem:Finiteness}} \noindent
We prove the theorem by reducing the halting problem to
	%\problemName{FinitenessReachablePart}.
	\problemName{FinitenessReachablePart} and to \problemName{FinitenessQueryResult}.

The halting problem asks whether a given Turing machine (TM) halts on a given input. % As Papadimitriou, we assume that the states and symbols of all TMs are integers~\cite{Papadimitriou93:ComputationalComplexityBook}. Furthermore, we adopt Papadimitriou's method to unambiguously describe TMs and their input by finite words .
For the reduction we assume an infinite Web of Linked Data $\WoD_\mathsf{TMs}$ which we define in the following. Informally, $\WoD_\mathsf{TMs}$ describes all possible computations of all TMs.
	%To define $\WoD_\mathsf{TMs}$ formally
	For a formal definition of $\WoD_\mathsf{TMs}$
we adopt the usual approach to unambiguously describe TMs and their input by finite words over the (finite) alphabet of a universal TM
	%(e.g.~\cite{Papadimitriou93:ComputationalComplexityBook}).
	(e.g.~[Pap93]).
Let $\mathcal{W}$ be the countably infinite set of all words that describe TMs. For each $w \in \mathcal{W}$ let $M(w)$ denote the machine described by $w$
	% and let $S(w)$ denote the finite set of states of $M(w)$. W.l.o.g.~we assume that the states are integers, i.e.~$\forall w \in \mathcal{W} : S(w) = \lbrace 0, 1 , ... \, , n_w \rbrace$, and that the start state of each machine is $0$. Based on these conventions we may ...
	and let $c^{w,x}$ denote the computation of $M(w)$ on input $x$.
Furthermore, let $\symURI_i^{w,x}$ denote an {\ID} for the $i$-th step in $c^{w,x}$. To denote the (infinite) set of all such {\ID}s we write $\symAllURIs_\mathsf{TMsteps}$. Using the {\ID}s $\symAllURIs_\mathsf{TMsteps}$ we may unambiguously identify each step in each possible computation of any TM on any given input.
However, if an {\ID} $\symURI \in \symAllURIs$ could potentially identify a computation step of a TM on some input (because $\symURI$ adheres to the pattern used for such {\ID}s) but the corresponding step may never exist, then $\symURI \notin \symAllURIs_\mathsf{TMsteps}$. For instance, if the computation of a particular TM $M(w_j)$ on a particular input $x_k$ halts with the $i'$-th step, then $\forall \, i \in \lbrace 1 , ... \, , i' \rbrace :\symURI_{i}^{w_j,x_k} \in \symAllURIs_\mathsf{TMsteps}$ and $\forall \, i \in \lbrace i'\!+\!1 , ... \rbrace :\symURI_{i}^{w_j,x_k} \not\in \symAllURIs_\mathsf{TMsteps}$.
Notice, while the set $\symAllURIs_\mathsf{TMsteps}$ is infinite, it is still countable because i)~$\mathcal{W}$ is countably infinite, ii)~the set of all possible input words for TMs is countably infinite, and iii)~$i$ is a natural number.

We now define $\WoD_\mathsf{TMs}$ as a Web of Linked Data $( D_\mathsf{TMs},\dataFct_\mathsf{TMs},$ $ \adocFct_\mathsf{TMs} )$ with the following elements:
$D_\mathsf{TMs}$ consists of $\left| \symAllURIs_\mathsf{TMsteps} \right|$ different {\LDdoc}s, each of which corresponds to one of the {\ID}s in $\symAllURIs_\mathsf{TMsteps}$ (and, thus, to a particular step in a particular computation of a particular TM).
Mapping $\adocFct_\mathsf{TMs}$ is bijective and maps each $\symURI_i^{w,x} \in \symAllURIs_\mathsf{TMsteps}$ to the corresponding $d_i^{w,x} \in D_\mathsf{TMs}$; hence, $\fctDom{\adocFct_\mathsf{TMs}} = \symAllURIs_\mathsf{TMsteps}$. We emphasize that mapping $\adocFct_\mathsf{TMs}$ is (Turing) computable because a universal TM may determine by simulation whether the computation of a particular TM on a particular input halts before a particular number of steps (i.e.~whether the $i$-th step in computation $c^{w,x}$ for a given {\ID} $\symURI_i^{w,x}$ may actually exist).
Finally, mapping $\dataFct_\mathsf{TMs}$ is computed as follows: The set $\dataFct_\mathsf{TMs}\bigl( d_i^{w,x} \bigr)$ of {\triple}s for an {\LDdoc} $d_i^{w,x}$ is empty if and only if $c^{w,x}$ halts with the $i$-th computation step. Otherwise, $\dataFct_\mathsf{TMs}\bigl( d_i^{w,x} \bigr)$ contains a single {\triple} $( \symURI_i^{w,x} \!, \mathsf{next} , \symURI_{i+1}^{w,x} )$ which associates the computation step $\symURI_i^{w,x}$ with the next step in $c^{w,x}$ ($\mathsf{next}$ denotes an {\ID} for this relationship).
Formally:
\begin{equation*}
	\dataFct_\mathsf{TMs}\bigl( d_i^{w,x} \bigr) \!=\!
	\begin{cases}
		\emptyset & \text{if $c^{w,x}$ halts with the $i$-th computation step,} \\
		\lbrace ( \symURI_i^{w,x} \!, \mathsf{next} , \symURI_{i+1}^{w,x} ) \rbrace & \text{else.}
	\end{cases}
\end{equation*}
We emphasize that mapping $\dataFct_\mathsf{TMs}$ is also (Turing) computable because a universal TM may determine by simulation whether the computation of a particular TM on a particular input halts after a given number of steps.

Before we come to the reduction we highlight
	%properties of $\WoD_\mathsf{TMs}$ that are
	a property of $\WoD_\mathsf{TMs}$ that is
important for our proof. Each {\triple}
	%of the form
$( \symURI_i^{w,x} \!, \mathsf{next} , \symURI_{i+1}^{w,x} )$ establishes a data link from $d_i^{w,x}$ to $d_{i+1}^{w,x}$. Due to such links we recursively may reach all {\LDdoc}s about all steps in a particular computation of any TM. Hence, for each possible computation $c^{w,x}$
	%of any
	%	%possible
	%TM
we have a (potentially infinite) simple path $\left( d_1^{w,x} \!, ... \, , d_i^{w,x} \!, ... \right)$ in the Web link graph
	%$G^{\WoD_\mathsf{TMs}}$
of $\WoD_\mathsf{TMs}$. Each such path is finite \IFF the corresponding computation halts. Finally, we note that each of these paths forms a separate subgraph of
	%$G^{\WoD_\mathsf{TMs}}$
	the Web link graph of $\WoD_\mathsf{TMs}$
because
	%we use a separate set of step identifiers for each computation and the {\triple}s in the corresponding {\LDdoc}s only mention steps from the same computation. 
	we use a separate set of step identifiers and {\LDdoc}s for each computation.

We now reduce the halting problem to \problemName{FinitenessReachablePart}. The input to the halting problem is a pair $(w,x)$ consisting of a TM description $w$ and a possible input word $x$. For the reduction we need a computable mapping $f_1$ that, given such a pair $(w,x)$, produces a tuple $(\WoD,\symSeedURIs,c,\symBQP)$ as input for \problemName{FinitenessReachablePart}.
We define $f_1$ as follows: Let
	$w$ be the description of a TM, let $x$ be a possible input word for $M(w)$,
	%$(w,x)$ be an input to the halting problem
and let $?v \in \symAllVariables$ be an arbitrary query variable,
then $f_1( w,x ) = \bigl( \WoD_\mathsf{TMs} , \symSeedURIs_{w,x}, \cAll, 
\symBQP_{w,x} \bigr)$ where $\symSeedURIs_{w,x} = \lbrace \symURI_1^{w,x} \rbrace$ and $\symBQP_{w,x} = \big\lbrace (\symURI_1^{w,x}, \mathsf{next} , ?v) \big\rbrace$. Given that $\cAll$ and $\WoD_\mathsf{TMs}$ are independent of $(w,x)$, it can be easily seen that $f_1$ is computable by TMs (including LD machines).

To show that \problemName{FinitenessReachablePart} is not LD machine decidable, suppose it were LD machine decidable. In such a case an LD machine could answer the halting problem for any input $(w,x)$ \removable{as follows}: $M(w)$ halts on $x$ if and only if the $(\symSeedURIs_{w,x},\cAll,\symBQP_{w,x})$-reachable part of $\WoD_\mathsf{TMs}$ is finite. However, we know the halting problem is undecidable for TMs (which includes LD machines). Hence, we have a contradiction and\removable{, thus,} \problemName{FinitenessReachablePart} cannot be LD machine decidable.

% \subsection{Proof of Theorem~\ref{Theorem:Problem:FinitenessQueryResult}}
The proof
	%is similar to that of Theorem~\ref{Theorem:Problem:FinitenessReachablePart} (cf.~\ref{Proof:Theorem:Problem:FinitenessReachablePart}).
	that \problemName{FinitenessQueryResult} is not LD machine decidable is similar to that for \problemName{FinitenessReachablePart}.
Hence, we only outline the idea: Instead of reducing the halting problem to \problemName{FinitenessReachablePart} based on mapping $f_1$, we now reduce the halting problem to \problemName{FinitenessQueryResult} using a mapping $f_2$ that differs from $f_1$ in the BQP it generates: $f_2(w,x) = \bigl( \WoD_\mathsf{TMs} , \symSeedURIs_{w,x}, \cAll, \symBQP_{w,x}' \bigr)$ where $\symBQP_{w,x}' = \big\lbrace (\symURI_1^{w,x}, \mathsf{next} , ?x) , (?y, \mathsf{next} , ?z) \big\rbrace$. Notice, the two triple patterns in $\symBQP_{w,x}'$ have no variable in common.
If \problemName{FinitenessQueryResult} were LD machine decidable then an LD machine could answer the halting problem for any $(w,x)$: $M(w)$ halts on $x$ if and only if $\queryRsltBScW{\symBQP_{w,x}'}{\symSeedURIs_{w,x}}{\cAll}{\WoD_\mathsf{TMs}}$ is finite.

\subsection{Proof of Lemma~\ref{Lemma:ComputabilityCriteria}} \label{Proof:Lemma:ComputabilityCriteria} \noindent
As a preliminary to prove Lemma~\ref{Lemma:ComputabilityCriteria} we introduce a specific LD machine for CLD queries:
\begin{definition} \label{Definition:CLDQueryMachine}
	Let $\queryFctBSc{\symBQP}{\symSeedURIs}{c}$ be a CLD query.
	The \definedTerm{$(\symBQP,\symSeedURIs,c)$-machine}
		%, denoted by $M^{(\symBQP,\symSeedURIs,c)}$,
	is an LD machine that implements Algorithm~\ref{Algorithm:CLDQueryMachine}. This algorithm
		makes use of
		%calls
	a
		%special
	subroutine
		called
	\code{lookup}. This subroutine, when called with an {\ID} $\symURI \in \symAllURIs$, i)~writes $\fctEncName(\symURI)$ to the right end of the word on the link traversal tape, ii)~enters the expand state, and iii)~performs the expand operation as specified in Definition~\ref{Definition:LDMachine}.
\end{definition}
\begin{algorithm}[h]
	\caption{\, The program of a $(\symBQP,\symSeedURIs,c)$-machine.} \label{Algorithm:CLDQueryMachine}
	\begin{algorithmic}[1]
			%\FORALL {$\symURI \in \symSeedURIs$} \STATE {Call \code{lookup} with $\symURI$.} \ENDFOR
			\STATE {Call \code{lookup} for each $\symURI \in \symSeedURIs$.}
		\label{Line:CLDQueryMachine:Init}

			\medskip
		\FOR { $j = 1, 2, ...$} \label{Line:CLDQueryMachine:LoopBegin}
			\STATE {Let $T_j$ denote the set of all {\triple}s currently encoded on the link traversal tape. Use the work tape to enumerate a set $\Omega_j$ that contains all valuations $\mu$ for which $\fctDom{\mu}=\fctVarsName(\symBQP)$ and $\mu[\symBQP]\subseteq T_j$.} \label{Line:CLDQueryMachine:Step1}
			\STATE {For each $\mu \in \Omega_j$ check whether $\mu$ is already encoded on the output tape; if not, then add $\fctEncName( \mu )$ to the output.} \label{Line:CLDQueryMachine:Step2}
			\STATE {Scan the link traversal tape for an {\triple} $t$ that contains a {\ID} $\symURI \in \fctIDsName(t)$ such that i)~$c(t,\symURI,P)=\true$ and ii)~the word on the link traversal tape neither contains \,$\fctEncName(\symURI) \, \fctEncName( \adocFct(\symURI) ) \, \sharp$\, nor \,$\fctEncName(\symURI) \, \sharp$. If such $t$ and $\symURI$ exist, call \code{lookup} for $\symURI$; otherwise halt the computation.} \label{Line:CLDQueryMachine:Step3}
		\ENDFOR \label{Line:CLDQueryMachine:LoopEnd}
	\end{algorithmic}
\end{algorithm}

\noindent
As can be seen in Algorithm~\ref{Algorithm:CLDQueryMachine}, the computation of each $(\symBQP,\symSeedURIs,c)$-machine
	(with a Web of Linked Data $\WoD$ encoded on its Web tape)
starts with an initialization (cf.~line~\ref{Line:CLDQueryMachine:Init}). After the initialization, the machine enters a (potentially non-terminating) loop. During each iteration of this loop, the machine generates valuations using all data that is currently encoded on the link traversal tape. The following proposition shows that these valuations are part of the corresponding query result (find the proof for Proposition~\ref{Proposition:Computability:Soundness} below in Section~\ref{Proof:Proposition:Computability:Soundness}):
\begin{proposition} \label{Proposition:Computability:Soundness}
	Let $M^{(\symBQP,\symSeedURIs,c)}$ be a $(\symBQP,\symSeedURIs,c)$-machine with a Web of Linked Data $\WoD$ encoded on its Web tape.
	During the execution of Algorithm~\ref{Algorithm:CLDQueryMachine} by $M^{(\symBQP,\symSeedURIs,c)}$
		%on (Web) input $\fctEncName(\WoD)$
	it holds $\forall \, j \in \lbrace 1,2,...\rbrace : \Omega_j \subseteq \queryRsltBScW{\symBQP}{\symSeedURIs}{c}{\WoD}$.
\end{proposition}
Proposition~\ref{Proposition:Computability:Soundness} presents the basis to prove the soundness of query results computed by Algorithm~\ref{Algorithm:CLDQueryMachine}.
To verify the completeness of these results it is important to note that $(\symBQP,\symSeedURIs,c)$-machines look up no more than one {\ID} per iteration (cf.~line~\ref{Line:CLDQueryMachine:Step3} in Algorithm~\ref{Algorithm:CLDQueryMachine}). Hence, $(\symBQP,\symSeedURIs,c)$-machines prioritize result construction over link traversal. Due to this feature we show that for each solution in a query result exists an iteration during which that solution is computed (find the proof for Proposition~\ref{Proposition:Computability:Completeness} below in Section~\ref{Proof:Proposition:Computability:Completeness}):
\begin{proposition} \label{Proposition:Computability:Completeness}
	Let $M^{(\symBQP,\symSeedURIs,c)}$ be a $(\symBQP,\symSeedURIs,c)$-machine with a Web of Linked Data $\WoD$ encoded on its Web tape.
	For each
		%solution
	$\mu \in \queryRsltBScW{\symBQP}{\symSeedURIs}{c}{\WoD}$ exists a $j_\mu \in \lbrace 1,2,...\rbrace$ such that during the execution of Algorithm~\ref{Algorithm:CLDQueryMachine} by $M^{(\symBQP,\symSeedURIs,c)}$
		%on (Web) input $\fctEncName(\WoD)$
	it holds $\forall \, j \!\in\! \lbrace j_\mu,j_\mu \!+\!1,...\rbrace : \mu \in \Omega_j$.
\end{proposition}

\noindent
So far our results verify that i)~the set of query solutions computed after any iteration is sound and ii)~that this set is complete after a particular (potentially infinite) number of iterations. We now show that the computation definitely reaches each iteration after a finite number of computation steps (find the proof for Proposition~\ref{Proposition:Computability:Finiteness} below in Section~\ref{Proof:Proposition:Computability:Finiteness}):
\begin{proposition} \label{Proposition:Computability:Finiteness}
	Let $M^{(\symBQP,\symSeedURIs,c)}$ be a $(\symBQP,\symSeedURIs,c)$-machine with a Web of Linked Data $\WoD$ encoded on its Web tape.
	For any possible iteration $it$ of the main processing loop in Algorithm~\ref{Algorithm:CLDQueryMachine}
		%(i.e.~lines~\ref{Line:CLDQueryMachine:LoopBegin} to~\ref{Line:CLDQueryMachine:LoopEnd})
	it requires only a finite number of computation steps before $M^{(\symBQP,\symSeedURIs,c)}$ starts $it$.
\end{proposition}
\todo{Point out that this machine is not meant to be efficient (
 We emphasize that 
-- although it is easy to see that this algorithm might be very inefficient. 
the repeated enumeration of already reported valuations 
-- However, efficiency is 
\\}%
We now prove Lemma~\ref{Lemma:ComputabilityCriteria}.
Let: \begin{itemize} \itemsep1mm
	\item $\WoD = (D,\dataFct,\adocFct)$ be a potentially infinite Web of Linked Data;
	\item $\queryFctBSc{\symBQP}{\symSeedURIs}{c}$ be a CLD query; and
	\item $\ReachPartScBW{\symSeedURIs}{c}{\symBQP}{W} = (\Reach{D},\Reach{\dataFct},\Reach{\adocFct} )$ denote the $(\symSeedURIs,c,\symBQP)$-reach\-able part of $\WoD$.
\end{itemize}

\noindent\textbf{If:} Let $\ReachPartScBW{\symSeedURIs}{c}{\symBQP}{W}$ be finite. Hence, $\queryRsltBScW{\symBQP}{\symSeedURIs}{c}{\WoD}$ is finite as well (cf.~Proposition~\ref{Proposition:Finding}).
We have to show that there exists an LD machine that computes $\queryRsltBScW{\symBQP}{\symSeedURIs}{c}{\WoD}$ and halts after a finite number of computation steps. Based on Propositions~\ref{Proposition:Computability:Soundness} to~\ref{Proposition:Computability:Finiteness} it is easy to verify that that the $(\symBQP,\symSeedURIs,c)$-machine (with $\fctEncName(W)$ on its Web tape) is such a machine: It computes $\queryRsltBScW{\symBQP}{\symSeedURIs}{c}{\WoD}$ and it is guaranteed to halt because $\ReachPartScBW{\symSeedURIs}{c}{\symBQP}{W}$ is finite.

\vspace{2ex} % Layout Adjustment

\noindent\textbf{Only if:} W.l.o.g., let $M$ be an LD machine (not necessarily a $(\symBQP,\symSeedURIs,c)$-machine) that computes $\queryRsltBScW{\symBQP}{\symSeedURIs}{c}{\WoD}$ and halts after a finite number of computation steps. We have to show that $\ReachPartScBW{\symSeedURIs}{c}{\symBQP}{W}$ is finite. We show this by contradiction, that is, we assume $\ReachPartScBW{\symSeedURIs}{c}{\symBQP}{W}$ is infinite. In this case $\Reach{D}$ is infinite. Since $M$ computes $\queryRsltBScW{\symBQP}{\symSeedURIs}{c}{\WoD}$, $M$ must (recursively) expand the word on its link traversal tape until it contains the encodings of (at least) each {\LDdoc} in $\Reach{D}$. Such an expansion is necessary to ensure that the computed query result is complete. Since $\Reach{D}$ is infinite the expansion requires infinitely many computing steps. However, we know that $M$ halts after a finite number of computation steps. Hence, we have a contradiction and, thus, $\ReachPartScBW{\symSeedURIs}{c}{\symBQP}{W}$ must be finite.

\subsection{Proof of Proposition~\ref{Proposition:Computability:Soundness}} \label{Proof:Proposition:Computability:Soundness} \noindent
Let: \begin{itemize} \itemsep1mm
	\item $\WoD = (D,\dataFct,\adocFct)$ be a Web of Linked Data;
	\item $M^{(\symBQP,\symSeedURIs,c)}$ be a $(\symBQP,\symSeedURIs,c)$-machine
		(cf.~Definition~\ref{Definition:CLDQueryMachine})
	with
		%$\WoD$ encoded
		$\fctEncName(W)$
	on its Web tape;
	%and
	\item $\ReachPartScBW{\symSeedURIs}{c}{\symBQP}{W}$ denote the $(\symSeedURIs,c,\symBQP)$-reach\-able part of $\WoD$.
\end{itemize}
To prove Proposition~\ref{Proposition:Computability:Soundness} we use the following result.
\begin{lemma} \label{SubLemma:Computability:Soundness}
	During the execution of Algorithm~\ref{Algorithm:CLDQueryMachine} by $M^{(\symBQP,\symSeedURIs,c)}$
		on (Web) input $\fctEncName(\WoD)$
	it holds $\forall \, j \in \lbrace 1,2,...\rbrace : T_j \subseteq \fctAllDataName\bigl( \ReachPartScBW{\symSeedURIs}{c}{\symBQP}{W} \bigr)$.
\end{lemma}
\begin{myproof}{of Lemma~\ref{SubLemma:Computability:Soundness}}
Let $w_j$ be the word on the link traversal tape of $M^{(\symBQP,\symSeedURIs,c)}$ when the $j$-th iteration of the main processing loop in Algorithm~\ref{Algorithm:CLDQueryMachine} (i.e.~lines~\ref{Line:CLDQueryMachine:LoopBegin} to~\ref{Line:CLDQueryMachine:LoopEnd}) starts.

To prove $\forall \, j \in \lbrace 1,2,...\rbrace : T_j \subseteq \fctAllDataName\bigl( \ReachPartScBW{\symSeedURIs}{c}{\symBQP}{W} \bigr)$ it is sufficient to show
		%(by induction over $j \in \lbrace 1,2,...\rbrace$)
for each $w_j$ (where $j \in \lbrace 1,2,...\rbrace$) exists a finite sequence $\symURI_1 ,...\,,\symURI_{n_j}$ of $n_j$ different {\ID}s ($\forall \,i \in [1,n_j]: \symURI_i \in \symAllURIs$) such that
i)~$w_j$ is\footnote{We assume $\fctEncName(\adocFct(\symURI_i))$ is the empty word if $\adocFct(\symURI_i)$ is undefined (i.e.~$\symURI_i \notin \fctDom{\adocFct}$).}
\begin{equation*}
	\fctEncName(\symURI_1) \, \fctEncName( \adocFct(\symURI_1) ) \, \sharp \, ... \, \sharp \, \fctEncName(\symURI_{n_j}) \, \fctEncName( \adocFct(\symURI_{n_j}) ) \, \sharp
\end{equation*}
and
ii)~for each $i \in [1,n_j]$ either $\symURI_i \notin \fctDom{\adocFct}$ (and, thus, $\adocFct(\symURI_i)$ is undefined) or $\adocFct(\symURI_i)$ is
	an {\LDdoc} which is
$(c,\symBQP)$-reachable from $\symSeedURIs$ in $\WoD$.
We use an induction over $j$ for this proof.

\vspace{1ex} \noindent
\textit{Base case} ($j = 1$): The computation of $M^{(\symBQP,\symSeedURIs,c)}$ starts with an empty link traversal tape. Due to the initialization, $w_1$ is a concatenation of sub-words \,$\fctEncName(\symURI) \, \fctEncName( \adocFct(\symURI) ) \, \sharp$\, for all $\symURI \in \symSeedURIs$ (cf.~line~\ref{Line:CLDQueryMachine:Init} in Algorithm~\ref{Algorithm:CLDQueryMachine}). Hence, we have a corresponding sequence $\symURI_1 ,...\,,\symURI_{n_1}$ where $n_1=\left|\symSeedURIs\right|$ and $\forall \,i \in [1,n_1]: \symURI_i \in \symSeedURIs$.
The order of the {\ID}s in that sequence depends on the order in which they have been looked up
	%(cf.~line~\ref{Line:CLDQueryMachine:Init} in Algorithm~\ref{Algorithm:CLDQueryMachine})
and is irrelevant for our proof.
For all $\symURI \in \symSeedURIs$ it holds either $\symURI_i \notin \fctDom{\adocFct}$ or $\adocFct(\symURI)$ is $(c,\symBQP)$-reachable from $\symSeedURIs$ in $\WoD$ (cf.~case~\ref{DefinitionCase:QualifiedReachability:IndBegin} in Definition~\ref{Definition:QualifiedReachability}).

\vspace{1ex} \noindent
\textit{Induction step} ($j>1$): Our inductive hypothesis is that there exists a finite sequence $\symURI_1 ,...\,,\symURI_{n_{j-1}}$ of $n_{j-1}$ different {\ID}s ($\forall \,i \in [1,n_{j-1}]: \symURI_i \in \symAllURIs$) such that i)~$w_{j-1}$ is
\begin{equation*}
	\fctEncName(\symURI_1) \, \fctEncName( \adocFct(\symURI_1) ) \, \sharp \, ... \, \sharp \, \fctEncName(\symURI_{n_{j-1}}) \, \fctEncName( \adocFct(\symURI_{n_{j-1}}) ) \, \sharp
\end{equation*}
and ii)~for each $i \in [1,n_{j-1}]$ either $\symURI_i \notin \fctDom{\adocFct}$ or $\adocFct(\symURI_i)$ is
	%an {\LDdoc} which is
$(c,\symBQP)$-reachable from $\symSeedURIs$ in $\WoD$.
In the ($j$-1)-th iteration the $(\symBQP,\symSeedURIs,c)$-machine finds a {\triple} $d$ encoded as part of $w_{j-1}$ such that $\exists \, \symURI \in \fctIDsName(t) : c(t,\symURI,\symBQP) = \true$ and \code{lookup} has not been called for $\symURI$. The machine calls \code{lookup} for $\symURI$, which changes the word on the link traversal tape to $w_j$. Hence, $w_j$ is equal to \,$w_{j-1} \, \fctEncName(\symURI) \, \fctEncName( \adocFct(\symURI) ) \, \sharp\,$ and, thus, our sequence of {\ID}s for $w_j$ is $\symURI_1 ,...\,,\symURI_{n_{j-1}}, \symURI$. It remains to show that if $\symURI_i \in \fctDom{\adocFct}$ then $\adocFct(\symURI)$ is $(c,\symBQP)$-reachable from $\symSeedURIs$ in $\WoD$.

Assume $\symURI \in \fctDom{\adocFct}$. Since {\triple} $t$ is encoded as part of $w_{j-1}$ we know, from our inductive hypothesis, that $t$ must be contained in the data of an {\LDdoc} $d^*$ that is $(c,\symBQP)$-reach\-able from $\symSeedURIs$ in $\WoD$ (and for which exists $i \in [1,n_{j-1}]$ such that $\adocFct(\symURI_i) = d^*$). Therefore, $t$ and $\symURI$ satisfy the requirements as given in case~\ref{DefinitionCase:QualifiedReachability:IndStep} of Definition~\ref{Definition:QualifiedReachability} and, thus, $\adocFct(\symURI)$ is $(c,\symBQP)$-reach\-able from $\symSeedURIs$ in $\WoD$.
	\hfill \,
\end{myproof}

\noindent
\Todo{Proposition~\ref{Proposition:Computability:Soundness} is an immediate consequence of Lemma~\ref{SubLemma:Computability:Soundness}.}{Be a bit more detailed.}

\subsection{Proof of Proposition~\ref{Proposition:Computability:Completeness}} \label{Proof:Proposition:Computability:Completeness} \noindent
Let: \begin{itemize} \itemsep1mm
	\item $\WoD = (D,\dataFct,\adocFct)$ be a Web of Linked Data;
	\item $M^{(\symBQP,\symSeedURIs,c)}$ be a $(\symBQP,\symSeedURIs,c)$-machine
		(cf.~Definition~\ref{Definition:CLDQueryMachine})
	with
		%$\WoD$ encoded
		$\fctEncName(W)$
	on its Web tape;
	%and
	\item $\ReachPartScBW{\symSeedURIs}{c}{\symBQP}{W}$ denote the $(\symSeedURIs,c,\symBQP)$-reach\-able part of $\WoD$.
\end{itemize}
To prove Proposition~\ref{Proposition:Computability:Completeness} we use the following result.
\begin{lemma} \label{SubLemma:Computability:Completeness}
	For each {\triple} $t \in \fctAllDataName\bigl( \ReachPartScBW{\symSeedURIs}{c}{\symBQP}{W} \bigr)$ exists a $j_t \in \lbrace 1,2,...\rbrace$ such that during the execution of Algorithm~\ref{Algorithm:CLDQueryMachine} by $M^{(\symBQP,\symSeedURIs,c)}$
		%on (Web) input $\fctEncName(\WoD)$
	it holds $\forall \, j \!\in\! \lbrace j_t,j_t\!+\!1,...\rbrace : t \in T_j$.
\end{lemma}
\begin{myproof}{of Lemma~\ref{SubLemma:Computability:Completeness}}
Let $w_j$ be the word on the link traversal tape of $M^{(\symBQP,\symSeedURIs,c)}$ when $M^{(\symBQP,\symSeedURIs,c)}$ starts the $j$-th iteration of the main processing loop in Algorithm~\ref{Algorithm:CLDQueryMachine} (i.e.~lines~\ref{Line:CLDQueryMachine:LoopBegin} to~\ref{Line:CLDQueryMachine:LoopEnd}).

W.l.o.g., let $t'$ be an arbitrary {\triple}
	%such that
$t' \in \fctAllDataName\bigl( \ReachPartScBW{\symSeedURIs}{c}{\symBQP}{W} \bigr)$.
	%Since $t \in \fctAllDataName\bigl( \ReachPartScBW{\symSeedURIs}{c}{\symBQP}{W} \bigr)$ there
	There
must exist an {\LDdoc} $d \in D$ such that i)~$t' \in \dataFct(d)$ and ii)~$d$ is $(c,\symBQP)$-reachable from $\symSeedURIs$ in $\WoD$.
Let $d'$ be such a document.
Since $M^{(\symBQP,\symSeedURIs,c)}$ only appends to the link traversal tape we prove that there exists a $j_{t'} \in \lbrace 1,2,...\rbrace$ with $\forall \, j \!\in\! \lbrace j_{t'},j_{t'}\!+\!1,...\rbrace : t' \in T_j$ by showing that there exists $j_{t'} \in \lbrace 1,2,...\rbrace$ such that $w_{j_{t'}}$ contains the sub-word \,$\fctEncName( d' )$.

Since $d'$ is $(c,\symBQP)$-reachable from $\symSeedURIs$ in $\WoD$, the Web link graph for $\WoD$ contains at least one finite path $(d_0, ... \, , d_n)$ of {\LDdoc}s $d_i$ where
i)~$n \in \lbrace 0,1,...\rbrace$,
i)~$\exists \, \symURI \in \symSeedURIs : \adocFct(\symURI) = d_0$,
ii)~$d_n = d'$,
and iii)~for each $i \in \lbrace 1,... \,,n \rbrace$ it holds:
\begin{equation} \label{Equation:Proof:Lemma:Computability:Completeness1}
% 	\begin{split}
% 		\exists \, t \!\in\! \dataFct(d_{i-1}) \!: \bigl( \, \exists \, \symURI \!\in\! \fctIDsName(t) : \, & \adocFct(\symURI)=d_{i} \,\,\land \\ & c(t,\symURI,\symBQP) \!=\! \true \, \bigr)
% 	\end{split}
		\exists \, t \in \dataFct(d_{i-1}) : \Bigl( \, \exists \, \symURI \in \fctIDsName(t) : \bigl( \adocFct(\symURI)=d_{i} \text{ and } c(t,\symURI,\symBQP) = \true \bigr) \Bigr)
\end{equation}
Let $(d_0^*, ... \,, d_n^*)$ be such a path. We use this path for our proof. More precisely, we show by induction over $i \in \lbrace 0,...,\,n \rbrace$ that there exists $j_t \in \lbrace 1,2,...\rbrace$ such that $w_{j_t}$ contains the sub-word \,$\fctEncName( d_n^* )$ (which is the same as \,$\fctEncName( d' )$\, because $d_n^* = d'$).

\vspace{1ex} \noindent
\textit{Base case} ($i=0$): Since $\exists \, \symURI \in \symSeedURIs : \adocFct(\symURI) = d_0^*$ it is easy to verify that $w_1$ contains the sub-word \,$\fctEncName( d_0^* )$.

\vspace{1ex} \noindent
\textit{Induction step} ($i>0$): Our inductive hypothesis is: There exists $j \in \lbrace 1,2,...\rbrace$ such that $w_{j}$ contains sub-word \,$\fctEncName( d_{i-1}^* )$. Based on this hypothesis we show that there exists a $j' \in \lbrace j,j\!+\!1,...\rbrace$ such that $w_{j'}$ contains the sub-word \,$\fctEncName( d_{i}^* )$. We distinguish two cases: either \,$\fctEncName( d_{i}^* )$\, is already contained in $w_{j}$ or it is not contained in $w_{j}$. In the first case we have $j'=j$; in the latter case we have $j'>j$. We have to discuss the latter case only.

Due to (\ref{Equation:Proof:Lemma:Computability:Completeness1}) exist $t^* \in \dataFct(d_{i-1}^*)$ and $\symURI^* \in \fctIDsName(t^*)$ such that $\adocFct(\symURI^*)=d_{i}^*$ and $c(t^*,\symURI^*,\symBQP) = \true$. Hence, there exists a $\delta \in \mathbb{N}^0$ such that $M^{(\symBQP,\symSeedURIs,c)}$ finds $t^*$ and $\symURI^*$ in the ($j$+$\delta$)-th iteration (cf.~line~\ref{Line:CLDQueryMachine:Step3} in Algorithm~\ref{Algorithm:CLDQueryMachine}). Since $M^{(\symBQP,\symSeedURIs,c)}$ calls \code{lookup} for $\symURI^*$ in that iteration, it holds that $w_{j+\delta+1}$ contains \,$\fctEncName( d_{i}^* )$\, and, thus, $j'=j+\delta+1$.
	\hfill \,
\end{myproof}

\noindent
\Todo{Proposition~\ref{Proposition:Computability:Completeness} is an immediate consequence of Lemma~\ref{SubLemma:Computability:Completeness}.}{Be a bit more detailed.}

\subsection{Proof of Proposition~\ref{Proposition:Computability:Finiteness}} \label{Proof:Proposition:Computability:Finiteness} \noindent
Let: \begin{itemize} \itemsep1mm
	\item $\WoD$ be a Web of Linked Data;
	\item $M^{(\symBQP,\symSeedURIs,c)}$ be a $(\symBQP,\symSeedURIs,c)$-machine
		(cf.~Definition~\ref{Definition:CLDQueryMachine})
	with
		%$\WoD$ encoded
		$\fctEncName(W)$
	on its Web tape.
\end{itemize}
To
	show that it requires only a finite number of computation steps before $M^{(\symBQP,\symSeedURIs,c)}$ starts any possible iteration of the main processing loop in Algorithm~\ref{Algorithm:CLDQueryMachine}
	%prove the lemma
we first emphasize that
1.)~each call of the subroutine \code{lookup} terminates because the encoding of $\WoD$ is ordered following the order of the {\ID}s used in $\WoD$
and that 2.)~the initialization in line~\ref{Line:CLDQueryMachine:Init} of Algorithm~\ref{Algorithm:CLDQueryMachine} finishes after a finite number of computation steps because $\symSeedURIs$ is finite.

Hence, it remains to show that each iteration of the loop also finishes after a finite number of computation steps:
	Let $\mathfrak{w}$ denote the word on the link traversal tape at any point in the computation. $\mathfrak{w}$ is always
	%At any point in the computation the word on the link traversal tape is
finite because $M^{(\symBQP,\symSeedURIs,c)}$ only gradually appends (encoded) {\LDdoc}s to the link traversal tape (one document per iteration) and the encoding of each document is finite (recall the set of {\triple}s $\dataFct(d)$ for each {\LDdoc} $d$ is finite). Due to the finiteness of
	$\mathfrak{w}$,
	%the word on the link traversal tape,
each $\Omega_j$ (for $j=1,2,...$) is
	%guaranteed to be
finite, resulting in a finite number of computation steps for lines~\ref{Line:CLDQueryMachine:Step1} and~\ref{Line:CLDQueryMachine:Step2} during any iteration. The scan in line~\ref{Line:CLDQueryMachine:Step3} also finishes after a finite number of computation steps because
	$\mathfrak{w}$
	%the word on the link traversal tape
is finite.

\subsection{Proof of Corollary~\ref{Corollary:ComputabilityTrivial}}
\noindent
Corollary~\ref{Corollary:ComputabilityTrivial} immediately follows from Lemma~\ref{Lemma:ComputabilityCriteria} and Fact~\ref{Fact:FinitenessQueryResult:TrivialCase} (for CLD queries that use an empty set $\symSeedURIs$ of seed {\ID}s) as well as from Lemma~\ref{Lemma:ComputabilityCriteria} and Proposition~\ref{Proposition:Finding}, case~\ref{Proposition:Finding:Case1} (for CLD queries under $\cNone$-semantics).

\subsection{Proof of Theorem~\ref{Theorem:Computability}} \noindent
To prove the theorem we only have to show that all CLD queries are at least eventually computable. Corollary~\ref{Corollary:ComputabilityTrivial} shows that some of them are even finitely computable.

To show that all CLD queries (using any possible reachability criterion) are at least eventually computable we use the notion of a $(\symBQP,\symSeedURIs,c)$-machine (cf.~Definition~\ref{Definition:CLDQueryMachine} in Section~\ref{Proof:Lemma:ComputabilityCriteria}) and show that all computations of $(\symBQP,\symSeedURIs,c)$-machines have the two properties as prescribed in Definition~\ref{Definition:EventuallyComputable}.

W.l.o.g., let $M^{(\symBQP,\symSeedURIs,c)}$ be an arbitrary $(\symBQP,\symSeedURIs,c)$-ma\-chine with an arbitrary Web of Linked Data $\WoD$ encoded on its Web tape; let $\ReachPartScBW{\symSeedURIs}{c}{\symBQP}{W}$ be the $(\symSeedURIs,c,\symBQP)$-reachable part of $\WoD$.
During the computation, $M^{(\symBQP,\symSeedURIs,c)}$ only writes to its output tape when it adds (encoded) valuations $\mu \in \Omega_j$ (for $j=1,2,...$). Since all these valuations are solutions for $\queryFctBSc{\symBQP}{\symSeedURIs}{c}$ in $\WoD$ (cf.~Proposition~\ref{Proposition:Computability:Soundness} in Section~\ref{Proof:Lemma:ComputabilityCriteria}) and line~\ref{Line:CLDQueryMachine:Step2} in Algorithm~\ref{Algorithm:CLDQueryMachine} ensures that the output is free of duplicates, we see that the word on the output tape is always a prefix of a possible encoding of $\queryRsltBScW{\symBQP}{\symSeedURIs}{c}{\WoD}$. Hence, the computation of $M^{(\symBQP,\symSeedURIs,c)}$ has the first property specified in Definition~\ref{Definition:EventuallyComputable}.
Property~\ref{DefinitionRequirement:EventuallyComputable:All} readily follows from Propositions~\ref{Proposition:Computability:Completeness} and~\ref{Proposition:Computability:Finiteness} (cf.~Section~\ref{Proof:Lemma:ComputabilityCriteria}).

%Finally, we use the empty BQP $\symBQP_0 = \emptyset$ as a (simple but sufficient) example to show that some CLD queries are finitely computable: Let $c$ be an arbitrary reachability criterion and let $M^{(\symBQP_0,c)}$ denote the corresponding $(\symBQP_0,c)$-machine. W.l.o.g., let $\WoD$ be an arbitrary Web of Linked Data. We consider the computation of $M^{(\symBQP_0,c)}$ on (Web) input $\fctEncName(W)$. Since $\symBQP_0$ is non-seeding, the set of {\LDdoc}s that are $c$-reachable from $\symBQP_0$ in $\WoD$ is empty (cf.~Fact~\ref{Fact:FinitenessQueryResult:TrivialCase}). Hence, the word on the link traversal tape of $M^{(\symBQP_0,c)}$ after the initialization (cf.~line~\ref{Line:CLDQueryMachine:Init} of Algorithm~\ref{Algorithm:CLDQueryMachine}) is empty. Consequently, the scan in line~\ref{Line:CLDQueryMachine:Step3} does not find an {\ID} to look up and, thus, $M^{(\symBQP_0,c)}$ halts in the first iteration of the main loop.

\subsection{Proof of Theorem~\ref{Theorem:Problem:ComputabilityDecision}} \noindent
We prove the theorem by reducing \problemName{FinitenessReachablePart} to \problemName{ComputabilityCLD}. For the reduction we use an identity function
	%$f_{\ref{Problem:ComputabilityDecision}}$
	$f_3$
that, for any Web of Linked Data $\WoD$, set $\symSeedURIs \subset \symAllURIs$ of seed {\ID}s, reachability criterion $c$, and BQP $\symBQP$, is defined
	%as:
	as \removable{follows}:
$f_3\bigl(\WoD,\symSeedURIs,c,\symBQP \bigr) = (\WoD,\symSeedURIs,c,\symBQP)$. Obviously, $f_3$ is computable by TMs (including LD machines).

To obtain a contradiction, we assume that \problemName{ComputabilityCLD} is LD machine decidable.
	%In this case
	If that were the case 
an LD machine could immediately use Lemma~\ref{Lemma:ComputabilityCriteria} to answer \problemName{FinitenessReachablePart} for any (potentially infinite) Web of Linked Data $\WoD$ and CLD query $\queryFctBSc{\symBQP}{\symSeedURIs}{c}$ where $\symSeedURIs$ is nonempty and $c$ is less restrictive than $\cNone$. Since we know \problemName{FinitenessReachablePart} is not LD machine decidable
	%(cf.~Theorem~\ref{Theorem:Problem:FinitenessReachablePart})
	(cf.~Theorem~\ref{Theorem:Problem:Finiteness})
we have a contradiction.
\subsection{Proof of Proposition~\ref{Proposition:PartSolAugmentationSoundness}} \noindent
Let: \begin{itemize} \itemsep1mm
	\item $\WoD$ be a Web of Linked Data;
	\item $\queryFctBSc{\symBQP}{\symSeedURIs}{c}$ be a CLD query;
	\item $\ReachPartScBW{\symSeedURIs}{c}{\symBQP}{W}$ denote the $(\symSeedURIs,c,\symBQP)$-reach\-able part of $\WoD$;
	\item $\DiscPartXX{W}$ be a discovered part of $\WoD$ \underline{and} an induced subweb of $\ReachPartScBW{\symSeedURIs}{c}{\symBQP}{W}$;
	\item $\sigma = (\symBQPpart,\mu)$ be a partial solution for $\queryFctBSc{\symBQP}{\symSeedURIs}{c}$ in $\WoD$; and
	%\item let $tp \in \symBQP \setminus \symBQPpart$ and $t \in \fctAllDataName\bigl( \DiscPartXX{W} \bigr)$.
	\item $\sigma' = (\symBQPpart',\mu')$ be a $(t,tp)$-augmentation of $\sigma$ in $\DiscPartXX{W}$.
\end{itemize}

To show that $\sigma'$ is a partial solution for $\queryFctBSc{\symBQP}{\symSeedURIs}{c}$ in $\WoD$, we have to show: (1)~$\symBQPpart' \subseteq \symBQP$ and (2)~$\mu'$ is a solution for CLD query $\queryFctBSc{\symBQPpart'}{\symSeedURIs}{c}$ in $\WoD$ (cf.~Definition~\ref{Definition:PartialSolution}).

(1)~holds because i)~$\sigma = (\symBQPpart,\mu)$ is a partial solution for $\queryFctBSc{\symBQP}{\symSeedURIs}{c}$ in $\WoD$ and, thus, $\symBQPpart \subseteq \symBQP$, and ii)~$\symBQPpart' = \symBQPpart \cup \lbrace tp \rbrace$ with $tp \in \symBQP \setminus \symBQPpart$ (cf.~Definition~\ref{Definition:PartSolAugmentation}).

To show (2) we note
	that
$\fctDom{\mu'} \!=\! \fctVarsName(\symBQPpart')$ (cf.~Definition~\ref{Definition:PartSolAugmentation}).
It remains to show $\mu'[\symBQPpart'] \!\subseteq\! \fctAllDataName\bigl( \ReachPartScBW{\symSeedURIs}{c}{\symBQP}{W} \bigr)$ (cf.~Definition~\ref{Definition:Solution}).
Due to Definition~\ref{Definition:PartSolAugmentation} we have $\mu'[\symBQPpart'] = \mu[\symBQPpart] \cup \lbrace t \rbrace$ with $t \in \fctAllDataName\bigl( \DiscPartXX{W} \bigr)$.
It holds $t \in \fctAllDataName\bigl( \ReachPartScBW{\symSeedURIs}{c}{\symBQP}{W} \bigr)$ because $\DiscPartXX{W}$ is an induced subweb of $\ReachPartScBW{\symSeedURIs}{c}{\symBQP}{W}$ and, therefore, $\fctAllDataName\bigl( \DiscPartXX{W} \bigr) \subseteq \fctAllDataName\bigl( \ReachPartScBW{\symSeedURIs}{c}{\symBQP}{W} \bigr)$. 
Furthermore, $\mu[\symBQPpart] \subseteq \fctAllDataName\bigl( \ReachPartScBW{\symSeedURIs}{c}{\symBQP}{W} \bigr)$ because $(\symBQPpart,\mu)$ is a partial solution for $\queryFctBSc{\symBQP}{\symSeedURIs}{c}$ in $\WoD$ and, thus, $\mu$ is a solution for $\queryFctBSc{\symBQPpart}{\symSeedURIs}{c}$ in $\WoD$.
Therefore, $\mu'[\symBQPpart'] \subseteq \fctAllDataName\bigl( \ReachPartScBW{\symSeedURIs}{c}{\symBQP}{W} \bigr)$.

\subsection{Proof of Proposition~\ref{Proposition:MatchBasedExpansionMonotonicity}} \label{Proof:MatchBasedExpansionMonotonicity} \noindent
Let: \begin{itemize} \itemsep1mm
	\item $\WoD = ( D,\dataFct,\adocFct )$ be a Web of Linked Data;
	\item $\DiscPartXX{W} = ( \Disc{D},$ $\Disc{\dataFct},\Disc{\adocFct} )$ be a discovered part of $\WoD$;
	\item $\mu$ be a valuation; and
	\item $\NewDiscPartXX{W} = (\NewDisc{D},\NewDisc{\dataFct},\NewDisc{\adocFct} )$ be the $\mu$-expansion of $\DiscPartXX{W}$.
\end{itemize}
To show that $\DiscPartXX{W}$ is an induced subweb of $\NewDiscPartXX{W}$ we have show that $\DiscPartXX{W}$ satisfies the three requirements in Definition~\ref{Definition:InducedSubWeb}
	w.r.t.
	%with respect to
$\NewDiscPartXX{W}$.

\vspace{1ex}
\noindent
For \textbf{requirement~\ref{DefinitionRequirement:InducedSubWeb:D}} we have to show $\Disc{D} \subseteq \NewDisc{D}$, which holds because $\NewDisc{D} = \Disc{D} \cup \ExpMDeltaMW{\mu}{\WoD}$ (cf.~Definition~\ref{Definition:MatchBasedExpansion}).

\vspace{1ex}
\noindent
For \textbf{requirement~\ref{DefinitionRequirement:InducedSubWeb:data}} we have to show:
\begin{equation} \label{Equation:Proof:MatchBasedExpansionMonotonicity:ToBeShownData}
	\forall d \in \NewDisc{D} : \NewDisc{\dataFct}(d) = \Disc{\dataFct}(d)
	\end{equation}
Since $\NewDiscPartXX{W}$ is an induced subweb of $\WoD$ (cf.~Definition~\ref{Definition:MatchBasedExpansion}) it holds:
\begin{align*}
	\forall d \in \NewDisc{D} &: \NewDisc{\dataFct}(d) = \dataFct(d) \\
	\intertext{and with $\Disc{D} \subseteq \NewDisc{D}$ (which we have shown before):}
	\forall d \in \Disc{D} &: \NewDisc{\dataFct}(d) = \dataFct(d) \\
	\intertext{$\DiscPartXX{W}$ is also an induced subweb of $\WoD$ (cf.~Definition~\ref{Definition:MatchBasedExpansion}). Hence:}
	\forall d \in \Disc{D} &: \Disc{\dataFct}(d) = \dataFct(d)
\end{align*}
and, thus, holds (\ref{Equation:Proof:MatchBasedExpansionMonotonicity:ToBeShownData}).

\vspace{1ex}
\noindent
For \textbf{requirement~\ref{DefinitionRequirement:InducedSubWeb:adoc}} we have to show:
\begin{equation*}
	\forall \, \symURI \in \lbrace \symURI \in \symAllURIs \,|\, \NewDisc{\adocFct}(\symURI) \in \Disc{D} \rbrace : \Disc{\adocFct}(\symURI) = \NewDisc{\adocFct}(\symURI)
\end{equation*}
Since $\DiscPartXX{W}$ is an induced subweb of $\WoD$ (cf.~Definition~\ref{Definition:DiscoveredPart}) it holds:
\begin{equation} \label{Equation:Proof:MatchBasedExpansionMonotonicity:ToBeShownADoc}
	\forall \, \symURI \in \lbrace \symURI \in \symAllURIs \,|\, \adocFct(\symURI) \in \Disc{D} \rbrace : \Disc{\adocFct}(\symURI) = \adocFct(\symURI)
\end{equation}
Furthermore, $\NewDiscPartXX{W}$ is an induced subweb of $\WoD$ (cf.~Definition~\ref{Definition:MatchBasedExpansion}). Hence:
\begin{equation*}
	\forall \, \symURI \in \lbrace \symURI \in \symAllURIs \,|\, \adocFct(\symURI) \in \NewDisc{D} \rbrace : \NewDisc{\adocFct}(\symURI) = \adocFct(\symURI)
\end{equation*}
Since $\Disc{D} \subseteq \NewDisc{D}$ (which we have shown before) we rewrite  (\ref{Equation:Proof:MatchBasedExpansionMonotonicity:ToBeShownADoc}) by using $\NewDisc{\adocFct}$ instead of $\adocFct$:
\begin{equation*}
	\forall \, \symURI \in \lbrace \symURI \in \symAllURIs \,|\, \NewDisc{\adocFct}(\symURI) \in \Disc{D} \rbrace : \Disc{\adocFct}(\symURI) = \NewDisc{\adocFct}(\symURI)
\end{equation*}

\subsection{Proof of Proposition~\ref{Proposition:MatchBasedExpansionClosedness}}
\noindent
Let: \begin{itemize} \itemsep1mm
	\item $\WoD = ( D,\dataFct,\adocFct )$ be a Web of Linked Data;
	\item $\DiscPartXX{W} = ( \Disc{D},$ $\Disc{\dataFct},\Disc{\adocFct} )$ be a discovered part of $\WoD$;
	\item $\mu$ be a valuation; and
	\item $\NewDiscPartXX{W} = (\NewDisc{D},\NewDisc{\dataFct},\NewDisc{\adocFct} )$ be the $\mu$-expansion of $\DiscPartXX{W}$.
\end{itemize}
To show that $\NewDiscPartXX{W}$ is a discovered part of $\WoD$ we have to show that $\NewDiscPartXX{W}$ is finite (cf.~Definition~\ref{Definition:DiscoveredPart}), which holds iff $\NewDisc{D}$ is finite.

We have $\NewDisc{D} = \Disc{D} \cup \ExpMDeltaMW{\mu}{\WoD}$ (cf.~Definition~\ref{Definition:MatchBasedExpansion}).
$\Disc{D}$ is finite because $\DiscPartXX{W}$ is a discovered part of $\WoD$. $\ExpMDeltaMW{\mu}{\WoD}$ is also finite because it contains at most as many elements as we have variables in $\fctDom{\mu}$, which is always a finite number.

\subsection{Proof of Proposition~\ref{Proposition:MatchBasedExpansionBoundedness}}
\noindent
Let: \begin{itemize} \itemsep1mm
	\item $\WoD = ( D,\dataFct,\adocFct )$ be a Web of Linked Data;
	\item $\queryFctBSc{\symBQP}{\symSeedURIs}{\cMatch}$ be a CLD query (under $\cMatch$-semantics);
	\item $\ReachPartScBW{\symSeedURIs}{\cMatch}{\symBQP}{W} = (\Reach{D},\Reach{\dataFct},\Reach{\adocFct} )$ denote the $(\symSeedURIs,\cMatch,\symBQP)$-reach\-able part of $\WoD$;
	\item $\DiscPartXX{W} = (\Disc{D},\Disc{\dataFct},\Disc{\adocFct} )$ be a discovered part of $\WoD$ \underline{and} an induced subweb of $\ReachPartScBW{\symSeedURIs}{\cMatch}{\symBQP}{W}$;
	\item $\sigma = (\symBQPpart,\mu)$ be a partial solution for $\queryFctBSc{\symBQP}{\symSeedURIs}{\cMatch}$ in $\WoD$; and
	\item $\NewDiscPartXX{W} = (\NewDisc{D},\NewDisc{\dataFct},\NewDisc{\adocFct} )$ be the $\mu$-expansion of $\DiscPartXX{W}$.
\end{itemize}
To show that $\NewDiscPartXX{W}$ is an induced subweb of $\ReachPartScBW{\symSeedURIs}{\cMatch}{\symBQP}{W}$ we have show that $\NewDiscPartXX{W}$ satisfies the three requirements in Definition~\ref{Definition:InducedSubWeb}
	%w.r.t.
	with respect to
$\ReachPartScBW{\symSeedURIs}{\cMatch}{\symBQP}{W}$.

\vspace{1ex} \noindent
For \textbf{requirement~\ref{DefinitionRequirement:InducedSubWeb:D}} we have to show $\NewDisc{D} \subseteq \Reach{D}$.
Due to Definition~\ref{Definition:MatchBasedExpansion} we have $\NewDisc{D} = \Disc{D} \cup \ExpMDeltaMW{\mu}{\WoD}$.
It also holds $\Disc{D} \subseteq \Reach{D}$ because $\DiscPartXX{W}$ is an induced subweb of $\ReachPartScBW{\symSeedURIs}{\cMatch}{\symBQP}{W}$. Hence, it remains to show $\ExpMDeltaMW{\mu}{\WoD} \subseteq \Reach{D}$.
We show $\ExpMDeltaMW{\mu}{\WoD} \subseteq \Reach{D}$ by contradiction, that is, we assume $\exists \, d \in \ExpMDeltaMW{\mu}{\WoD} : d \notin \Reach{D}$.

According to the definition of $\ExpMDeltaMW{\mu}{\WoD}$ must exist $v' \in \fctDom{\mu}$ such that $\mu(v') \in \symAllURIs$ and $\adocFct\bigl( \mu(v') \bigr) = d$ (cf.~Definition~\ref{Definition:MatchBasedExpansion}).

Since $\sigma \!=\! (\symBQPpart,\mu)$ is a partial solution for $\queryFctBSc{\symBQP}{\symSeedURIs}{\cMatch}$ in $\WoD$, we know that $\mu$ is a solution for $\queryFctBSc{\symBQPpart}{\symSeedURIs}{\cMatch}$ in $\WoD$~(cf. Definition~\ref{Definition:PartialSolution}) and, thus, $\mu[\symBQPpart] \subseteq \fctAllDataName( \ReachPartScBW{\symSeedURIs}{\cMatch}{\symBQP}{W} )$ (cf.~Definition~\ref{Definition:Solution}). Together with $v' \in \fctDom{\mu}$ (see above) we have $\exists \, tp' \in \symBQPpart : v' \in \fctVarsName(tp')$ and $\exists \, t' \in \fctAllDataName( \ReachPartScBW{\symSeedURIs}{\cMatch}{\symBQP}{W} ) : \mu[tp']=t'$. Since $\mu(v') \in \symAllURIs$ and $v' \in \fctVarsName(tp')$ it must hold $\mu(v') \in \fctIDsName(t')$.

Because of $t' \!\in\! \fctAllDataName( \ReachPartScBW{\symSeedURIs}{\cMatch}{\symBQP}{W} )$ we also have $\exists \, d' \!\in\! \Reach{D} : t' \in \dataFct(d')$. Notice, $d'$ is $(\cMatch,\symBQP)$-reach\-able (from $\symSeedURIs$ in $\WoD$).
Furthermore, it must hold $\cMatch(t',\mu(v'),\symBQP) = \true$ because $t'$ matches $tp'\in \symBQPpart \subseteq \symBQP$.

Putting everything together, we have $d' \in D$, $t' \in \dataFct(d')$, and $\mu(v') \in \fctIDsName(t')$, and we know that i)~$d'$ is $(\cMatch,\symBQP)$-reach\-able
	from $\symSeedURIs$ in $\WoD$,
ii)~$\cMatch(t',\mu(v'),\symBQP) = \true$, and iii)~$\adocFct\bigl( \mu(v') \bigr) = d$. Thus, $d$ must be $(\cMatch,\symBQP)$-reach\-able from $\symSeedURIs$ in $\WoD$ (cf.~Definition~\ref{Definition:MatchBasedExpansion}); i.e. $d \in \Reach{D}$. This contradicts our assumption $d \notin \Reach{D}$.

\vspace{1ex} \noindent
We omit showing that $\NewDiscPartXX{W}$ satisfies \textbf{requirement~\ref{DefinitionRequirement:InducedSubWeb:data}} and \textbf{requirement~\ref{DefinitionRequirement:InducedSubWeb:adoc}} w.r.t.~$\ReachPartScBW{\symSeedURIs}{\cMatch}{\symBQP}{W}$; the proof ideas are the same as those that we use in the proof of Proposition~\ref{Proposition:MatchBasedExpansionMonotonicity} (cf.~\ref{Proof:MatchBasedExpansionMonotonicity}).

\subsection{Proof of Theorem~\ref{Theorem:MatchBasedExec}} \label{Proof:Theorem:MatchBasedExec}
\noindent
As a basis for proving the soundness we use the following lemma, which may be verified based on Propositions~\ref{Proposition:PartSolAugmentationSoundness}, \ref{Proposition:MatchBasedExpansionClosedness}, and~\ref{Proposition:MatchBasedExpansionBoundedness} (find the proof for the following lemma below in Section~\ref{Proof:Lemma:MatchBasedExec:Soundness}).
\begin{lemma} \label{Lemma:MatchBasedExec:Soundness}
	Let $\WoD$ be a Web of Linked Data and let $\queryFctBSc{\symBQP}{\symSeedURIs}{\cMatch}$ be a CLD query.
	During an (arbitrary) execution of $ltbExec(\symSeedURIs,\symBQP,W)$ it always holds:
	i)~each $\sigma \in \mathfrak{P}$ is a partial solution for $\queryFctBSc{\symBQP}{\symSeedURIs}{\cMatch}$ in~$\WoD$
	and ii)~$\mathfrak{D}$ is a discovered part of $\WoD$ and an induced subweb of $\ReachPartScBW{\symSeedURIs}{\cMatch}{\symBQP}{W}$.
\end{lemma}

\noindent
Analogous to Lemma~\ref{Lemma:MatchBasedExec:Soundness}, the following lemma provides the basis for our proof of completeness (find the proof for the following lemma below in Sections~\ref{Proof:Lemma:MatchBasedExec:Completeness-i} and~\ref{Proof:Lemma:MatchBasedExec:Completeness-ii}).
\begin{lemma} \label{Lemma:MatchBasedExec:Completeness}
	Let
		%$\WoD$
		$\WoD = (D,\dataFct,\adocFct)$
	be a Web of Linked Data and let $\queryFctBSc{\symBQP}{\symSeedURIs}{\cMatch}$ be a CLD query.
	i)~For each
		%{\LDdoc} $d$
		$d \in D$
	that is $(\cMatch,\symBQP)$-reach\-able from $\symSeedURIs$ in $\WoD$ there will eventually be an iteration in \emph{any} execution of $ltbExec(\symSeedURIs,\symBQP,W)$ after which $d$ is part of $\mathfrak{D}$.
	ii)~For each partial solution $\sigma$ that may exist for $\queryFctBSc{\symBQP}{\symSeedURIs}{\cMatch}$ in $\WoD$, there will eventually be an iteration in \emph{any} execution of $ltbExec(\symSeedURIs,\symBQP,W)$ after which $\sigma \in \mathfrak{P}$.
\end{lemma}

\noindent
We now use Lemmas~\ref{Lemma:MatchBasedExec:Soundness} and~\ref{Lemma:MatchBasedExec:Completeness} to prove Theorem~\ref{Theorem:MatchBasedExec}.
Let: \begin{itemize} \itemsep1mm
	\item $\WoD$ be a Web of Linked Data;
	\item $\queryFctBSc{\symBQP}{\symSeedURIs}{\cMatch}$ be a CLD query (under $\cMatch$-semantics);
	\item $\ReachPartScBW{\symSeedURIs}{\cMatch}{\symBQP}{W}$ denote the $(\symSeedURIs,\cMatch,\symBQP)$-reachable part of $\WoD$;
	\item $\mathfrak{P}$ be the set of partial solutions (for $\queryFctBSc{\symBQP}{\symSeedURIs}{\cMatch}$ in $\WoD$) that is used in $ltbExec(\symSeedURIs,\symBQP,W)$; and
	\item $\mathfrak{D}$ be the discovered part of $\WoD$ that is used in $ltbExec(\symSeedURIs,\symBQP,W)$.
\end{itemize}

\noindent \textbf{Soundness:}
W.l.o.g., let $\mu^*$ be a valuation that an \removable{arbitrary} execution of $ltbExec(\symSeedURIs,\symBQP,W)$ reports in some iteration $it_j$.
	%To show that $\mu^* \in \queryRsltBScW{\symBQP}{\symSeedURIs}{\cMatch}{\WoD}$ we use the following lemma:
	We have to show $\mu^* \in \queryRsltBScW{\symBQP}{\symSeedURIs}{\cMatch}{\WoD}$.
$\mu^*$ originates from the pair $(\symBQPpart^*\!,\mu^*)$ that the execution of $ltbExec(\symSeedURIs,\symBQP,W)$ constructs and adds to $\mathfrak{P}$ in iteration $it_j$. Since $(\symBQPpart^*\!,\mu^*)$ is a partial solution for $\queryFctBSc{\symBQP}{\symSeedURIs}{\cMatch}$ in $\WoD$ (cf.~Lemma~\ref{Lemma:MatchBasedExec:Soundness}) and $ltbExec$ reports $\mu^*$ only if $\symBQPpart^* \! = \symBQP$ (cf.~line~\ref{Line:ReportSolution} in Algorithm~\ref{Procedure:MatchBasedExec}), it holds that $\mu^*$ is a solution for $\queryFctBSc{\symBQP}{\symSeedURIs}{\cMatch}$ in $\WoD$ (cf.~Definition~\ref{Definition:PartialSolution}); i.e.~$\mu^* \in \queryRsltBScW{\symBQP}{\symSeedURIs}{\cMatch}{\WoD}$.

\vspace{1ex}
\noindent \textbf{Completeness:}
% 	To show that any $\mu \in \queryRsltBScW{\symBQP}{\symSeedURIs}{\cMatch}{\WoD}$ will (eventually) be reported by any execution of $ltbExec(\symSeedURIs,\symBQP,W)$, we first show that i)~any execution of $ltbExec(\symSeedURIs,\symBQP,W)$ will eventually discover any {\LDdoc} that is $(\cMatch,\symBQP)$-reachable from $\symSeedURIs$ in $\WoD$ (Lemma~\ref{Lemma:MatchBasedExec:Completeness1}), and that ii)~any execution of $ltbExec(\symSeedURIs,\symBQP,W)$ will eventually add each partial solution (for $\queryFctBSc{\symBQP}{\symSeedURIs}{\cMatch}$ in $\WoD$) to $\mathfrak{P}$ (Lemma~\ref{Lemma:MatchBasedExec:Completeness2}).
% 
	W.l.o.g., let $\mu^*$ be an arbitrary solution for $\queryFctBSc{\symBQP}{\symSeedURIs}{\cMatch}$ in $\WoD$; i.e.~$\mu^* \in \queryRsltBScW{\symBQP}{\symSeedURIs}{\cMatch}{\WoD}$.
	We have to show that any execution of $ltbExec(\symSeedURIs,\symBQP,W)$ will eventually report $\mu^*$.
	For $\mu^*$ exists a partial solution $\sigma^* = (\symBQPpart^*\!,\mu^*)$ (for $\queryFctBSc{\symBQP}{\symSeedURIs}{\cMatch}$ in $\WoD$) such that $\symBQPpart^* \! = \symBQP$. Due to Lemma~\ref{Lemma:MatchBasedExec:Completeness} we know that during any execution of $ltbExec(\symSeedURIs,\symBQP,W)$ there will be an iteration in which this partial solution $\sigma^*$ is constructed and added to $\mathfrak{P}$. This iteration will report $\mu^*$ because $\symBQPpart^* \! = \symBQP$ (cf.~line~\ref{Line:ReportSolution} in Algorithm~\ref{Procedure:MatchBasedExec}).

\subsection{Proof of Lemma~\ref{Lemma:MatchBasedExec:Soundness}} \label{Proof:Lemma:MatchBasedExec:Soundness}
\noindent
Let $\WoD$ be a Web of Linked Data and let $\queryFctBSc{\symBQP}{\symSeedURIs}{\cMatch}$ be a CLD query (under $\cMatch$-semantics).
We show Lemma~\ref{Lemma:MatchBasedExec:Soundness} by induction over the iterations of the main processing loop (lines~\ref{Line:WhileBegin} to~\ref{Line:WhileEnd} in Algorithm~\ref{Procedure:MatchBasedExec}) in $ltbExec(\symSeedURIs,\symBQP,W)$.

\vspace{1ex} \noindent
\textbf{Base case} ($i=0$):
Before the first iteration, $ltbExec(\symSeedURIs,\symBQP,W)$ initializes $\mathfrak{P}$ as a set containing a single element: $\sigma_0 = (\symBQPpart_0,\mu_0)$ where $\symBQPpart_0 = \emptyset$ (cf.~line~\ref{Line:InitP} in Algorithm~\ref{Procedure:MatchBasedExec}). $\sigma_0$ is a partial solution for $\queryFctBSc{\symBQP}{\symSeedURIs}{\cMatch}$ in $\WoD$ because it holds:
\begin{itemize}
	\item $\symBQPpart_0 \subseteq \symBQP$,
	\item $\fctDom{\mu_0} = \emptyset = \fctVarsName(\symBQPpart_0)$, and
	\item $\mu_0[\symBQPpart_0] = \emptyset \subseteq \fctAllDataName\bigl( \ReachPartScBW{\symSeedURIs}{\cMatch}{\symBQP}{W} \bigr)$.
\end{itemize}

\noindent
$\mathfrak{D}$ is initialized with $\mathfrak{D}_\mathsf{init}^{\symSeedURIs,W} = ( D_0,\dataFct_0,\adocFct_0)$ (cf.~line~\ref{Line:InitD} in Algorithm~\ref{Procedure:MatchBasedExec}). Recall the definition of $D_0$ (cf.~(\ref{Equation:D_0}) in Section~\ref{Subsection:ExecModel:Preliminaries}):
\begin{equation*}
	D_0 = \big\lbrace \adocFct(\symURI) \,\big|\, \symURI \in \symSeedURIs \text{ and } \symURI \in \fctDom{\adocFct} \big\rbrace
\end{equation*}
Hence, $D_0$ contains at most $\left| \symSeedURIs \right|$ {\LDdoc}s. Therefore, $\mathfrak{D}_\mathsf{init}^{\symSeedURIs,W}$ is finite and, thus, a discovered part of $\WoD$.
$\mathfrak{D}_\mathsf{init}^{\symSeedURIs,W}$ is also an induced subweb of $\ReachPartScBW{\symSeedURIs}{\cMatch}{\symBQP}{W}$ because each $d \in D_0$ satisfies case~\ref{DefinitionCase:QualifiedReachability:IndBegin} in Definition~\ref{Definition:QualifiedReachability}.

\vspace{1ex} \noindent
\textbf{Induction step} ($i > 0$):
Our inductive hypothesis is that after the ($i$-1)-th iteration it holds i)~each $\sigma \in \mathfrak{P}$ is a partial solution for $\queryFctBSc{\symBQP}{\symSeedURIs}{\cMatch}$ in $\WoD$ and ii)~$\mathfrak{D}$ is a discovered part of $\WoD$ and an induced subweb of $\ReachPartScBW{\symSeedURIs}{\cMatch}{\symBQP}{W}$. We show that these two assumptions still hold after the $i$-th iteration.
Let $(\sigma,t,tp)$ be the open {\AEtask} selected in the $i$-th iteration (cf.~line~\ref{Line:SelectTask} in Algorithm~\ref{Procedure:MatchBasedExec}).

$ltbExec(\symSeedURIs,\symBQP,W)$ extends $\mathfrak{P}$ by adding $(\symBQPpart',\mu')$ (cf.~line~\ref{Line:ChangeP}), the $(t,tp)$-augmentation of $\sigma$ in $\mathfrak{D}$. According to Proposition~\ref{Proposition:PartSolAugmentationSoundness}, $(\symBQPpart',\mu')$ is a partial solution for $\queryFctBSc{\symBQP}{\symSeedURIs}{\cMatch}$ in $\WoD$ because $\mathfrak{D}$ is an induced subweb of $\ReachPartScBW{\symSeedURIs}{\cMatch}{\symBQP}{W}$ (inductive hypothesis) and $\sigma$ is a partial solution for $\queryFctBSc{\symBQP}{\symSeedURIs}{\cMatch}$ in $\WoD$ (cf.~Definition~\ref{Definition:AETask}).

Furthermore, the result of the $\mu'$-expansion $\expMWD{\mu'}{\WoD}{\mathfrak{D}}$ of $\mathfrak{D}$ becomes the new $\mathfrak{D}$ (cf.~line~\ref{Line:ChangeD}).
According to Proposition~\ref{Proposition:MatchBasedExpansionClosedness}, $\expMWD{\mu'}{\WoD}{\mathfrak{D}}$ is again a discovered part of $\WoD$; and, according to Proposition~\ref{Proposition:MatchBasedExpansionBoundedness}, it is also an induced subweb of $\ReachPartScBW{\symSeedURIs}{\cMatch}{\symBQP}{W}$.

\subsection{Proof of Assertion i) in Lemma~\ref{Lemma:MatchBasedExec:Completeness}} \label{Proof:Lemma:MatchBasedExec:Completeness-i}
\noindent
Let $\WoD = ( D, \dataFct, \adocFct )$ be a Web of Linked Data and let $\queryFctBSc{\symBQP}{\symSeedURIs}{\cMatch}$ be a CLD query (under $\cMatch$-semantics).
At any point in the execution of $ltbExec(\symSeedURIs,\symBQP,W)$ let $\Disc{D}$ denote the set of {\LDdoc}s in the currently discovered part $\mathfrak{D}$ of $\WoD$.

W.l.o.g., let $d^*$ be an arbitrary {\LDdoc} that is $(\cMatch,\symBQP)$-reach\-able from $\symSeedURIs$ in $\WoD$.
We have to show that during any possible execution of $ltbExec(\symSeedURIs,\symBQP,W)$ there will eventually be an iteration after which $d^* \in \Disc{D}$.
In correspondence to Definition~\ref{Definition:QualifiedReachability} we distinguish two cases: 1.)~$\exists \, \symURI \in \symSeedURIs : \adocFct(\symURI) = d^*$ and 2.)~$\neg \, \exists \, \symURI \in \symSeedURIs : \adocFct(\symURI) = d^*$.

\vspace{1ex} \noindent
\textbf{Case 1.)}
Before the first iteration, any execution of $ltbExec(\symSeedURIs,\symBQP,W)$ initializes $\mathfrak{D}$ with $\mathfrak{D}_\mathsf{init}^{\symSeedURIs,W} = ( D_0,\dataFct_0,\adocFct_0)$ (cf.~line~\ref{Line:InitD} in Algorithm~\ref{Procedure:MatchBasedExec}). Recall the definition of $D_0$ (cf.~(\ref{Equation:D_0}) in Section~\ref{Subsection:ExecModel:Preliminaries}):
\begin{equation*}
	D_0 = \big\lbrace \adocFct(\symURI) \,\big|\, \symURI \in \symSeedURIs \text{ and } \symURI \in \fctDom{\adocFct} \big\rbrace
\end{equation*}
Since $\exists \, \symURI \in \symSeedURIs : \adocFct(\symURI) = d^*$ it holds $d^* \in D_0$. Due to the initialization $\Disc{D} = D_0$ we have $d^* \in \Disc{D}$ before the first iteration.

\vspace{1ex} \noindent
\textbf{Case 2.)}
If $\neg \, \exists \, \symURI \in \symSeedURIs : \adocFct(\symURI) = d^*$, it must hold that the Web link graph for $\WoD$ contains at least one finite path $(d_0, ... \, , d_n)$ of $(\cMatch,\symBQP)$-reach\-able {\LDdoc}s $d_i$ where
i)~$\exists \, \symURI \in \symSeedURIs : \adocFct(\symURI) = d_0$
ii)~$d_n = d^*$,
and iii)~for each $i \in \lbrace 1,... \,,n \rbrace$ it holds:
\begin{equation} \label{Equation:Proof:Lemma:Proof:MatchBasedExec:Completeness1}
	%\begin{split}
	%	\exists \, t \!\in\! \dataFct(d_{i-1}) \!: \bigl( \, \exists \, \symURI \!\in\! \fctIDsName(t) : \, & \adocFct(\symURI)=d_{i} \,\,\land \\ & \cMatch(t,\symURI,\symBQP) \!=\! \true \, \bigr)
	%\end{split}
	\exists \, t \in \dataFct(d_{i-1}) : \Bigl( \, \exists \, \symURI \in \fctIDsName(t) : \bigl( \adocFct(\symURI)=d_{i} \,\land\, \cMatch(t,\symURI,\symBQP) = \true \, \bigr) \Bigr)
\end{equation}
Let $(d_0^*, ... \,, d_n^*)$ be such a path. In the following, we show by induction over $i \in \lbrace 0,... \,,n \rbrace$ that there will eventually be an iteration (during any possible execution of $ltbExec(\symSeedURIs,\symBQP,W)$) after which $\Disc{D}$ contains $d_n^* = d^*$.

\vspace{1ex} \noindent
\textit{Base case} ($i=0$): We have already shown for case 1.) that $d_0^* \in \Disc{D}$ before the first iteration in any possible execution of $ltbExec(\symSeedURIs,\symBQP,W)$.

\vspace{1ex} \noindent
\textit{Induction step} ($i>0$):
W.l.o.g., for the following discussion we assume a particular execution of $ltbExec(\symSeedURIs,\symBQP,W)$.
Our inductive hypothesis is that during this execution there will eventually be an iteration $it_j$ after which $d_{i-1}^* \in \Disc{D}$. Based on this hypothesis we show that there will be an iteration $it_{j+\delta}$ after which $d_i^* \in \Disc{D}$. We distinguish two cases: either after iteration $it_j$ it already holds $d_i^* \in \Disc{D}$ or it still holds $d_i^* \notin \Disc{D}$. We have to discuss the latter case only.

Due to (\ref{Equation:Proof:Lemma:Proof:MatchBasedExec:Completeness1}) exist $t^* \in \dataFct(d_{i-1}^*)$ and $\symURI^* \in \fctIDsName(t^*)$ such that $\adocFct(\symURI^*)=d_{i}^*$ and $\cMatch(t^*,\symURI^*,\symBQP) = \true$. Hence, there must be at least one {\TP} $tp \in \symBQP$ such that $t^*$ matches $tp$. Let $tp^* \in \symBQP$ be such a {\TP}. Since $t^*$ matches $tp^*$, there exists a partial solution $\sigma^* = ( \lbrace tp^* \rbrace, \mu^* )$ with $\mu^*[tp^*] = t^*$ and $\exists \, ?v \in \fctDom{\mu^*} : \mu^*(?v) = \symURI^*$.
After iteration $it_j$ this $\sigma^*$ has either been constructed (and added to $P$) or there exists an open {\AEtask} $( \sigma_0,t^*,tp^* )$ which will eventually be executed in some iteration $it_{j+\delta}$, resulting in the construction of $\sigma^*$. Let $it_{j'}$ be the iteration in which $\sigma^*$ has been or will be constructed. In this iteration $ltbExec(\symSeedURIs,\symBQP,W)$ expands $\mathfrak{D}$ to $\expMWD{\mu^*}{\WoD}{\mathfrak{D}}$. This expansion results in adding each $d \in \ExpMDeltaMW{\mu^*}{\WoD}$ to $\Disc{D}$ (cf.~Definition~\ref{Definition:MatchBasedExpansion}).
Since $\exists \, ?v \in \fctDom{\mu^*} : \mu^*(?v) = \symURI^*$ and $\adocFct(\symURI^*)=d_{i}^*$ it holds $d^*_i \in \ExpMDeltaMW{\mu^*}{\WoD}$. Hence, $d^*_i$ will be added to $\Disc{D}$ in iteration $it_{j'}$ (if it has not been added before).

\subsection{Proof of Assertion ii) in Lemma~\ref{Lemma:MatchBasedExec:Completeness}} \label{Proof:Lemma:MatchBasedExec:Completeness-ii}
\noindent
Let $\WoD = ( D, \dataFct, \adocFct )$ be a Web of Linked Data and let $\queryFctBSc{\symBQP}{\symSeedURIs}{\cMatch}$ be a CLD query (under $\cMatch$-semantics).
At any point in the execution of $ltbExec(\symSeedURIs,\symBQP,W)$ let $\Disc{D}$ denote the set of {\LDdoc}s in the currently discovered part $\mathfrak{D}$ of $\WoD$.

W.l.o.g., let $\sigma^*=( \symBQPpart^*, \mu^* )$ be an arbitrary partial solution for $\queryFctBSc{\symBQP}{\symSeedURIs}{\cMatch}$ in $\WoD$.
The construction of $\sigma^*$ comprises the iterative construction of a finite sequence $\bigl( \sigma_0 = (\symBQPpart_0,\mu_0), ...\, , \sigma_n = (\symBQPpart_n,\mu_n) \bigr)$ of partial solutions where i)~$\sigma_0$ is the empty partial solution (cf.~Section~\ref{Subsection:ExecModel:Constructing}), ii)~$\sigma_n=\sigma^*$, and iii)~for each $i \in \lbrace 1,... \,,n \rbrace$ it holds
\begin{align*}
	\exists \, tp \in \symBQP \setminus \symBQPpart_{i-1} : \symBQPpart_i = \symBQPpart_{i-1} \cup \lbrace tp \rbrace
	&&& \text{ and } &&&
	\mu_{i-1}[\symBQPpart_{i-1}] = \mu_{i}[\symBQPpart_{i-1}]
\end{align*}
We show by induction over $i \in \lbrace 0,...\,,n \rbrace$ that there will eventually be an iteration (during any possible execution of $ltbExec(\symSeedURIs,\symBQP,W)$) after which $\mathfrak{P}$ contains $\sigma_n=\sigma^*$.

\vspace{1ex} \noindent
\textbf{Base case} ($i=0$): Any execution of $ltbExec(\symSeedURIs,\symBQP,W)$ adds $\sigma_0$ to $\mathfrak{P}$ before it starts the first iteration (cf.~line~\ref{Line:InitP} in Algorithm~\ref{Procedure:MatchBasedExec}).

\vspace{1ex} \noindent
\textbf{Induction step} ($i>0$):
W.l.o.g., for the following discussion we assume a particular execution of $ltbExec(\symSeedURIs,\symBQP,W)$.
Our inductive hypothesis is that there will eventually be an iteration $it_j$ after which $\sigma_{i-1} \in \mathfrak{P}$. Based on this hypothesis we show that there will be an iteration $it_{j+\delta}$ after which $\sigma_i \in \mathfrak{P}$. We distinguish two cases: either after iteration $it_j$ it already holds $\sigma_i \in \mathfrak{P}$ or it still holds $\sigma_i \notin \mathfrak{P}$. We have to discuss the latter case only.

Let $tp^* \in \symBQP$ be the {\TP} for which $\symBQPpart_i = \symBQPpart_{i-1} \cup \lbrace tp^* \rbrace$. Since $\sigma_i=(\symBQPpart_i,\mu_i)$ is a partial solution for $\queryFctBSc{\symBQP}{\symSeedURIs}{\cMatch}$ in $\WoD$, it holds that $\mu_i$ is a solution for $\queryFctBSc{\symBQPpart_i}{\symSeedURIs}{\cMatch}$ in $\WoD$ and, thus, there exists a $(\cMatch,\symBQP)$-reach\-able {\LDdoc} $d^* \in D$ such that $\mu_i[tp^*]=t^* \in \dataFct(d^*)$. According to Assertion i) in Lemma~\ref{Lemma:MatchBasedExec:Completeness} there will eventually be an iteration after which $d^* \in \Disc{D}$. By then, $\sigma_i$ has either already been constructed and added to $\mathfrak{P}$ or there exists an open {\AEtask} $(\sigma_{i-1},t^*,tp^*)$. In the latter case, this task will eventually be executed, resulting in the construction and addition of $\sigma_i$.

\subsection{Proof of Lemma~\ref{Lemma:IteratorFiniteness}} \label{Proof:Lemma:IteratorFiniteness}
\noindent
Let $\symSeedURIs \subset \symAllURIs$ be a finite set of seed {\ID}s and
let $\symBQP = \lbrace tp_1 , ... \, , tp_n \rbrace$ be a BQP such that $\queryFctBSc{\symBQP}{\symSeedURIs}{\cMatch}$ is a CLD query (under $\cMatch$-semantics).
Furthermore, let $\WoD = ( D, \dataFct, \adocFct )$ be the Web of Linked Data over which $\queryFctBSc{\symBQP}{\symSeedURIs}{\cMatch}$ has to be executed using the iterator based implementation of link traversal based query execution that we introduce in~\cite{Hartig09:QueryingTheWebOfLD}.

As a preliminary for proving Lemma~\ref{Lemma:IteratorFiniteness} we
	%provide a more detailed discussion of the iterator based execution approach, including a more detailed interpretation in terms of our query execution model.
	introduce the iterator based implementation approach using the concepts and the formalism that is part of our query execution model.

For the iterator based execution of $\queryFctBSc{\symBQP}{\symSeedURIs}{\cMatch}$ over $\WoD$ we assume an order for the triple patterns in $\symBQP$; w.l.o.g.~let this order be denoted by the indices of the symbols that denote the triple patterns; i.e.~$tp_i \in \symBQP$ precedes $tp_{i+1} \in \symBQP$ for all $i \in \lbrace 1, ... \, , n-1 \rbrace$.
Accordingly, we write $P_k$ to denote the subset of $\symBQP$ that contains the first $k$ {\TP}s in the ordered $\symBQP$, that is, for all $k \in \lbrace 1, ... \, , n \rbrace$ holds $P_k=\lbrace tp_i \in \symBQP \,|\, 1 \leq i \leq k \rbrace$.

Furthermore, let $I_0, I_1 , ... \, , I_n$ be the chain of iterators used for the iterator based execution of $\queryFctBSc{\symBQP}{\symSeedURIs}{\cMatch}$ over $\WoD$. Iterator $I_0$ is a special iterator that provides a single, empty partial solution $\sigma_0$ (cf.~Section~\ref{Subsection:ExecModel:Constructing}). For all $k \in \lbrace 1, ... \, , n \rbrace$ iterator $I_k$ is responsible for triple pattern $tp_k$ from the ordered BQP. We shall see that each $I_k$ provides partial solutions $(\symBQPpart,\mu)$ (for $\queryFctBSc{\symBQP}{\symSeedURIs}{\cMatch}$ in $\WoD$) for which $\symBQPpart = \symBQPpart_k$, that is, the valuation $\mu$ of each such partial solution is a solution for CLD query $\queryFctBSc{\symBQPpart_{k}}{\symSeedURIs}{\cMatch}$ (in $\WoD$). As a consequence, for each partial solution $(\symBQPpart_n,\mu)$ provided by the last iterator $I_n$, valuation $\mu$ can be reported as a solution for $\queryFctBSc{\symBQP}{\symSeedURIs}{\cMatch}$ in $\WoD$.

During query execution all iterators access and change the (currently) discovered part $\mathfrak{D}$ of the queried Web of Linked Data $\WoD$. Before the execution, $\mathfrak{D}$ is initialized as $\mathfrak{D}_\mathsf{init}^{\symSeedURIs,\WoD}$ (cf.~Section~\ref{Subsection:ExecModel:Preliminaries}). This initialization may be performed in the \texttt{Open} function of the aforementioned special iterator $I_0$.

Algorithm~\ref{Algorithm:BlockingGetNext} presents the \texttt{GetNext} function%
\footnote{From the three versions of the iterator based implementation approach that we introduce in~\cite{Hartig09:QueryingTheWebOfLD}, Algorithm~\ref{Algorithm:BlockingGetNext} corresponds to the first, most naive version. That is, Algorithm~\ref{Algorithm:BlockingGetNext} neither applies the idea of URI prefetching nor the idea of non-blocking iterators~\cite{Hartig09:QueryingTheWebOfLD}.}
implemented by each iterator $I_k$ (for all $k \in \lbrace 1, ... \, , n \rbrace$).
In order to compute partial solutions iterator $I_k$ first consumes a partial solution $\sigma_\mathsf{pred}=(\symBQPpart_{k-1},\mu_\mathsf{pred})$ from its predecessor $I_{k-1}$ (cf.~line~\ref{Line:ConsumeIS} in Algorithm~\ref{Algorithm:BlockingGetNext}).
Lines~\ref{Line:ConstructPartialSolutions}
	%to~\ref{Line:ExpandEnd}
	and~\ref{Line:Expand}
may be understood as a (combined) performance of multiple (open) {\AEtask}s: For each {\triple} $t^*$ that i)~is contained in the data of all {\LDdoc}s discovered so far and that ii)~matches {\TP} $tp_k' = \mu_\mathsf{pred}[tp_k]$, iterator $I_k$ adds a partial solution $\sigma_{t^*}$ to $M_k$; each $\sigma_{t^*}$ is the $(t^*,tp_k')$-augmentation of $\sigma_\mathsf{pred}$ in $\mathfrak{D}$ (cf.~line~\ref{Line:ConstructPartialSolutions}). Due to the construction of $tp_k'$ from $tp_k$ (cf.~line~\ref{Line:ApplicationOfPrevMu}), any {\triple} $t^*$ that matches $tp_k'$ also matches $tp_k$ and, thus, each $\sigma_{t^*}$ is also the $(t^*,tp_k)$-augmentation of $\sigma_\mathsf{pred}$ (in $\mathfrak{D}$).
After populating $M_k$, iterator $I_k$ uses all $\sigma_{t^*} \in M_k$ to expand the currently discovered part of $\WoD$ incrementally (cf.~line~\ref{Line:Expand}). Hence, lines~\ref{Line:ConstructPartialSolutions}
	%to~\ref{Line:ExpandEnd}
	and~\ref{Line:Expand}
may be understood as a (combined) performance of all those (open) {\AEtask}s $\bigl( \sigma,t,tp \bigr)$ for which $\sigma = \sigma_\mathsf{pred}$, $tp = tp_k$, and $t^*$ is a {\triple} that
	%i)~is contained in the data of all {\LDdoc}s discovered so far and that ii)~matches {\TP} $tp_k' = \mu_\mathsf{pred}[tp_k]$.
	has the aforementioned properties.
Due to the finiteness of $\mathfrak{D}$ (cf.~Definition~\ref{Definition:DiscoveredPart} and Proposition~\ref{Proposition:MatchBasedExpansionClosedness}) there is only a finite number of such {\triple}s $t^*$ and, thus, the number of {\AEtask}s iterator $I_k$ performs for $\sigma_\mathsf{pred}$ is also finite. As a consequence, to prove Lemma~\ref{Lemma:IteratorFiniteness} it suffices to show that each iterator $I_k$ only consumes a finite number of partial solutions from its predecessor $I_{k-1}$. Hence, it suffices to prove the following lemma.

\begin{algorithm}[t]
	\caption{ ~ \texttt{GetNext} function for iterator $I_k$ in our iterator based implementation of link traversal based query execution~\cite{Hartig09:QueryingTheWebOfLD}.}
\label{Algorithm:BlockingGetNext}
\begin{algorithmic}[1]
	\REQUIRE ~
		\par-- a triple pattern $tp_k$;
		\par-- a predecessor iterator $I_{k-1}$; % that provides valuations which are solutions for CLD query $\queryFctBSc{\symBQPpart_{k-1}}{\symSeedURIs}{\cMatch}$ where $\symBQPpart_k=\lbrace tp_1 , ... \, , tp_{k-1} \rbrace$;
		\par-- the currently discovered part $\mathfrak{D}$ of the queried Web of Linked Data $\WoD$ (note, all iterators have access to $\mathfrak{D}$);
		\par-- an initially empty set $M_k$ that allows the iterator to keep (precomputed) partial solutions between calls of this \texttt{GetNext} function
		\medskip
		\WHILE {$M_k = \emptyset$}
			\STATE $\sigma_\mathsf{pred} := I_{k-1}$.\texttt{GetNext} \label{Line:ConsumeIS} \hspace{56mm} \COMMENT{consume partial solution from direct predecessor $I_{k-1}$}

			\IF {$\sigma_\mathsf{pred}$ = \textsc{\scriptsize EndOfFile}}
				\RETURN \textsc{\scriptsize EndOfFile}
			\ENDIF

			\medskip
			\STATE $tp_k' := \mu_\mathsf{pred}[tp_k]$ \label{Line:ApplicationOfPrevMu} \hspace{66mm} \COMMENT{$\mu_\mathsf{pred}$ is the valuation in $\sigma_\mathsf{pred} = (\symBQPpart_\mathsf{pred},\mu_\mathsf{pred})$}

			\medskip
% 			\STATE $T_k := \big\lbrace \, ( \sigma_\mathsf{pred}, t^*, tp_k ) \,\big|\, t^* \text{ matches } tp_k' \text{ and } t^* \in \fctAllDataName(\mathfrak{D}) \, \big\rbrace$ \label{Line:GenerateOpenTasks} \hspace{1mm} \COMMENT{generate open {\AEtask}s}
% 			\STATE $M_k := \big\lbrace \, \Aug{t}{tp}{\sigma}{\mathfrak{D}} \,\big|\, (\sigma,t,tp) \in T_k \, \big\rbrace$ \label{Line:ConstructPartialSolutions} \hspace{1mm} \COMMENT{construct partial solutions for all {\AEtask}s in $T_k$}

			\STATE $M_k := \big\lbrace \, \Aug{t^*}{tp_k'}{\sigma_\mathsf{pred}}{\mathfrak{D}} \,\big|\, t^* \text{ matches } tp_k' \text{ and } t^* \in \fctAllDataName(\mathfrak{D}) \, \big\rbrace$ \label{Line:ConstructPartialSolutions} \hspace{1mm} \COMMENT{construct partial solutions}

% 			\FORALL { $\sigma' = ( \symBQPpart' \!, \mu' ) \in M_k$ } \label{Line:ExpandBegin}
% 				\STATE $\mathfrak{D} := \expMWD{\mu'}{\WoD}{\mathfrak{D}}$
% 			\ENDFOR \label{Line:ExpandEnd}
			\STATE \textbf{for all } $\sigma' = ( \symBQPpart' \!, \mu' ) \in M_k$ \textbf{ do } $\mathfrak{D} := \expMWD{\mu'}{\WoD}{\mathfrak{D}}$ \textbf{ end for} \label{Line:Expand} \hspace{13.5mm} \COMMENT{expand $\mathfrak{D}$ using all newly constructed partial solutions}

			\medskip
		\ENDWHILE

		\medskip
		\STATE $\sigma' := $ an element in $M_k$ \label{Line:ObtainPartialSolution}
		\STATE $M_k := M_k \setminus \lbrace \sigma' \rbrace$
		\RETURN $\sigma'$ \label{Line:Return}
	\end{algorithmic}
\end{algorithm}

\begin{lemma} \label{SubLemma:Proof:Lemma:IteratorFiniteness}
	The overall number of partial solutions provided by each iterator via its \texttt{GetNext} function is finite.
\end{lemma}

\begin{myproof}{of Lemma~\ref{SubLemma:Proof:Lemma:IteratorFiniteness}}
	We prove the lemma by induction over the chain of iterators $I_0, I_1 , ... \, , I_n$.

	\vspace{1ex} \noindent
	\textit{Base case ($I_0$):} The special iterator provides a single partial solution $\sigma_0$.

	\vspace{1ex} \noindent
	\textit{Induction step ($I_k$ for $k \in \lbrace 1, ... \, , n \rbrace$):} Our inductive hypothesis is that iterator $I_{k-1}$ provides a finite number of partial solutions via its \texttt{GetNext} function. Based on this hypothesis we show that iterator $I_{k}$ provides a finite number of partial solutions via its \texttt{GetNext} function. Due to our inductive hypothesis it is sufficient to show that for each partial solution which $I_k$ consumes from $I_{k-1}$, $I_k$ provides a finite number of partial solutions.
	Let $\sigma_\mathsf{pred} = (\symBQPpart_\mathsf{pred},\mu_\mathsf{pred})$ be such a partial solution that $I_k$ consumes from $I_{k-1}$ (line~\ref{Line:ConsumeIS} in Algorithm~\ref{Algorithm:BlockingGetNext}). $I_k$ applies $\mu_\mathsf{pred}$ to its triple pattern $tp_k$ (line~\ref{Line:ApplicationOfPrevMu}) and uses the resulting triple pattern $tp_k' = \mu_\mathsf{pred}[tp_k]$ to generate set $M_k$ (line~\ref{Line:ConstructPartialSolutions}). Hence, this set contains exactly those partial solutions that $I_k$ provides based on $\sigma_\mathsf{pred}$ (lines~\ref{Line:ObtainPartialSolution} to~\ref{Line:Return}). However, $M_k$ is finite because $I_k$ generates $M_k$ on a particular snapshot of the discovered part $\mathfrak{D}$ of $\WoD$ and $\mathfrak{D}$ is finite at any point during query execution.
	\hfill \,
\end{myproof}

\subsection{Proof of Theorem~\ref{Theorem:Iterator}}
\noindent
The guarantee for termination is a direct consequence of Lemma~\ref{Lemma:IteratorFiniteness}. The whole chain of iterators performs a finite number of {\AEtask}s only. The performance of each {\AEtask} terminates because all operations in Algorithm~\ref{Algorithm:BlockingGetNext} are synchronized and are guaranteed to terminate.

It remains to show that the set of valuations reported by any iterator based execution is always a finite subset of the corresponding query result:
Let $\symSeedURIs \subset \symAllURIs$ be a finite set of seed {\ID}s and
let $\symBQP = \lbrace tp_1 , ... \, , tp_n \rbrace$ be a BQP such that $\queryFctBSc{\symBQP}{\symSeedURIs}{\cMatch}$ is a CLD query (under $\cMatch$-semantics).
Furthermore, let $\WoD = ( D, \dataFct, \adocFct )$ be the Web of Linked Data over which $\queryFctBSc{\symBQP}{\symSeedURIs}{\cMatch}$ has to be executed.

For the iterator based execution of $\queryFctBSc{\symBQP}{\symSeedURIs}{\cMatch}$ over $\WoD$ we assume an order for the triple patterns in $\symBQP$; w.l.o.g.~let this order be denoted by the indices of the symbols that denote the triple patterns; i.e.~$tp_i \in \symBQP$ precedes $tp_{i+1} \in \symBQP$ for all $i \in \lbrace 1, ... \, , n-1 \rbrace$.
Furthermore, let $I_0, I_1 , ... \, , I_n$ be the chain of iterators as introduced in the proof for Lemma~\ref{Lemma:IteratorFiniteness} (cf.~Section~\ref{Proof:Lemma:IteratorFiniteness}).

% In Section~\ref{Proof:Lemma:IteratorFiniteness} we interpret Algorithm~\ref{Algorithm:BlockingGetNext} in terms of our query execution model and show that each valuation reported by an iterator may be understood as a valuation that is part of a partial solution constructed by the iterator. Hence, based on the same proof idea with which we show the soundness result in Theorem~\ref{Theorem:MatchBasedExec} (cf.~Section~\ref{Proof:Theorem:MatchBasedExec}), we may show that each valuation reported by iterator $I_n$ is a solution for $\queryFctBSc{\symBQP}{\symSeedURIs}{\cMatch}$ over $\WoD$.

For our proof we use the following lemma.
\begin{lemma} \label{SubLemma:Proof:Theorem:IteratorIncompleteness}
	For any partial solution $\sigma=(\symBQPpart,\mu)$ provided by the \texttt{GetNext} function of iterator $I_n$ holds $\symBQPpart=\symBQPpart_n$.
\end{lemma}
\begin{myproof}{of Lemma~\ref{SubLemma:Proof:Theorem:IteratorIncompleteness}}
	We prove the lemma by induction over the chain of iterators $I_0, I_1 , ... \, , I_n$.

	\vspace{1ex} \noindent
	\textit{Base case ($I_0$):} The special iterator provides a single partial solution $\sigma_0 = (\symBQPpart_0,\mu_0)$ which covers the empty part $\symBQPpart_0 = \emptyset$ of $\symBQP$.

	\vspace{1ex} \noindent
	\textit{Induction step ($I_k$ for $k \in \lbrace 1, ... \, , n \rbrace$):} Our inductive hypothesis is that for any partial solution $\sigma=(\symBQPpart,\mu)$ provided by the \texttt{GetNext} function of iterator $I_{k-1}$ holds $\symBQPpart=\symBQPpart_{k-1}$. Based on this hypothesis we show that for any partial solution $\sigma'=(\symBQPpart',\mu')$ provided by the \texttt{GetNext} function of iterator $I_{k}$ holds $\symBQPpart'=\symBQPpart_{k}$. However, this is easily checked in Algorithm~\ref{Algorithm:BlockingGetNext}: As we discuss in Section~\ref{Proof:Lemma:IteratorFiniteness}, each partial solution $\sigma'=(\symBQPpart',\mu')$ added to (any $\sigma_\mathsf{pred}$-specific version of) $M_k$ (cf.~line~\ref{Line:ConstructPartialSolutions}) and returned later (cf.~line~\ref{Line:Return}) is a $(t^*,tp_k)$-augmentation of some partial solution $\sigma_\mathsf{pred}=(\symBQPpart_\mathsf{pred},\mu_\mathsf{pred})$ consumed from $I_{k-1}$. According to our inductive hypothesis $\symBQPpart_\mathsf{pred} = \symBQPpart_{k-1}$. Therefore, $\symBQPpart' = \symBQPpart_{k-1} \cup \lbrace tp_k \rbrace = \symBQPpart_{k}$ (cf.~Definition~\ref{Definition:PartSolAugmentation}.
	\hfill \,
\end{myproof}

\noindent
Lemma~\ref{SubLemma:Proof:Theorem:IteratorIncompleteness} shows that each partial solution $(\symBQPpart_n,\mu)$ computed by the last iterator of the chain of iterators covers the whole BQP of the executed CLD query $\queryFctBSc{\symBQP}{\symSeedURIs}{\cMatch}$ (recall, $\symBQP = P_n$). Hence, each valuation $\mu$ that the iterator based execution reports from such a partial solution $(\symBQPpart_n,\mu)$, is a solution for $\queryFctBSc{\symBQP}{\symSeedURIs}{\cMatch}$ over $\WoD$.

It remains to show that the iterator based execution may always only report a finite number of such solutions. This result, however, is a direct consequence of Lemma~\ref{SubLemma:Proof:Lemma:IteratorFiniteness} (cf.~Section~\ref{Proof:Lemma:IteratorFiniteness}).

\end{document}